\documentclass[%
 reprint,
 nofootinbib,
 amsmath,amssymb,
 aps,
]{revtex4-2}

\usepackage{graphicx}
\usepackage{dcolumn}
\usepackage{bm}
\usepackage[colorlinks=true,citecolor=blue,urlcolor=blue,linkcolor=red]{hyperref}
\usepackage{physics}
\usepackage{siunitx}
\usepackage{pgf}
\usepackage[normalem]{ulem}
\usepackage{xspace}

\newcommand\Xe{X_{\rm e}}
\newcommand\zmin{z_{\rm min}}
\newcommand\zmax{z_{\rm max}}

\newcommand\me{m_{\rm e}}
\newcommand\mfp{\ell_{\rm mfp}}

\newcommand{\changed}[1]{{\textcolor{black}{#1}}}

\newcommand{\planck}{{\it Planck}}
\newcommand{\lcdm}{\mbox{$\Lambda$CDM}}
\newcommand{\modrec}{\texttt{ModRec}}
\newcommand{\sptthreeg}{SPT-3G}

\newcommand{\mnras}{Mon. Not. Roy. Astron. Soc.}
\newcommand{\apjl}{Astrophys. J. Lett.}
\newcommand{\physrep}{Physical Reports}
\newcommand{\baas}{Bull. Am. Astron. Soc.}
\DeclareUnicodeCharacter{2212}{-}

\newcommand{\shoes}{SH0ES\xspace}

\makeatletter
\newcommand\thefontsize[1]{{#1 The current font size is: \f@size pt\par}}
\makeatother

\begin{document}

\preprint{APS/123-QED}

\title{Reconstructing the recombination history by combining \\early and late cosmological probes}

\author{Gabriel P. Lynch$^1$ }
\email{gplynch@ucdavis.edu}
\author{Lloyd Knox$^1$}
\author{Jens Chluba$^2$}

\affiliation{
$^1$Department of Physics and Astronomy, University of California, Davis, CA, USA 95616
\\
$^2$Jodrell Bank Centre for Astrophysics, University of Manchester, Manchester M13 9PL, UK
}

\date{\today}

\begin{abstract}

We develop and apply a new framework for reconstructing the ionization history during the epoch of recombination with combinations of cosmic microwave background (CMB), baryon acoustic oscillation (BAO) and supernova data. We find a wide range of ionization histories that are consistent with current CMB data, and also that cosmological parameter constraints are significantly weakened once freedom in recombination is introduced. BAO data partially break the degeneracy between cosmological parameters and the recombination model, and are therefore important in these reconstructions. The 95\% confidence upper limits on $H_0$ are 80.1 (70.7) km/s/Mpc given CMB (CMB+BAO) data, assuming no other changes are made to the standard cosmological model. Including Cepheid-calibrated supernova data in the analysis drives a preference for non-standard recombination histories with visibility functions that peak early and exhibit appreciable skewness. Forthcoming measurements from SPT-3G will reduce the uncertainties in our reconstructions by about a factor of two.

\end{abstract}

\maketitle

\section{\label{sec:intro}Introduction}
Precision measurements of anisotropies in the cosmic microwave background (CMB) have enabled cosmologists to place increasingly tight constraints on the energy content of the universe, the spectrum of primordial fluctuations, and the optical depth to Thomson scattering in the reionized intergalactic medium, with many of these determinations reaching percent level uncertainties \citep{Planck:2018vyg}. Underpinning the analysis of these measurements is a detailed theoretical understanding of the cosmological recombination process, as the temperature and polarization power spectra of the CMB are highly sensitive to the physical conditions present in the primordial plasma during this epoch \citep{Hu1995, Chluba:2005uz, Lewis2006, Jose2010}.

The sensitivity of the data to details of recombination also provides us with an opportunity: we can reconstruct those details from the data themselves. In this paper, we use CMB, baryon acoustic oscillation (BAO) and supernova data to reconstruct the history of the ionization fraction, $\Xe(z)$, over a redshift range spanning the epoch of hydrogen recombination. We work out the uncertainties in such a reconstruction and examine its consistency with the \lcdm\ predictions. We also examine the impact of freedom in $\Xe(z)$ on cosmological parameter inferences, and the Hubble constant in particular.  

Responses of CMB angular power spectra to changes in recombination have previously been exploited to constrain elements of the recombination process. For example, the \planck\ collaboration used measurements of the temperature and polarization power spectra to constrain the atomic $2s \to 1s$ two-photon transition with 7\% uncertainty \cite{Planck:2015fie}. Additionally, phenomenological modifications to the recombination history have been constrained under the assumption that the angular power spectra respond linearly to changes in $\Xe(z)$, what we will refer to henceforth as a linear response approximation, or LRA \citep{Farhang:2011pt, Planck:2018vyg, Hart:2019gvj}. This sensitivity has also been used to constrain physical models with energy injection into the primordial plasma \citep{Adams:1998nr, Chen2004, Galli:2013dna, Slatyer:2016qyl, Finkbeiner:2011dx}, varying fundamental constants \citep{Planck:2014ylh, Hart:2017ndk, Hart2020H0, Sekiguchi:2020teg}, and small-scale clumping of baryons during recombination \citep{Jedamzik2020Relieving, Thiele2021, Rashkovetskyi:2021rwg, Galli:2021mxk}. The CMB is sensitive to all of these models primarily through the changes induced in the recombination process around $z\simeq 10^3$.

In our reconstruction of the ionization history, we avoid employing an LRA as we have found that nonlinear responses have significant impact on the reconstructions. However, an LRA does enable a motivated form of parameter space dimension reduction in the form of a principal component analysis (PCA) of recombination perturbations \citep{Farhang:2011pt, Farhang:2012jz, Hart:2019gvj}. Without this reduction traditional Markov Chain Monte Carlo (MCMC) methods can be prohibitively slow in exploring the extended parameter space of cosmological and modified recombination parameters. \citet{Lee:2022gzh} similarly introduce an LRA to explore modified recombination histories, also motivated by a desire to avoid computationally prohibitive MCMC inference. To render such an analysis tractable without an LRA, we turn to the use of an Einstein-Boltzmann solver emulator \citep{Nygaard:2022wri} to quickly provide theoretical predictions for our model, significantly accelerating the inference process covering non-perturbative modifications. 

Whether or not we use an LRA impacts the inferred values and uncertainties of cosmological parameters. PCA-based approaches to perturbatively modified recombination, which make this approximation, generally find little to no shift in parameter values and negligible loosening of constraints, even when up to four eigenmodes are retained \citep{Hart:2019gvj, Planck:2018vyg}. \changed{However, it should be remarked that a PCA has not been applied directly to the combination of CMB and Cepheid-calibrated supernovae data, and doing so might hint towards a modified recombination history.}

Conversely, studies of specific physical models with a modified recombination epoch, such as models containing time-varying fundamental constants \citep{Hart2020H0, Sekiguchi:2020teg}, do see significant shifts in cosmological parameters with possible implications for the Hubble tension. These models make specific assumptions about how recombination differs from the standard scenario, and thus allow for a full treatment of CMB angular power spectral responses without relying on approximations. Our analysis aims to combine the advantages inherent in each approach. Our phenomenological model frees us from making specific physical-model-dependent assumptions about how recombination is modified, and our abandonment of a LRA allows for a full treatment of CMB responses to these modifications. We find, in contrast to previous phenomenological approaches, that parameter uncertainties are significantly increased once $\Xe(z)$ is varied.

Our reconstruction of $\Xe(z)$ and the associated uncertainty allows us to explore modified recombination histories that are nonetheless consistent with the data. We have found that including BAO data is helpful in this reconstruction, and we discuss why. The importance of BAO data in constraining modified recombination has previously been recognized \citep{Planck:2015fie, Chiang:2018xpn, Hart2020H0, Lee:2022gzh}, and our work reinforces this conclusion even when one moves away from specific physical models and simplifying approximations. 

We pay special attention to the possibility that a non-standard $\Xe(z)$ could resolve the Hubble tension (see \citep{Verde:2023lmm} for a recent review). This possibility was first illustrated by \citet{Chiang:2018xpn} by parameterizing the position and width of the visibility function. Due to their modifications of the recombination history, models with a time-varying electron mass \citep{Hart2020H0, Sekiguchi:2020teg, Lee:2022gzh} were also found to provide sufficient freedom to reduce the Hubble tension, as were models with early-structure formation due to primordial magnetic fields (PMFs) \cite{Jedamzik2020Relieving, Thiele2021, Galli:2021mxk, Rashkovetskyi:2021rwg, Lucca:2023cdl}. The attempt by \citet{Cyr-Racine:2021oal} to solve the $H_0$ problem by exploiting a symmetry of the Einstein-Boltzmann equations and initial conditions also relies on changes to the ionization history. 

This work is organized as follows. In Sec.~\ref{sec:Methodology} we introduce our parameterization of $\Xe(z)$ and discuss the priors on the function space to be explored. We also present our emulator design choices and validation. In Sec.~\ref{sec:constraints} we present our reconstructions of $\Xe(z)$ conditioned on different data sets and discuss the constraints these imply on modifications to recombination. In Sec.~\ref{sec:Hubble} we use these reconstructions to explore recombination histories which either fully or partially alleviate the Hubble tension. In Sec.~\ref{sec:Forecasts} we discuss the sensitivity of \sptthreeg{} to signals from the recombination era, and the impact that upcoming measurements will have on both the precision of recombination reconstruction and the status of modified recombination as a solution to the Hubble tension. We conclude in Sec.~\ref{sec:conclusion}.

\vspace{-3mm}\section{\label{sec:Methodology}Methodology}
\vspace{-2mm}
Estimating the ionization fraction $\Xe(z)$ from data is a problem of functional inference in which one is attempting to infer an infinite number of parameters, in this case the values of $\Xe(z)$ over a given redshift range. In practice it is necessary to parameterize the problem so that only a finite number of parameters are inferred, and the rest of the function is constructed according to a given prescription, such as via interpolation or as a linear combination of basis functions. This choice of parameterization is an implicit prior on the functional space explored in the analysis and restricts the possible functional forms $\Xe(z)$ can take. In general, a higher dimensional parameterization allows for a more general functional space to be explored. This poses a problem for traditional methods of inference such as random walk Markov Chain Monte Carlo (MCMC) methods, which are inefficient in correlated, high-dimensional parameter spaces and generally leads to vastly increased errors given the finite information that is available.

A common method for addressing this trade-off uses a linear response approximation to either reduce the dimension of the parameter space \citep{Finkbeiner:2011dx, Farhang:2012jz, Hart:2019gvj} or to facilitate quasi-analytic approaches \citep{Lee:2022gzh}. In the former case, by using a LRA it is possible to use a linear (Fisher) approximation to the full likelihood in a PCA to identify combinations of basis functions which, through their correlated effect on CMB observables, are most-constrained by the data. This enables a motivated form of dimensional reduction in which only the amplitudes of these most constrained modes are retained as new parameters. In the latter case, the problem is linearized in order to analytically compute changes in $\chi^2$ as a smooth function is varied, avoiding a costly MCMC. It is important to recognize that a LRA is only accurate in the limit of small perturbations, and one is vulnerable to inconsistent results outside of this regime. 

One option to accurately include larger perturbations is to iteratively solve the inference problem using a LRA and then update the model around a new fiducial cosmology \citep[see Sec.~4.4 of][]{Farhang:2011pt}. However, this is not guaranteed to converge. To avoid the need for a LRA in our analysis, we instead accelerate the inference process using an emulator. The remainder of this section discusses the details of our parameterization of $\Xe(z)$, as well as our choices in emulator and analysis design.

\subsection{Details of the modified recombination model}
The ionization fraction is the ratio of the number density of free electrons, $n_{\rm e}(z)$, to the number density of hydrogen nuclei, $n_\mathrm{H}(z) = n_{\mathrm{HI}}(z) + n_{\mathrm{HII}}(z)$; i.e.,
\begin{equation}
    \Xe(z) \equiv \frac{n_e(z)}{n_\mathrm{H}(z)}.
\end{equation}
This quantity can be calculated using recombination codes such as \texttt{CosmoRec}\citep{Chluba:2010ca} and \texttt{HyRec} \citep{2011PhRvD..83d3513A}, as is commonly done during calculations of CMB power spectra using Einstein-Boltzmann solvers.  We model departures from the standard ionization fraction via the introduction of $\Delta \Xe(z) = \Xe(z) - \Xe^{\mathrm{std}}(z)$, where $\Xe^{\mathrm{std}}(z)$ is the standard ionization fraction as computed from the above codes for a given set of cosmological parameters. The dependence of $\Xe^{\mathrm{std}}(z)$ on the standard cosmological parameters is small, varying at most by about 2\% at redshifts below $z \approx 800$ among models in the \planck\ posterior assuming \lcdm\, and by less at higher redshifts. We parameterize $\Delta \Xe (z)$ via control points with an amplitude $\tilde{q}_i$ placed at pivot redshifts $z_i$, and use a cubic spline interpolation to fill in the rest of the function.  

The definition of $\Xe(z)$ given above does not include helium nuclei in the denominator. As such, $\Xe(z)$ can exceed 1, as is the case prior to helium recombination. Neglecting the possibility of ionizing HeII, the maximum value of the ionization fraction is given by
\begin{equation}
    \Xe^{\rm max} = 1 + \frac{Y_{\rm p} }{2}(1-Y_{\rm p})
\end{equation}
where $Y_{\rm p}$ is the primordial helium abundance. For the standard big bang nucleosynthesis value of $Y_p = 0.24$, $\Xe^{\text{\rm max}} \approx 1.09$. 

This means that any physical $\Delta \Xe(z)$ must fall within the following bounds:
\begin{equation} \label{eqn:dxe_bounds}
\Delta \Xe(z) \in \left[-\Xe^{\text{std}}(z), \Xe^{\text{\rm max}} - \Xe^{\text{std}}(z)\right] 
\end{equation}
It is not enough to ensure the $\tilde{q}_i$ parameters obey this restriction, as interpolants between control points can overshoot the values of the control points themselves. This is if often the case with cubic spline interpolation, although it could be remedied with other interpolation schemes. Instead, to enforce this restriction at all $z$ and not just at the control points, we first map the control points $\tilde{q}_i$ from the closed interval given by Eq. \ref{eqn:dxe_bounds} to values $q_i$ in an unbounded interval $(-\infty, \infty)$. We perform the interpolation between the $q_i$ values to obtain a function $f(z)$, and use the inverse mapping at all $z$ to obtain $\Delta \Xe(z)$ subject to the electron conservation restriction. The explicit $\tilde{q}_i \to q_i$ mapping we use is

\begin{equation*}
    q_i = \mathrm{logit}\left(\frac{\tilde{q}_i + \Xe^{\rm std}(z_i)}{\Xe^{\rm max}}\right) - p(z_i) 
\end{equation*}
with inverse mapping
\begin{equation}
    \label{eqn:transformation}
    \Delta \Xe (z) = \Xe^{\text{\rm max}}\mathrm{expit}\left(f(z) + p(z)\right) - \Xe^{\text{std}}(z)
\end{equation}
where
\begin{equation*}
p(z) \equiv \text{logit}\left( \frac{\Xe^{\text{std}}(z) }{\Xe^{\text{\rm max}}} \right).
\end{equation*}
We use the logistic sigmoid map $\mathrm{expit}(x)$ and its inverse $\mathrm{logit}(x)$
\begin{equation*}
    \text{expit}(x) = \frac{1}{1 + \exp(-x)} \qand \mathrm{logit}(x) = \ln\left(\frac{x}{1-x} \right)
\end{equation*}
due to their simple mapping between closed and unbounded intervals. The choice of $p$ is guided by a desire to have $\Delta \Xe(z) = 0$ if $f(z) = 0$.  

\begin{figure}[t]
    \centering
    \includegraphics{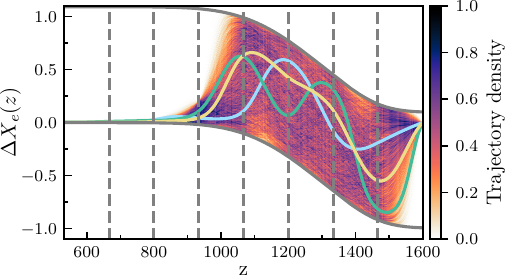}
    \caption{\label{fig:prior_delta_xe} Samples from the prior distribution for $\Delta \Xe(z)$ used in this work. The colors in the shaded region denote the density of $\Delta \Xe(z)$ curves in the sample, normalized so that the densest point at each $z$ has a value of 1. Three individual functions are presented to demonstrate the variety of deviations possible. The solid lines indicate the maximum and minimum deviations allowed by electron conservation. Dashed vertical lines indicate the pivot redshifts. }
\end{figure}

\changed{\subsection{Choice of control points}}

Our choice of how many control points and their placement constitutes a prior on the functional space of possible ionization histories. We use 7 control points placed evenly between $\zmin = 533$ and $\zmax = 1600$ exclusive, and for $z \leq \zmin$ and $z \geq \zmax$, $\Delta \Xe(z) = 0$. One implication of this is that in our model, helium recombination, which ends at $z\simeq 1700$ \citep{Switzer2007I, Kholupenko2007, Rubino-Martin:2007tua}, is left untouched, as are the freeze-out residual ionization fraction and the reionization history.  We chose these values for $\zmin$ and $\zmax$ in order to allow for deviations over the entire redshift range for which the standard hydrogen recombination visibility function has substantial support, which is where the effect on CMB anisotropies is expected to be the most prominent. 

After constructing a $\Delta \Xe(z)$, we add it to the standard $\Xe^{\rm std}(z)$ for a given model, and propagate this change through a full solution of the Einstein-Boltzmann equations governing the evolution of CMB anisotropies. We perform these calculations using a modified version of \texttt{CLASS} \cite{Blas:2011rf}.

In Fig.~\ref{fig:prior_delta_xe}, we show some possible $\Delta \Xe(z)$ functional forms attainable with this model, drawn from the prior distribution outlined in Table~\ref{tab:priors}. Vertical lines indicate the redshifts at which we place control points, and the solid gray lines show $\Delta \Xe^{\rm min}(z)$ and $\Delta \Xe^{\rm max}(z)$ from Eq.~\ref{eqn:dxe_bounds}. To illustrate the wide range of ionization histories possible within this model, we draw from the prior for the $\tilde{q}_i$ parameters 5000 times, and show the density of trajectories through any given point using shading. We choose a narrower prior for $\tilde{q}_1, \tilde{q}_2$ and $\tilde{q}_3$ than for the higher redshift control points. We found that large variations in this region often caused \texttt{CLASS} to fail, so excluding such extreme variations with our priors aided the emulator training process discussed in the following section. Additionally, this is the redshift range that the angular power spectra are most sensitive to, and therefore we expect such extreme variations to be excluded by the data. This narrower prior still allows for a wide range of ionization histories, and as we see in Sec.~\ref{sec:constraints} these variations are further constrained by data.

\begin{figure}
    \centering
    \includegraphics{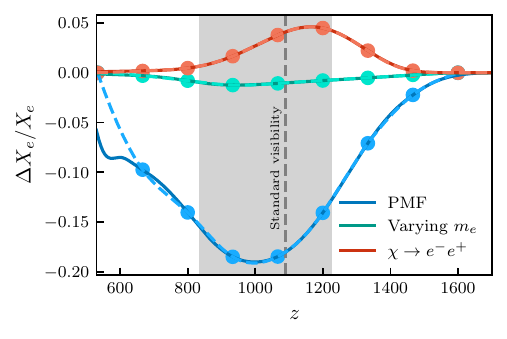}
    \caption{\label{fig:approximate_physical} Approximations to physically-arising $\Delta \Xe(z)$'s within our model of modified recombination. Solid lines indicate the target $\Delta \Xe(z)$ and dashed are approximations with 7 control points. The gray band indicates the 95\% probability interval for the standard visibility function --- this is the region where poor approximations have the most impact on the resulting $\chi^2$ once changes in recombination are propagated through to observables. The PMF scenario mimics the treatment of \citet{Galli:2021mxk}, whereas the varying $\me$ and decaying dark matter ionization fractions are computed using \texttt{CLASS}.}
\end{figure}

The choices defined above restrict the space of possible ionization histories obtainable in our phenomenological model. Ideally, the allowed space of ionization histories includes a sufficiently good approximation to any that might arise in some physical model. To assess our model for this ability to reproduce ionization histories from physical models, we approximate several such $\Delta \Xe(z)$'s with 7 control points and compare the $\chi^2$ that results once those ionization histories are propagated through to observables. For this comparison we chose a model containing primordial magnetic fields \citep{Galli:2021mxk}, one with time varying electron mass, and one with dark matter decay to electron/positron pairs. In each case, we have chosen parameter values which are not ruled out by existing data. For the varying electron mass model we use a value of $\Delta m_e / m_e = 10^{-3}$, which is consistent with \planck\ 2018 data \citep{Hart2020H0}. For the decaying dark matter we use a lifetime of $\tau = 10^{12} \text{ s}$ and decay fraction $f_{\rm frac} = 8.6 \cdot 10^{-11}$, which are below the 95\% confidence exclusion limit from a combination of  \planck\ and FIRAS data \citep{Lucca:2019rxf}. Finally, for the PMF model we adopt a clumping factor of $b=0.5$, which corresponds to the 95\% upper limit from \planck\ data \cite{Galli:2021mxk}.

For each comparison, we first obtain $\Delta \Xe(z)$ from the physical model. These models introduce other effects beyond changes in $\Xe(z)$ which can affect the final $C_\ell$'s, which we do not want to consider here. To remove these effects in the comparison, we mimic the $\Delta \Xe(z)$ from the physical model with 50 control points evenly placed between $\zmin=500$ and $\zmax=3000$ to closely trace the target functional shape. We insert this target $\Delta \Xe(z)$ into \texttt{CLASS} to compute the power spectra resulting from this recombination history, absent other physical effects from the models. We do the same for our approximation with 7 control points between $\zmin = 533$ and $\zmax = 1600$, and use the change in quality of fit $\Delta \chi^2$ to \planck\ 2018 TTTEEE+low-$\ell$ TT+low-$\ell$ EE to assess the quality of our approximation. The cosmological parameters are the same for all models, corresponding to the \planck\ 2018 mean parameters, which we hereafter refer to as the fiducial cosmology. The target and approximate $\Delta \Xe(z)$'s are shown in Fig.~\ref{fig:approximate_physical}.

For all of these models,  $\Delta \chi^2 < 1$, and additionally the differences between the target $\Delta X_e(z)$ and our approximation to it are significantly smaller than the reconstruction uncertainties presented in Sec.~\ref{sec:constraints}. We conclude from this assessment that sufficiently accurate approximations to realistic $\Delta \Xe$'s are within our prior function space.

To summarize, our model of modified recombination has seven control points placed at $z_i \in \{666, 800, 933, 1066, 1200,1333, 1466\}$, and outside of $\zmin = 533$ and $\zmax=1600$ the ionization fraction has its standard value. In total our model has 13 free parameters once the standard 6 \lcdm\ parameters $\{\omega_b, \omega_{\rm cdm}, \tau_{\rm reio}, n_s, A_s, H_0\}$ are included. We also include one species of massive neutrino with $m_\nu = 0.06$ eV. We refer to this model throughout as the ``\modrec" model. 

\subsection{Accelerated inference using an emulator}
Parameter inference using Markov-Chain Monte Carlo with the Metropolis-Hastings algorithm is a popular choice to infer cosmological parameters from CMB data. However, it is difficult to find an efficient generating function for the Metropolis-Hastings proposal step in higher dimensional parameter spaces, as in general posteriors become more non-Gaussian as more degeneracy directions open up, along with the geometric volume effects of increasing dimension. This results in inefficient sampling as many proposed steps are rejected, each of which requires the evaluation of the likelihood function. Both the evaluation of the likelihood itself as well as the required theoretical calculation to do so can be computationally expensive, especially as the latter involves (in the context of constraints using CMB data) the computation of angular power spectra $C^{X}_\ell$ for the candidate set of parameters. Depending on the data sets used, $X \in \{TT, TE, EE, \phi \phi \}$. This presents a considerable computational burden: a typical calculation of these quantities using \texttt{CLASS} at the necessary precision up to $\ell_{\rm max} = 5000$ takes $\mathcal{O}(1\text{ min})$ on a single CPU. This $\ell_{\rm max}$ is necessary if we wish to enable forecasting for future CMB experiments which will measure these small angular scales. 

We accelerate this inference process by creating an {\em emulator} of the Einstein-Boltzmann solver. An ideal emulator takes a given input and reproduces the output that would come from an Einstein-Boltzmann solver, and does so much more rapidly. This approach has gained popularity as a method to accelerate the analysis of cosmological data. There now exist many emulators such as \citep{Bolliet:2023sst, Mootoovaloo:2021rot, Arico:2021izc,Albers:2019rzt, Gunther:2022pto}, as well as publicly available frameworks to create custom emulators \citep{SpurioMancini:2021ppk, Nygaard:2022wri}.

We here used the \texttt{CONNECT} framework of \citet{Nygaard:2022wri} to create an emulator for the \modrec\ model. \texttt{CONNECT} generates training data and uses it to train a neural network which can reproduce the output of the full Einstein-Boltzmann calculation $10^3-10^4$ times faster than the Einstein-Boltzmann solver itself. The training data set consists of pairs of input parameter vectors and the corresponding output observable vectors, which must be pre-computed. Carefully selecting which input vectors to be included in the training set can reduce the overall size of the set needed to reach a fixed emulator accuracy. Popular strategies such Latin Hypercube Sampling \citep{Tang:1993} or Fisher Sphere Sampling \citep{Schneider:2011} offer significant improvement over grid-based approaches.

\texttt{CONNECT} uses an iterative strategy to construct the training data, so that it is concentrated near the region of high posterior probability. The observation underpinning this strategy is that the final emulator is unlikely to be used far away from the maximum of the posterior, as a MCMC sampler spends negligible time there after the initial burn-in phase. Therefore it is inefficient and unnecessary to include points from these regions in the training set. To estimate the region of high posterior probability (which is unknown prior to inference), \texttt{CONNECT} begins by training an initial emulator on a Latin hypercube with $\mathcal{O}(10^4)$ points, with the edges of the cube chosen so that the region of high likelihood is contained within the cube. The initial emulator is used to provide theoretical observables during a high-temperature Metropolis-Hastings random walk of the parameter space. While the accuracy of the emulator at this stage may be low, it is sufficient to bias the random walk towards regions of higher likelihood. Once the MCMC has converged, selected points from the chain are added to the training set and an Einstein-Boltzmann solver is used to compute the true theoretical outputs for those points. A new emulator is trained on the expanded training set, at which point the process begins a new iteration and is terminated after a convergence criterion is met. See \citet{Nygaard:2022wri} for full details.

\begin{figure}[ht]
    \centering
    \includegraphics{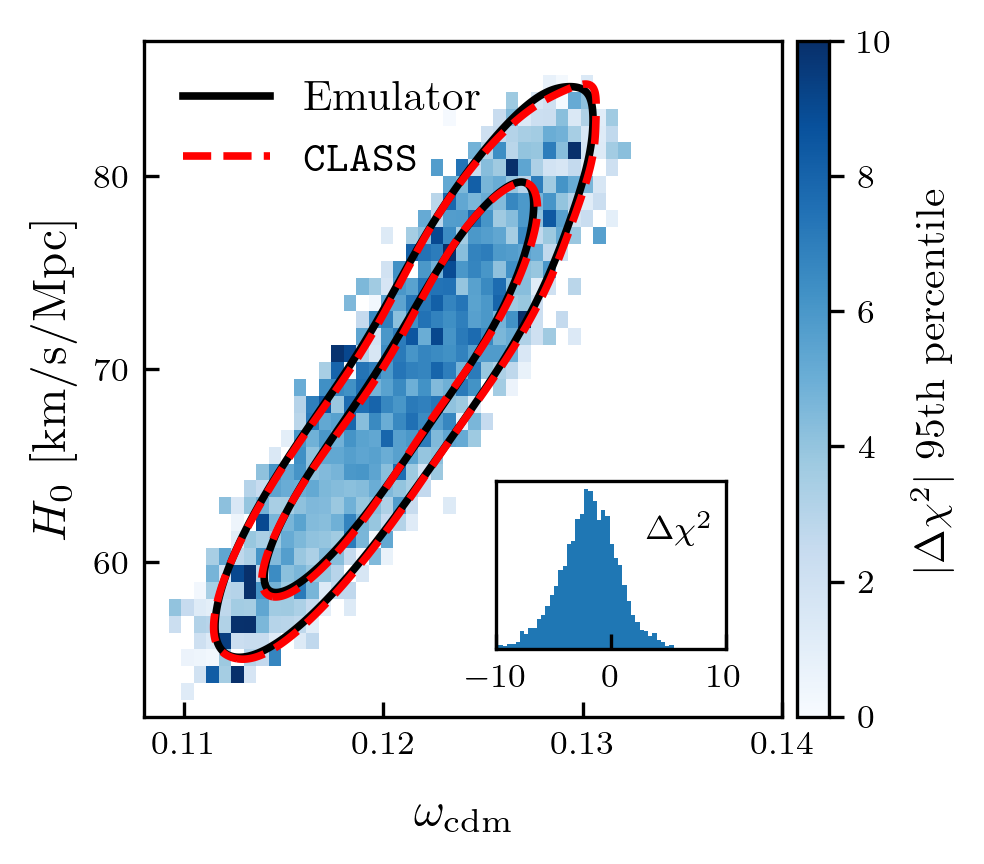}
    \caption{\label{fig:reweighted} Contours from a MCMC chain using the final trained emulator for the \modrec\ model as well as the reweighted contours using \texttt{CLASS} computed spectra. We used Planck 18 high-$\ell$ TTTEEE, low-$\ell$ TT and EE, and lensing data for this iterative process. We also bin all points from the chain and show the 95th percentile of $|\Delta \chi^2| = | \chi^2_{\text{emulator}} - \chi^2_{\text{CLASS}} |$. The $\chi^2$ values in each case were calculated using only the high-$\ell$ nuisance parameter marginalized likelihood \cite{Planck:2019nip, Prince:2019hse}. \textit{Inset:} A histogram of $\Delta \chi^2$ values for all points in the chain, indicating that on this set of points the emulator predictions marginally skew towards better fits than the \texttt{CLASS} computed likelihoods, as might be expected since the chain was generated using the emulator.}
\end{figure}

We modify the \texttt{CONNECT} code to include additional observables in addition to CMB power spectra in the training set. Our emulator is trained to reproduce the following quantities:
\begin{enumerate}
    \item The TT, TE, EE, and $\phi \phi$ CMB power spectra computed for $\ell_{\rm min} = 2$ and $\ell_{\rm max}=5000$. The default \texttt{CLASS} settings are insufficient to ensure accuracy at high $\ell$, so increased precision settings are required. We use settings adopted from \cite{Bolliet:2023sst}, modified to optimize performance in our use case while maintaining accuracy. A full description of the settings used can be found in Appendix \ref{app:emulator_appendix}. To reduce the length of the output vector, the power spectra are saved on a grid of $\ell$ values: the emulator is trained on these values, and other values are obtained via interpolation.
    \item The ionization fraction $\Xe(z)$, visibility function $g(z)$, $D_A(z)$, $H(z)$, and $\sigma_8(z)$. The latter three quantities are the angular diameter distance, Hubble parameter, and amplitude of matter fluctuations on $8h^{-1}$ Mpc scales, and are required in order to place constraints using BAO measurements. These quantities are also saved on a grid of $z$ values, and interpolation is used with the emulated result to recover function values at other redshifts. Our output grid for the $D_A(z)$, $H(z)$, and $\sigma_8(z)$ covers a redshift range from $z=0$ to $z=5$, sufficient for current and upcoming data which measure the BAO feature.
    \item Other derived parameters: the current density parameter of dark energy, $\Omega_\Lambda$; $\sigma_8$; the primordial helium fraction, $Y_{\rm p}$; the redshift of the baryon drag epoch, $z_{\rm drag}$; the sound horizon at baryon drag, $r_s^{\rm drag}$; the redshift at which $\Xe(z)=0.5$, $z_{\rm rec}$; the conformal time of recombination, $\tau_{\rm drag}$; the angular diameter distance to $z_{\rm rec}$, $D_A^{\rm rec}$; the redshift of reionization, $z_{\rm reio}$; the redshift of last scattering, $z_\star$; the conformal time of last scattering $\tau_\star$; the sound horizon at last scattering, $r_s^{\star}$; the angular diameter distance to last scattering, $D_A^{\star}$; and the angular size of the sound horizon at last scatter, $100 \theta_\star$. Some of these do not enter into this analysis, but were included for possible future use.
\end{enumerate}

\changed{These quantities are concatenated to form the output vector which the neural network is trained to reproduce, given the corresponding input vector of the 13 model parameters, and this vector is normalized using the ``standardization" option in \texttt{CONNECT} \citep[see][]{Nygaard:2022wri}. The neural network had 6 hidden layers with 512 fully connected nodes per layer. We trained for 150 epochs with a batch size of 256 using a mean square error loss function. We terminated the iterative process once the 95th percentile errors on angular power spectra were below 1\%. As an additional post hoc assessment of the emulator accuracy, we importance sampled an MCMC chain produced with this emulator, with weights set by the likelihood as computed using \texttt{CLASS}-calculated power spectra. This re-weighted distribution showed little change as compared to the original distribution obtained using the emulator, indicating that the emulator is sufficiently accurate for inference in regions of appreciable likelihood. A comparison of the original and re-weighted distributions in the $H_0 - \omega_b$ plane is shown in Fig.~\ref{fig:reweighted}. We also show the 95th percentile $\Delta \chi^2$ values for points drawn from the posterior in this region. See Appendix \ref{app:emulator_appendix} for further details on the emulator and additional error quantification.}

\changed{The final training set contained 182640 models, of which 5\% were reserved for test data. These are the only $\texttt{CLASS}$ executions required for the entire analysis, which is the same order of magnitude as the number of accepted steps in a single chain within the \modrec\ model, although additional computations would be required for rejected steps. Furthermore, since we are emulating the theoretical quantities directly as opposed to the likelihoods, the same emulator can be used with a variety of data combinations. As long as some observable is included among or can be computed from the emulated quantities, the emulator can be used in combination with these data at no additional cost. Avoiding these computations with the trained emulator represents a significant acceleration of the inference process.}

\section{\label{sec:constraints}Reconstructions of the ionization fraction history}
Using a trained emulator, we run a MCMC analysis using the Metropolis-Hastings algorithm as implemented in the \texttt{COBAYA} sampler \cite{Torrado:2020dgo}. Plots are made using the GetDist package \cite{Lewis:2019xzd}. For every data combination considered below, we use the same priors which are compiled in Table~\ref{tab:priors}. We use the the \planck\ 2018 high-$\ell$ and low-$\ell$ data for temperature and polarization, as well as \planck\ lensing data, in our reconstruction of the ionization history using CMB data alone. For the high-$\ell$ likelihood we use the nuisance-parameter marginalized likelihood detailed in \citet{Planck:2019nip}. We then combine this with BAO data from the eBOSS \citep{eBOSS:2020yzd} DR16 release and SDSS DR7 Main Galaxy Sample (MGS) \citep{Ross:2014qpa}. We also consider the impact of including the $H_0$ determination from the \shoes team, inferred from Cepheid calibrated supernovae. We begin with a brief overview of the origin of constraints on $\Xe$ from CMB data.

\begin{table}[h]
\caption{ \label{tab:priors}
Priors used throughout this analysis for the standard cosmological parameters, as well as the new \modrec\ control point parameters. All parameters have a uniform prior between the specific ranges. For the control point parameters, these values were chosen as $\tilde{q}_i \in [-0.9 \Xe^{\rm fid}(z_i), \min(10 \Xe^{\rm fid}(z_i), \Xe^{\rm max}(z_i)]$, where the fiducial ionization history is computed using {\it Planck} and assuming $\Lambda$CDM.}
\begin{ruledtabular}
\begin{tabular}{lcl}
\textrm{Name}  & Prior range & Notes \\
\colrule
$\omega_b$  & [0.0177, 0.0271] & -- \\
$\omega_{cdm}$  & [0.1028, 0.1374] & -- \\
$n_s$  & [0.8784, 1.0533] & -- \\
$\tau_{reio}$  & [0.0276, 0.1] & -- \\
$\ln(10^{10}A_s)$  & [2.837,  3.257] & -- \\
$H_0$  & [50, 85] & --\\
\colrule
$\tilde{q}_1$  & [-0.00117, 0.0131] & $z_1 = 666.6$ \\
$\tilde{q}_2$  & [-0.00321, 0.0356] & $z_2 = 800.0$ \\
$\tilde{q}_3$  & [-0.0182, 0.203] & $z_3 = 933.3$\\
$\tilde{q}_4$  & [-0.0939, 0.988] & $z_4 = 1066.6$\\
$\tilde{q}_5$  & [-0.290, 0.770] & $z_5 = 1200.0$\\
$\tilde{q}_6$  & [-0.581, 0.447] & $z_6 = 1333.3$\\
$\tilde{q}_7$  & [-0.827, 0.174] & $z_7 = 1466.6$\\
\end{tabular}
\end{ruledtabular}
\end{table}

\subsection{\label{sec:cmb_information}Information about $\Xe(z)$ from CMB data}

All of the CMB constraints on the ionization history arise from the sensitivity of the photon scattering rate to the ionization history. At fixed baryon fraction and helium fraction, $X_{\rm e}(z)$ is the only unknown quantity influencing the average photon scattering rate:
\begin{equation}
    \label{eqn:scattering_rate}
    \Gamma_{\rm T}(z) = X_{\rm e}(z) n_{\rm H}(z) \sigma_T
\end{equation}
where $\sigma_T$ is the Thomson scattering cross section, and 
\begin{equation}
    n_{\rm H}(z) = \left( \frac{\Omega_b h^2}{0.022} \right) \left( \frac{1-Y_p}{0.76} \right) \frac{1.8\cdot 10^{-7}}{\text{cm}^3}  (1+z)^3
\end{equation}
is the density of hydrogen nuclei ($n_{\rm H} \equiv n_{\rm HI} + n_{\rm HII}$).

In the following we discuss the impact of the Thomson scattering rate on i) the redshift of last scattering, ii) photon diffusion damping, and iii) polarization generation. We argue that photon diffusion and polarization generation effects on the power spectra are sufficiently high-dimensional to enable constraints on the free function $X_{\rm e}(z)$.

\subsubsection{Preliminaries}

The rapid drop in $\Xe(z)$ during recombination leads to a proportional drop in the scattering rate. As $\Gamma_{\rm T}(z)/ H(z) \lesssim 1$ the radiation component effectively decouples from the matter and free-streams to the present day, where it is observed as the CMB \cite{Hu:1994uz, Bond:1984}. This notion can be made precise by introducing the visibility function
\begin{equation}
    \label{eqn:visibility}
    g(z) \equiv \dv{e^{-\tau(z)}}{\eta} 
    = -\tau'(z) e^{-\tau(z)}.
\end{equation}
where $\tau$ is the optical depth to Thomson scattering and $\tau'=\dd\tau/{\rm d}\eta=-\Gamma_{\rm T} a$ is its conformal time derivative.

The quantity $g(z)\dd z$  gives the probability that a CMB photon observed today last scattered between redshift $z$ and $z + \dd z$, so the region of support of $g(z)$ defines the epoch during which we are able to observe the plasma. The peak of the visibility function is often used to define $z_\star$ and $\eta_\star$, the redshift and conformal time of last scattering\footnote{\texttt{CLASS} defines $z_\star$ as $\tau(z_\star) = 1$. While the two definitions result in similar $z_\star$ for the standard recombination scenario, they can lead to differences within modified recombination.}. This redshift is used to define $r_s^\star$, the comoving size of the sound horizon at last scattering, and $D_A^\star$, the comoving distance to the last-scattering surface:
\begin{align}
\label{eqn:rs_star}
r_s^{\star} &= \int_{z_{\star}}^\infty \frac{c_s(z)}{H(z)} dz
\\[1mm]
\label{eqn:DA_star}    
D_A^{\star} &= \int_0^{z_\star} \frac{\dd z}{H(z)}
\end{align}
where $c_s(z)$ is the sound speed in the plasma. Although for precise definitions like that of $r_s^\star$ or $D_A^\star$ it is convenient to use `the' redshift of last scattering, the process is of course extended, and as we shall see much of the information on $X_e(z)$ from the CMB arises from features of angular power spectra generated during this extended decoupling.

\subsubsection{Redshift of last scattering}

Modified recombination can shift $z_\star$, altering $r_{\rm s}^\star$ and therefore the phase of oscillation at decoupling for any given mode with wavenumber $k$.  In order to avoid changing the angular size of the sound horizon, $\theta_\star = r_s^\star / D_A^\star$, which is measured to within $0.03\%$ precision by \planck\ \citep{Planck:2018vyg}, there must be an equivalent fractional change to $D_A^\star$. This shift can be accomplished by adjusting $\Omega_\Lambda$ or, equivalently, $H_0$ \cite{Chiang:2018xpn}.  As we discuss below, BAO data are sensitive to this adjustment and its alteration to the low-$z$ distance-redshift relation, and can thus contribute indirectly to $X_{\rm e}(z)$ reconstruction.

\subsubsection{Diffusion damping}

At sufficiently early times, the plasma can be approximated as a tightly-coupled baryon-photon fluid at scales above the mean free path of the photons, $k \mfp \ll 1$, where $\mfp = 1/\left(a \Gamma_{\rm T} \right)$ is the comoving mean free path. For perturbations at or below this scale, the diffusive effects of thermal conduction and viscosity must be taken into account \citep{Hu:2001bc}. To second order in $k \mfp$ these effects lead to exponential damping with a characteristic scale \citep{Hu:1996mn, Zaldarriaga:1995gi}
\begin{equation}
    \label{eqn:damping}
    k_D^{-2} (z) = \int_{z}^{\infty} \frac{1}{6 H(z')\Gamma_{\rm T}(z')}\left(\frac{R^2}{1+R} + \frac{16}{15} \right)  \frac{dz'}{(1+R)}
\end{equation}
where $R(z) = 3 \omega_b / [4 \omega_\gamma (1+z)]$ is the baryon-to-photon energy ratio at a redshift $z$. We use this expression to explicitly show how the damping scale depends on $z$ and the scattering rate. 

If the effects of damping were completely captured by a single damping scale, e.g. $k_D(z_\star)$, then from damping effects we could only extract one number for constraining $X_e(z)$. However, due to the finite width of the visibility function, there is sensitivity to $k_D(z)$ over a range of redshifts.
The cumulative damping in a given mode $k$ can be approximated by averaging over $g(z)$ \citep{Hu:1996mn}
\begin{equation}
    \label{eqn:vis_avg_damping}
    \mathcal{D}(k) \sim \int_0^{\eta_0} g(\eta) \exp\left(-k^2 / k_D^2(\eta) \right) \dd \eta.
\end{equation}
This sensitivity to the damping scale across the shape of the visibility function allows for damping effects to be a high-dimensional source of information about the ionization history, providing information on skewness and higher order moments of the visibility function.

\begin{figure}
    \centering
    \includegraphics{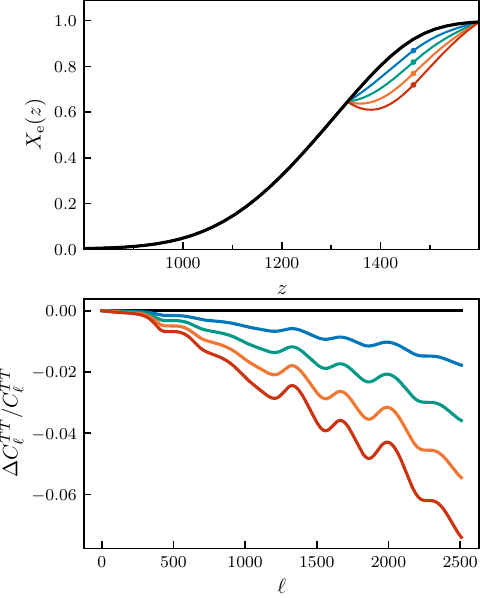}
    \caption{\label{fig:high_z_xe_changes} High redshift ($1333 \lesssim z < 1600$) changes to the ionization fraction and corresponding effects on the temperature power spectrum.}
\end{figure}

CMB data are also sensitive, via damping effects, to changes in the ionization history at redshifts above where the visibility function has appreciable support. Such changes to the scattering rate cumulatively affect $k^{D}(z)$ at all lower redshifts, via the integral in Eq.~\ref{eqn:damping}, including at redshifts where $g(z)$ is non-zero. We illustrate this in Fig.~\ref{fig:high_z_xe_changes} by varying $\Xe(z)$ only above where the standard $g(z)$ has support. Decreasing $\Xe(z)$ in this redshift range, relative to some fiducial, results in a decreased scattering rate and therefore an increase in diffusion damping for modes which have entered the horizon by this point. The net result is an increase in damping, as shown in the bottom panel. High redshift changes to $\Xe(z)$ only impact CMB power spectra through this effect on damping. We will see below that this is an important degree of freedom in our model.

One should bear in mind that Eq.~\ref{eqn:damping} is derived from a tight coupling approximation. This is a very relevant caveat as we are using the expression to guide our intuition through a regime where tight coupling completely breaks down. We suspect that our qualitative conclusion, about sensitivity to damping over a range of redshifts resulting in high-dimensional sensitivity to $\Xe(z)$, still holds. 

For more on the performance of this approximation through decoupling, see \citep{Pan:2016zla}. Of course in Einstein-Boltzmann solvers the tight coupling approximation is abandoned sufficiently early to preserve the accuracy of model power spectra \citep[e.g.][]{Blas:2011rf}.

\subsubsection{Polarization generation}

Polarization is generated from unpolarized light when an electron scatters incident radiation with a quadrupolar structure \citep{Zaldarriaga:1995gi}. Quadrupoles are generated during decoupling by free-streaming from the spatially varying monopole. Photons scattering off an electron during decoupling previously scattered an average distance $\mfp$ away, so for a given mean free path polarization is generated by modes $k \mfp \approx 2$. Viewed from the present day, these $k$ modes are converted by free streaming to anisotropies at angular scales $\ell_{\rm p} \sim k D_A^\star = 2 D_A^\star / \mfp$. Small angle polarization is sourced when the mean free path is small at early times, and larger scales are sourced as the mean free path grows during recombination. Thus, in principle one could  use polarization to probe the ionization history to arbitrarily high redshifts; however, at sufficiently early times the modes that generate polarization are inaccessible due to damping. The correspondence between polarization at a given angular scale and the epoch at which it is generated is another high-dimensional source of information regarding the ionization history: measurements of $C_\ell^{EE}$ can be used to constrain the free function $\Xe(z)$.

It should be noted as well that the amplitude of the polarization signal from a single $k$ mode is proportional to the width of the last-scattering surface \citep{Zaldarriaga:1995gi}, and the ratio of the temperature and polarization power spectra can constrain the width of the visibility function \citep{Hadzhiyska:2018mwh}.

\begin{figure*}[ht]
    \centering
    \includegraphics{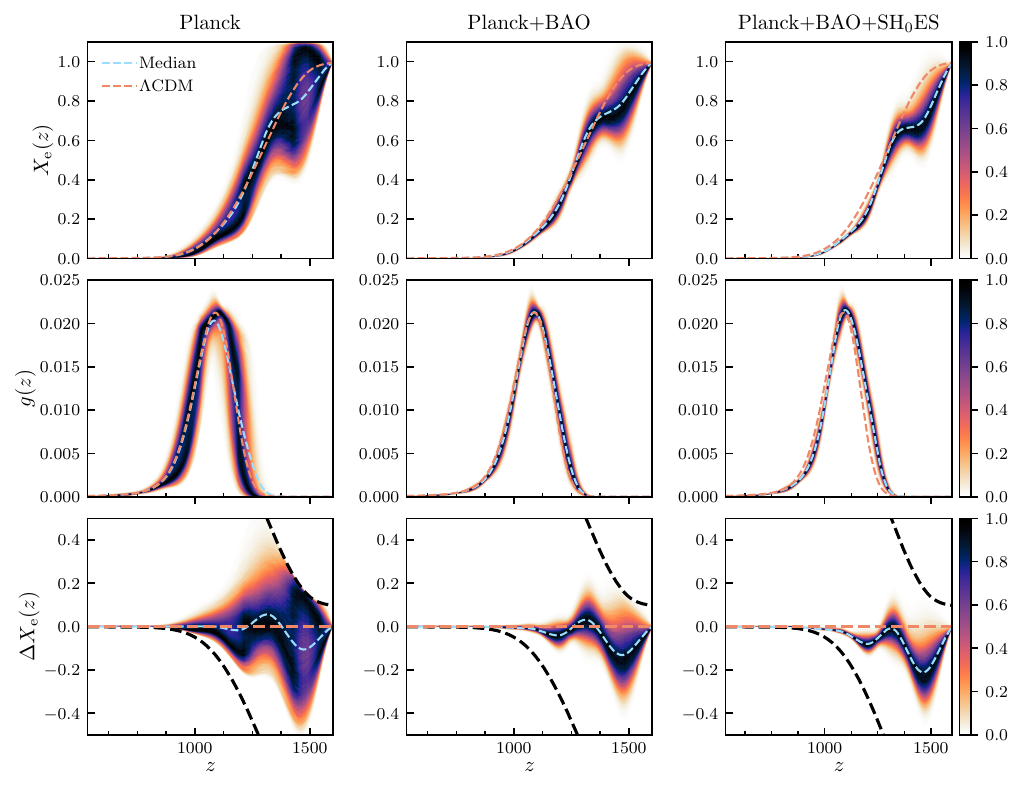}
    \caption{\label{fig:reconstructions} Reconstruction of recombination quantities for combinations of \planck\ 2018 data, eBOSS BAO data, and the \shoes\ distance-ladder $H_0$ measurement. The color indicates the density of trajectories through a given point in the plane, normalized so that the point passed through by the most trajectories at a given redshift has a color value of 1. The emulator was used to compute the plotted quantities for every model in the chain. The light blue line indicates the pointwise mean within the chain, and the light red line indicates the value for that quantity using the fiducial \textit{Planck} mean cosmology and assuming \lcdm. In the last row, $\Delta \Xe(z) = \Xe(z) - \Xe^{\rm fid}(z)$. The black dashed lines indicate the bounds from enforcing electron conservation.}
\end{figure*}

\begin{figure*}[t]
    \centering
    \includegraphics{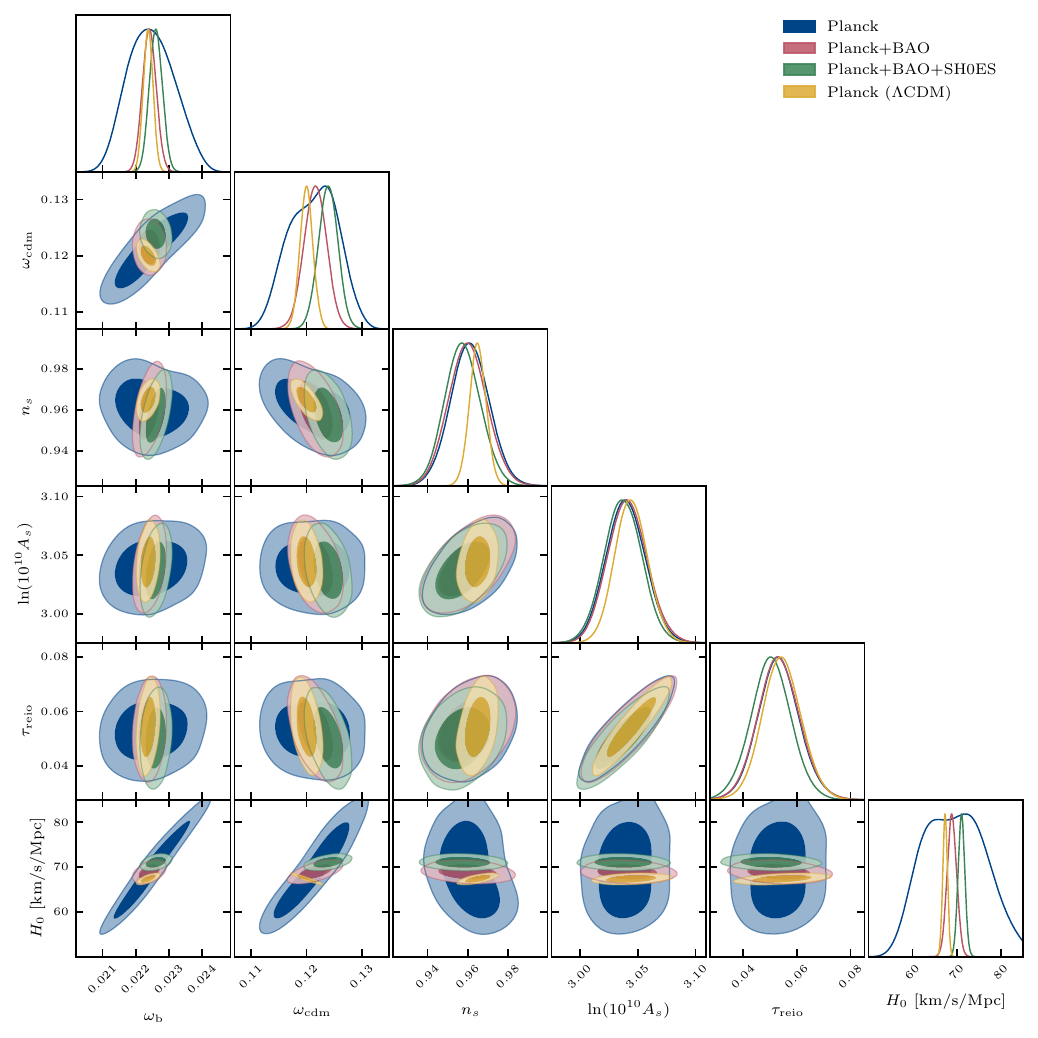}
    \caption{\label{fig:cosmo_constraints}  Marginalized posterior distributions from the analyses presented in this work. Here, \planck\ refers to \planck\ 2018, BAO is eBOSS DR16 BAO+RSD measurements as well as SDSS MGS. }
\end{figure*}

\begin{figure*}[t]
    \centering
    \includegraphics{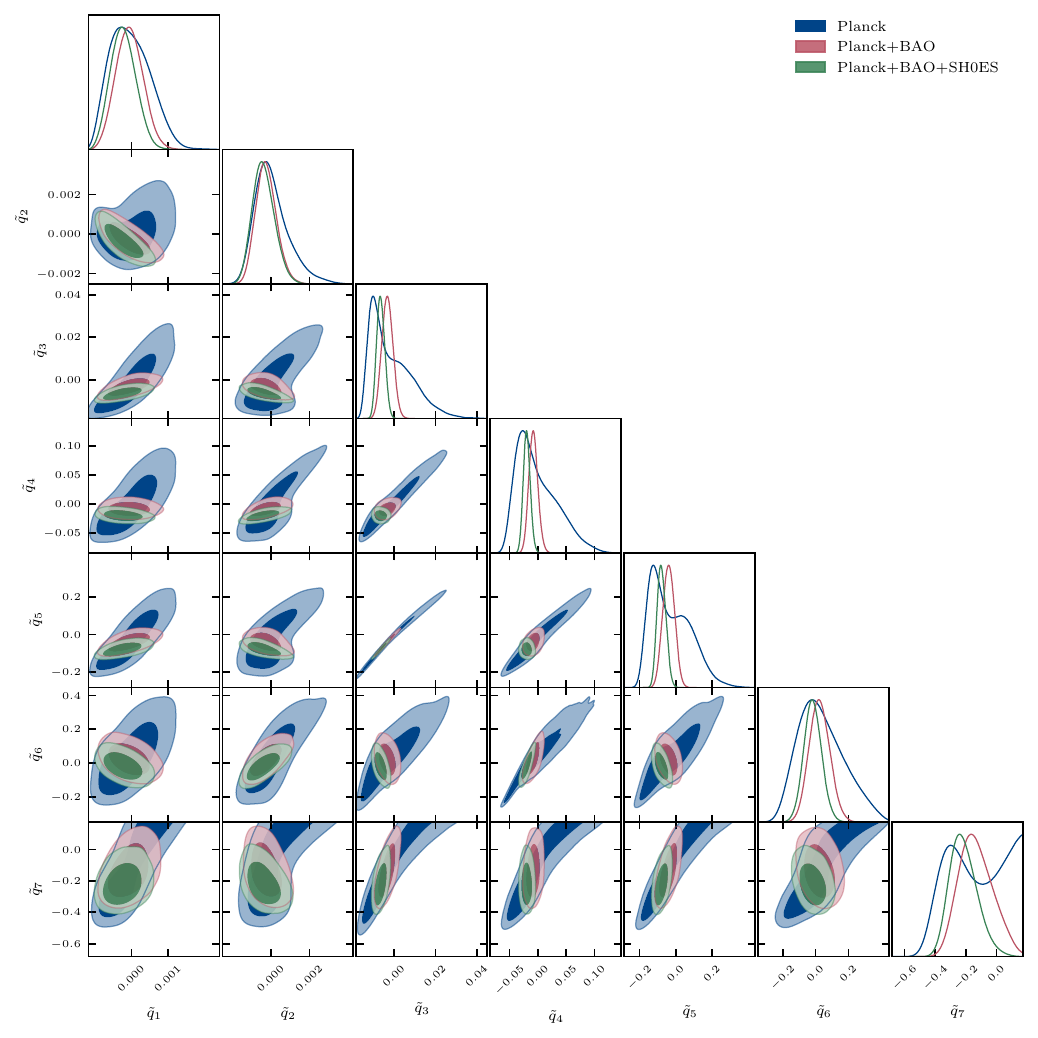}
    \caption{\label{fig:control_constraints}  Marginalized posterior distributions for the control point parameters.}
\end{figure*}

\begin{table*}[t]
\caption{\label{tab:parameter_constraints}
Marginalized 68\% confidence intervals model parameters using different data combinations. The first column are reference constraints assuming \lcdm\ and using the \planck\ baseline data. Note that within the \modrec\ model many parameters have non-Gaussian posteriors, hence the quoted results are the two-tailed 68\% confidence limits. The final row shows the 95\% one-tailed upper limits on $H_0$ in each chain.
}
\begin{ruledtabular}
\begin{tabular} { l c c c c}

 Parameter & Planck (\lcdm)& Planck &  +BAO &  +SH0ES\\
\hline
$H_0         $ [km/s/Mpc] &$67.36\pm 0.54             $ & $69.2\pm 6.6               $ & $68.9\pm 1.1               $ & $71.06\pm 0.77             $\\

$\omega_{\rm b} $& $0.02237\pm 0.00015        $ & $0.02245^{+0.00066}_{-0.00076}$ & $0.02241\pm 0.00021        $ & $0.02261\pm 0.00020        $\\

$\omega_{\rm cdm}$ & $0.1200\pm 0.0012          $ & $0.1212^{+0.0047}_{-0.0042}$ & $0.1216\pm 0.0020          $ & $0.1239\pm 0.0018          $\\

$\ln(10^{10}A_s)$ & $3.044\pm 0.014            $ & $3.040\pm 0.017            $ & $3.041\pm 0.017            $ & $3.037\pm 0.016            $\\

$n_s            $& $0.9649\pm 0.0042          $ & $0.9609\pm 0.0096          $ & $0.9600\pm 0.0095          $ & $0.9574\pm 0.0089          $\\

$\tau_{\rm reio}$ & $0.0544\pm 0.0073          $&  $0.0533\pm 0.0077          $ & $0.0533\pm 0.0078          $ & $0.0503\pm 0.0075          $\\
\hline 
$\tilde{q}_1    $& --  & $-0.00003^{+0.00045}_{-0.00065}$ & $-0.00005\pm 0.00037       $ & $-0.00022^{+0.00032}_{-0.00037}$\\

$\tilde{q}_2    $& -- & $0.00008^{+0.00056}_{-0.0011}$ & $-0.00024^{+0.00048}_{-0.00059}$ & $-0.00038^{+0.00050}_{-0.00063}$\\

$\tilde{q}_3    $& -- & $-0.0012^{+0.0052}_{-0.012}$ & $-0.0032\pm 0.0027         $ & $-0.0064^{+0.0018}_{-0.0021}$\\

$\tilde{q}_4    $& -- & $-0.001^{+0.021}_{-0.046}  $ & $-0.0078^{+0.0073}_{-0.0084}$ & $-0.0194^{+0.0056}_{-0.0064}$\\

$\tilde{q}_5    $& -- & $-0.027^{+0.084}_{-0.14}   $ & $-0.040\pm 0.032           $ & $-0.079^{+0.022}_{-0.026}  $\\

$\tilde{q}_6    $ & --& $0.04^{+0.11}_{-0.17}      $ & $0.024\pm 0.062            $ & $-0.017\pm 0.054           $\\

$\tilde{q}_7    $& -- & $> -0.257                  $ & $-0.134^{+0.096}_{-0.13}   $ & $-0.214^{+0.074}_{-0.10}   $\\

\hline
$H_0$ [km/s/Mpc] (95\% u.l.) & $<68.2$ & $<80.1$ & $<70.7$ & $<72.3$
\end{tabular}
\end{ruledtabular}
\end{table*}

\subsection{Reconstruction from \textit{Planck} CMB data}
 
Our reconstruction using \planck\ data is presented in the left-hand column of Fig.~\ref{fig:reconstructions}. We find that the reconstructed ionization fraction localizes around the standard $\Xe(z)$, which can be viewed as validation of the standard \lcdm\ picture of hydrogen recombination: out of quite general possible recombination histories, the fiducial history is clearly consistent with CMB data. The absolute uncertainty in the reconstruction increases sharply above $z\approx 1350$, which is also where the visibility function has vanishing support. Essentially none of the CMB signal is sourced from these early redshifts, and consequently the constraining power of anisotropy measurements is weakened. The constraints that are present are due to the high redshift sensitivity due to damping effects discussed in Sec.~\ref{sec:cmb_information}.

The overall constraining power of CMB data is emphasized by comparing the posterior distribution of $\Delta \Xe(z)$ in the final row to the prior distribution illustrated in Fig.~\ref{fig:prior_delta_xe}. We find that fractional uncertainty in this reconstruction of $\Xe(z)$ is to within roughly 45\% at 68\% confidence (c.f Fig.~\ref{fig:fractional_reconstruction}).

Parameter constraints are significantly affected by the new freedom in recombination. Table~\ref{tab:parameter_constraints} shows the 1D marginalized 68\% confidence intervals for all parameters. This is most pronounced with $H_0$, which we find to be $H_0 = 69.2 \pm 6.6$ km/s/Mpc, compared to the \planck\ constraint assuming \lcdm\ of $67.36 \pm 0.54$ km/s/Mpc. The 1D marginalized posterior for $H_0$ in this chain shows a large region, from 63 to 75 km/s/Mpc, of roughly equal posterior probability, which highlights the broad degeneracy between variations in the ionization fraction and $H_0$.  We see only minor shifts in mean values for parameters compared to the baseline \planck\ mean values assuming \lcdm.

The increased uncertainty in standard parameters is in contrast to previous perturbative phenomenological studies which have found that constraints are not greatly loosened \cite{Planck:2018vyg, Farhang:2011pt}. The increased uncertainty is a result of our choice to avoid the LRA and to instead fully treat the impact of a modified $\Xe(z)$ on the final radiation anisotropies. Our results are consistent with other studies focused on particular physical models which do see an effect on parameter constraints \citep[e.g.,][]{Hart2020H0}, albeit in a model-independent way.

\subsection{Reconstruction from CMB and BAO data}
The same initial conditions and dynamics that lead to a series of peaks in the CMB power spectrum result in a series of peaks in the matter power spectrum, and, equivalently, a peak in the matter correlation function 
\citep{Hu:1995en, Eisenstein:1997ik, Eisenstein:1998tu, Weinberg:2013, eBOSS:2020yzd}. This BAO feature is observed as an excess of clustering in the large-scale galaxy distribution on comoving scales roughly equal to the sound horizon at the redshift $z_{\rm drag}$ at which baryons decouple from the radiation.\footnote{\changed{The drag epoch ends when the optical depth for baryons drops below unity, i.e. when $\tau_{\rm d}(\eta_{\rm d}) \equiv \int_{\eta_{\rm d}}^{\eta_0} \dot{\tau}_{\rm d} \dd \eta < 1$. Here $\dot{\tau}_{\rm d} = \tau/R$ and $z_{\rm drag}$ is the redshift corresponding to $\eta_{\rm d}$.}} This occurs at a slightly later redshift than last scattering due to the low baryon-to-photon ratio, and is precisely defined conventionally as the redshift at which the baryon optical depth passes unity. The sound horizon at this epoch is
\begin{equation}
    \label{eqn:rs_drag}
    r_s^{\text{drag}} \equiv \int_{z_{\text{drag}}}^\infty \frac{c_s(z)}{H(z)} {\rm d}z.
\end{equation}

Analyses of galaxy survey data can determine the redshift interval that this feature covers along the line-of-sight, and in the transverse direction can determine the angle that it subtends:
\begin{equation}
    \label{eqn:bao_feature}
    \frac{1}{\Delta z} = \frac{D_H(z)}{r_s^{\rm drag}} \qand \frac{1}{\Delta \theta} = \frac{D_A(z)}{r_s^{\rm drag}}.
\end{equation}
Here the distances are the Hubble distance and the comoving angular diameter distance  
\begin{equation}
    \label{eqn:distances}
    D_H(z) = \frac{c}{H(z)} \qand D_A(z) = c \int_0^z \frac{\dd z'}{H(z')}
\end{equation}
as we are assuming $\Omega_k=0$. 

Without calibrating $r_s^{\rm drag}$ from other early universe measurements such as CMB anisotropies or BBN, these measurements constrain the shape of the low-redshift expansion history. As we shall shortly see, including BAO data in our analysis breaks the geometric degeneracy between cosmological parameters and variations in the ionization history (referred to in Sec.~\ref{sec:cmb_information}), tightening our reconstructions and constraints on parameters.

We use measurements of the BAO feature from the eBOSS DR16 release \citep{eBOSS:2020yzd}, including luminous red galaxies (LRG), emission line galaxies (ELG), quasars, and Lyman-$\alpha$ forest samples. We also include the BOSS galaxy sample \citep{BOSS:2016wmc} as well as the SDSS main galaxy sample (MGS). In total these data provide measurements of the low redshift distances in Eq.~\ref{eqn:distances} in seven bins between $0.15<z<2.33$. We include redshift space distortion measurements in our analysis, \changed{although we find that they are not very constraining}. \changed{The robustness of these measurements in the presence of a modified recombination era is discussed in Appendix~\ref{app:BAO_appendix}.}

\changed{The middle column of Fig.~\ref{fig:reconstructions} shows the reconstructions of the ionization history using \planck+BAO data. The inclusion of BAO data sharpens the reconstructions and correspondingly tightens cosmological parameter constraints, which are shown in the red contours in Fig.~\ref{fig:cosmo_constraints}. Consistent with the discussion above, constraints on $H_0$ are tightened from $H_0 = 69.2 \pm 6.6$ km/s/Mpc in the case of \planck\ data alone to $ H_0 = 68.9\pm 1.1$ km/s/Mpc when including BAO data. This is consistent with the conclusions in \citep{Hart2020H0} for a varying $\me$ model. }

\begin{figure}[t]
    \centering
    \includegraphics{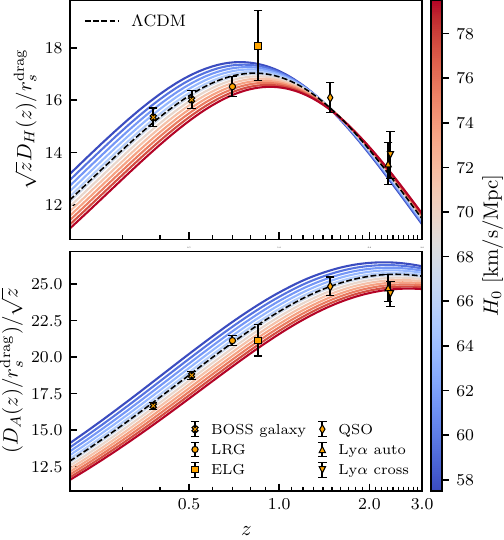}
    \caption{\label{fig:bao_tension} \textit{Top panel:} The low redshift Hubble distance relative to $r_s^{\text{drag}}$ for models with a given value of $H_0$, drawn from the CMB data only chain. Each curve is the mean of the corresponding distance measure in a unit bin centered on the labeled values of $H_0$. Only bins with a center value within the 95\% limits for $H_0$ within the \planck\ chain are shown here. Also shown are the eBOSS and BOSS data used in this analysis. \textit{Bottom panel:} Similar to the top panel, but for $D_A(z)/r_s^{\text{drag}}$. We do not show the MGS data point as it is reported in terms of a spherically-averaged distance $D_V(z)$ and converting would require assuming a model. Redshift-space distortion measurements of $f\sigma_8$ are also used in the analysis but not presented here as they are only weakly constraining. }
\end{figure}

In Fig.~\ref{fig:bao_tension}, we show the BAO measurements along with low-redshift distances computed using models from our \planck-only chain. For a given curve, we bin models drawn from the chain centered on integer values of $H_0$ with unit bin width, in units of km/s/Mpc, and plot the mean of the resulting set of curves. Each curve is therefore representative of models that have $H_0$ near the labeled values. We only show curves representing $H_0$ values within the 95\% confidence limits for $H_0$ in the \planck\ chain.

As can be seen in the top panel of Fig.~\ref{fig:bao_tension}, $D_H(z) / r_s^{\rm drag}$ is nearly independent of $H_0$ for $z\gtrsim 1$. This is because $D_H(z)$ only depends on $\omega_m$ at these redshifts, and this value is tightly constrained due to its relationship to the radiation driving envelope effect \citep{Hu:2001bc} in the CMB power spectra. At $z\lesssim 1$, dark energy significantly contributes to the expansion rate, and the value of $\Omega_\Lambda$ (or equivalently $H_0$) affects these low-redshift distances. 

A similar level of convergence is not seen in the bottom panel of Fig.~\ref{fig:bao_tension}, although by $z\approx 3$ the trend has started to become apparent. This is because the low-redshift differences in $D_H(z)/r_s^{\rm drag}$, when integrated according to Eq.~\ref{eqn:distances}, result in a spread of predictions for $D_A(z)/r_s^{\rm drag}$ at intermediate redshifts. However, all of these predictions must still converge at $z \approx 1100$ to be consistent with the precise measurement of $\theta_s^{\star}$.

This illustrates why BAO data are so effective at constraining models which alter recombination-era physics. A fractional change to the sound horizon, $\Delta r_s^{\rm drag} / r_s^{\rm drag}$, must be matched by comparable changes in $\Delta D_H(z) / D_H(z)$ and $\Delta D_A(z) / D_A(z)$ in order to maintain values consistent with the BAO data. Without introducing other model changes affecting low-redshift distances, $H_0$ is the only free parameter available to adjust that can affect these distances. However, it has already been adjusted to maintain consistency with $\theta_s^\star$, and so there is little room for further adjustment to maintain the low-redshift values demanded by the BAO measurements. These constraints can be evaded by introducing an additional late-time degree of freedom to the model, as was done, e.g., by \citet{Sekiguchi:2020teg} who introduced mean spatial curvature in addition to a time-varying electron mass.

\section{The $H_0$ tension with modified recombination}\label{sec:Hubble}
Distance ladder based measurements of the expansion rate are in significant tension with inferences using early universe data and assuming \lcdm. The most recent \shoes \ measurement, using Cepheid-calibrated supernovae, yields a value of $H_0 = 73.04 \pm 1.04$ km/s/Mpc \cite{Riess:2021jrx}, to be compared with the indirect measurement from \planck\ assuming \lcdm, which yields $67.36 \pm 0.54$ km/s/Mpc \cite{Planck:2018vyg}. Although the tension is often summarized in terms of tension between \planck\ and \shoes, it is in fact robust to changes in CMB dataset \cite{Huang:2018xle, Aylor:2018drw, SPT-3G:2022hvq, ACT:2020gnv}, and is even present, albeit at lower significance, when CMB data are removed altogether in favor of BBN-calibrated BAO measurements of the sound horizon \cite{Addison:2017fdm}. See \cite{Verde:2023lmm} for a detailed review of the current status of the $H_0$ tension.

This robustness has made it exceedingly difficult to resolve the tension through simple extensions to \lcdm\, and thus more radical departures have enjoyed increased attention \citep{Valentino2021H0, Schoneberg:2021qvd}. Much of this attention has been focused on resolving the tension by modifying \lcdm\ in the two decades of scale factor expansion preceding recombination \cite{Aylor:2018drw, Knox:2019rjx}. Models modifying this epoch are perhaps uniquely capable of re-calibrating the early universe standard ruler used to infer $H_0$ from CMB data and galaxy clustering surveys while leaving late time distances unaffected. 

Modified recombination falls within this category proposed solutions to the Hubble tension, and different variations have been previously studied. For example, \citet{Chiang:2018xpn} explored changes to the position and width of the visibility function as one possible solution, and \citet{Lee:2022gzh} explored more general perturbative changes to the ionization fraction. The modified recombination history was able to partially (in the former case) or fully (in the latter case) eliminate the tension when only CMB data was considered, but lacked the flexibility to evade the constraints from late-time distance measurements, discussed in Sec~\ref{sec:constraints}, once such data are included in the analysis. \citet{Hart2020H0} first recognized the role that a time-varying $m_e$ might play in resolving the Hubble tension, and placed constraints using \planck\ data. \citet{Sekiguchi:2020teg} reached a similar conclusion studying modified recombination via a time-varying $\me$, but found that concordance with the late-time distance measurements could be preserved by freeing $\Omega_k$, the mean spatial curvature.

Using our reconstructions of the ionization history, we examine the extent to which non-perturbatively modified recombination can accommodate higher values of $H_0$. This extends existing phenomenological studies \citep[e.g.,][]{Farhang:2012jz, Hart:2019gvj} to non-perturbative changes to the recombination history in a physical model-independent way. We are also able to study where in redshift space modifications to $\Xe$ become important in terms of resolving the tension and identify specific modifications which do a good job of alleviating the tension while maintaining a fit to the data. This may be useful to model builders seeking to alleviate the $H_0$ tension through models which primarily affect the ionization history. We begin with a brief review of why modified recombination is able to alleviate the tension at all, and then study which recombination histories can achieve this while maintaining a good fit to the data.

\subsection{How modified recombination affects inferences of $H_0$}
Assuming a flat universe, and ignoring massive neutrinos for simplicity, we can write the expansion rate as
\begin{equation}
\label{eqn:friedmann}
       h(z) = \sqrt{\omega_m (1+z)^3 + \omega_r (1+z)^4 + \omega_\Lambda}
\end{equation}
where $h(z) \equiv H(z) / \left(100 \text{ km/s/Mpc} \right) $ and $\omega_\Lambda = h^2 - \omega_m - \omega_r$. Increasing $H_0$ while keeping the physical densities $\omega_r$ and $\omega_m$ fixed therefore only changes $\omega_\Lambda$, which has no effect on $H(z)$ at redshifts $z \gtrsim 1$. As such, a high value of $H_0$ implies a decreased distance to the last-scattering surface but has no implications for $r_s^\star$, due to the different redshift ranges over which Eqs.~\ref{eqn:DA_star} and ~\ref{eqn:rs_star} are integrated. The net result is an increase in the angular size of the sound horizon $\theta_s^\star$, and the same argument also applies to other angular scales.

\begin{figure}[t]
    \centering
    \includegraphics{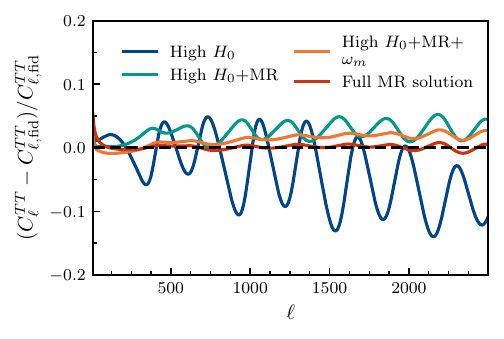}
    \caption{\label{fig:mr_compensation} Temperature power spectrum residuals for the different adjustments described in the text. The line labeled ``High $H_0$" corresponds to the case in which only $H_0$ is adjusted to have a high value: all other parameters are fixed to their fiducial value. Likewise, the ``MR" line has the fiducial value of all cosmological parameters (including a low $H_0$), but a modified recombination history. The line labeled ``MR+high $\omega_m$" has the same recombination history as the previous, but now $\omega_m$ has been adjusted to tune the peak heights. Finally, a full MR solution is presented which has a high $H_0$, high $\omega_m$, modified recombination, and minor shifts in $\ln10^{10}A_s$, $\tau_{\text{reio}}$, and $n_s$.}
\end{figure}

In the power spectrum, these changes correspond to a decrease in the spacing between acoustic peaks as well as a shift in the onset of damping to lower multipoles, features which are both tightly constrained by \planck. To maintain the fit with the data while inferring a higher $H_0$, these changes in the power spectra must be addressed in some way by any alternative model.

Within \lcdm, these effects can be partially compensated by allow the physical matter density $\omega_m$ to vary. Increasing the matter density at fixed $H_0$ decreases the sound horizon at $z_\star$ and also affects the height of the acoustic peaks, partially accommodating the decreased distance to last scattering and change in damping scale introduced by raising $H_0$. However, as pointed out by \cite{Knox:2019rjx}, there is no value of $\omega_m$ which can simultaneously accommodate local measurements of $H_0$, BAO constraints, and \planck\ data. We are therefore motivated to find another way to adjust the sound horizon and damping scale in order to maintain the fit with the data. As discussed in Sec.~\ref{sec:constraints}, both $k_D$ and $r_s^\star$ are sensitive to changes in $\Xe(z)$ and it is therefore possible to adjust the recombination history in such a way as to accommodate different values of $H_0$.

As a concrete example, in Fig.~\ref{fig:mr_compensation} we demonstrate these effects in the temperature power spectrum. We begin with the mean parameters from \planck\ assuming \lcdm\, and artificially adjust $H_0$ to 73.4 km/s/Mpc, resulting in the blue curve. This value was chosen as it corresponds to the $H_0$ value from the best fitting model in a bin centered at $H_0=73.0$ km/s/Mpc from the \planck\ only chain. This curve has both an oscillatory feature, following from the change in angular-diameter distance to last scattering, as well as a downwards tilt at smaller scales as a result of the changed damping scale. We then allow the recombination history to vary in an attempt to preserve the fit to the data, as indicated by the green curve. The modified recombination history is one in which the visibility function peaks earlier and has an increased width (see Fig.~\ref{fig:xe_colorcoded_H0}), which is able to partially restore the fit to the data. However, modifications to recombination are not completely degenerate with changes to $H_0$, and in general we find that $\omega_m$ must be increased to further adjust the peak heights, as shown in the red curve. Minor adjustments to other parameters further restore the fit with the data, which we call the ``full solution".

\subsection{High $H_0$ recombination histories}

\begin{figure}[t]
    \centering
    \includegraphics{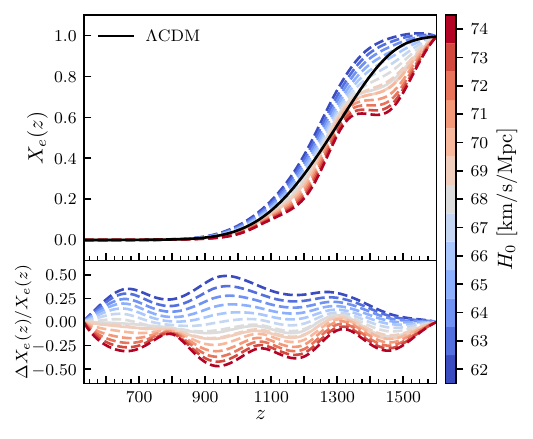}
    \caption{\label{fig:xe_colorcoded_H0} \textit{Top panel:} Recombination histories that allow for the labeled values of $H_0$ while maintaining a good fit to the data. \textit{Bottom panel:} Residuals of the top panel with respect to $\Xe(z)$ in the \lcdm\ model, using the \textit{Planck} best-fit cosmology. }
\end{figure}

\begin{figure}[t]
    \centering
    \includegraphics{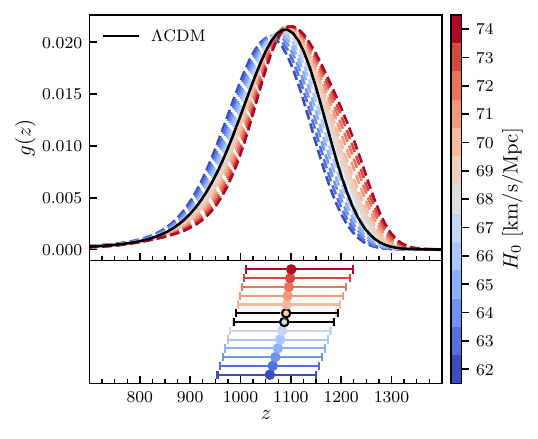}
    \caption{\label{fig:g_colorcoded_H0} \textit{Top panel:} Similar to Fig.~\ref{fig:xe_colorcoded_H0} but for the visibility function. \textit{Bottom panel:} Visibility function maximum, $z_\star$, and FWHM. Points outlined in black are from the \planck+BAO chain.}
\end{figure}

We now examine in more detail those recombination histories which can alleviate the Hubble tension. As a criterion to identify models which maintain a fit to the data, we select models from the chains with $\Delta \chi^2 \equiv \chi^2_{\modrec} - \chi^2_{\Lambda{\rm CDM, bf}} < 0$. Here, $\chi^2_{\Lambda{\rm CDM, bf}}$ is $\chi^2$ of the best fit to the \planck+BAO data combination, assuming \lcdm. 

We bin these well-fitting models according to their values of $H_0$, with bins centered at integer values and unit bin widths. We discard bins centered at values of $H_0$ above or below the marginalized 68\% confidence levels in each chain. In the \planck-only chain, the 68\% two-tailed lower and upper limits are $H_0 = 61.77$ km/s/Mpc and $H_0 = 75.92$ km/s/Mpc, and in the \planck+BAO chain they are $H_0 = 67.82$ km/s/Mpc and $H_0=69.97$ km/s/Mpc. Within each bin, $\Xe(z)$ and $g(z)$ are computed for each model using the emulator, and the pointwise means are plotted in Figs.~\ref{fig:xe_colorcoded_H0} and ~\ref{fig:g_colorcoded_H0}. Dashed lines indicate quantities computed from the \planck-only chain and solid lines indicate quantities computed from the \planck+BAO chain. 

In Fig.~\ref{fig:xe_colorcoded_H0}, we show the ionization fractions for this selection of models. Allowing for non-perturbative changes to recombination, a wide range of $H_0$ values are consistent with \planck\ data, including values as high as the \shoes \citep{Riess:2021jrx} measurement. For these high $H_0$ models, the required departures from the standard recombination scenario are as large as $\sim 40\%$ near $z\sim900$, and have an oscillatory redshift dependence. We note that a similar feature was identified by \citet{Lee:2022gzh} in the context of perturbative changes introduced by a time-varying $\me$. However, these oscillations are not an essential feature to the overall fit: smoothing them in an appropriate manner degrades the fit by $\Delta \chi^2 \lesssim 5$ \citep{DESIfollowup}.

One notable feature of these ionization histories is the trend seen in the highest control point $\tilde{q}_7$, placed at $z=1466$. This control point almost exclusively impacts redshifts where the visibility function has no support, which, following the discussion in Sec.~\ref{sec:cmb_information}, primarily affects the amount of damping. This feature can therefore be understood as a way for the \modrec\ model to compensate for the larger physical damping scale implied by the decreased $D_A^\star$ associated with a high $H_0$. The return to ionization fraction values near the standard $\Xe(z)$ at $z \approx 1300$ is because this is where $g(z)$ begins to be non-zero, and as such constraints arising from polarization begin to be important.

As expected following the discussion in Sec.~\ref{sec:constraints}, including BAO data restricts the range of $H_0$ values which can be obtained while maintaining a good fit to all of the data. The estimate for $H_0$ from this chain is $H_0 = 68.9 \pm 1.1$ km/s/Mpc, indicating a reduction of the tension to a $2.7 \sigma$ level. The ionization fraction is restricted to near its standard value, with the maximum departure among models satisfying our selection criterion being around 15\% near $z=900$. This restriction is once again a consequence of the sensitivity of BAO data to low-redshift distance measures, as highlighted in Fig.~\ref{fig:bao_tension} and Sec.~\ref{sec:constraints} \citep[see also the discussion in][]{Hart2020H0}. 

We show the corresponding visibility functions in Fig.~\ref{fig:g_colorcoded_H0}. There is a slight increase in $z_\star$ for high-$H_0$ models, resulting in the lower $r_s^\star$ needed to accommodate the decreased distance to last scattering that comes with a high $H_0$. The broadening of the visibility function for these models is not symmetric, and in particular the high-$z$ tail of the visibility function is increased relative to the standard recombination scenario. As we will see below, this seems to be a common feature in recombination models with varying electron mass or PMFs, pointing towards a new degree of freedom that has been excited beyond just a change of the mean and width.

\subsection{A FFAT Scaling Perspective}

In \cite{Cyr-Racine:2021oal}, a new path toward a possible solution of the Hubble tension was introduced that also required a change to the Thomson scattering rate and is thus worth remarking upon here. Central to that work was the discovery of a symmetry of dimensionless observables under a scaling transformation of all the rates in the relevant Einstein-Boltzmann equations and the amplitude of the initial power spectrum, a scaling transformation more thoroughly explored in \cite{Ge:2022qws}. They called this FFAT scaling, for free-fall, amplitude, and Thomson since the rates are gravitational free-fall rates ($\sqrt{G\rho_i}$ for each component $i$) and the Thomson scattering rate. Note that the expansion rate scales as well given the Friedmann equation (and an assumption of zero mean curvature). 

This symmetry can, in principle, be exploited to address the Hubble tension. To boost $H_0$ by 8\%, one can FFAT transform from the \planck\ best-fit \lcdm\ model, with all rates boosted by 8\%, and have no impact on the statistical properties of CMB temperature and polarization maps or BAO observables. The search for models that can implement this scaling, or approximate it well enough, is non-trivial. One cannot implement the free-fall rate scaling directly because of FIRAS constraints on $\sqrt{G\rho_\gamma}$ but it can be mimicked with the introduction of atomic dark matter. It is possible that a time-varying fine structure constant could deliver the required boost in $\Gamma_{\rm T}$ \cite{Zhang:2022ujw}. A single correlated variation of fine structure constant and electron mass was shown to work in \cite{Greene:2023cro, Greene:2024qis}. In \cite{Cyr-Racine:2021oal} and \cite{Ge:2022qws} the helium abundance was lowered to boost $\Gamma_{\rm T}$. All of these solutions face challenges with light element abundances from big bang nucleosynthesis. To date, the intriguing possibility of an FFAT-scaling solution to the Hubble tension has raised more questions than it has answered. 

Note that such a solution to the Hubble tension is quite different to the ones we have referenced already and seen emerge in our $X_{\rm e}(z)$ reconstructions. These have recombination occur at higher redshift in order to reduce $r_{\rm s}^*$. These are thus {\em reductions} in $\Gamma_{\rm T}$, whereas FFAT scaling solutions require a boost. 

We do not see this boosted $\Gamma_{\rm T}$ solution arise in our work here because the \lcdm\ + \modrec\ model (combined with our implicit incorporation of the FIRAS constraint) does not allow for the scaling of all free-fall rates. It would be interesting to explore what is needed, in practice, from the ionization history, for a solution of the Hubble tension, in a model that includes \modrec\ {\em and} allows for a (mimicked) scaling of free-fall rates. We have kept such exploration beyond the scope of this paper.

\subsection{Inclusion of SH0ES data}

We now include SH0ES data in our analysis to illustrate how distance-ladder based measurements of $H_0$ drive a preference for non-standard recombination. We adopt the \shoes\ measurement of $H_0$ as the mean and standard deviation of a simple Gaussian likelihood for $H_0$, and include this in a chain along with the \planck\ and BAO data sets. Fig.~\ref{fig:control_constraints} shows the constraints on the control point parameters for each data combination considered in this work. We see that including the SH0ES measurement drives recombination quantities away from their \lcdm\ values in an attempt to reconcile high $H_0$ with the early universe data. 

The final column of Fig.~\ref{fig:reconstructions} shows our reconstruction of recombination quantities when \shoes data are included. These recombination histories differ substantially from the \lcdm\ predictions, with the means for $\tilde{q}_3$,$\tilde{q}_4$, and $\tilde{q}_5$  being $3.3\sigma$, $3.1\sigma$, and $3.3\sigma$ away from 0 deviation respectively. One interpretation of this is that within our model space, if distance ladder based determinations of $H_0$ are taken at face value, then \planck\ and BAO data favor a non-standard recombination over the redshift range $933 \lesssim z \lesssim 1200$. This is an analogous result to previous studies which have also identified that non-standard recombination-era physics is preferred, if one includes the \shoes result in the analysis \citep[e.g.][]{Hart2020H0, Shimon:2020dvb, Lee:2022gzh}. Our result shows this independently of the assumed alternative model, and also identifies the recombination history that best accommodates both early and late universe measurements. 

\begin{figure}[t]
    \centering
    \includegraphics{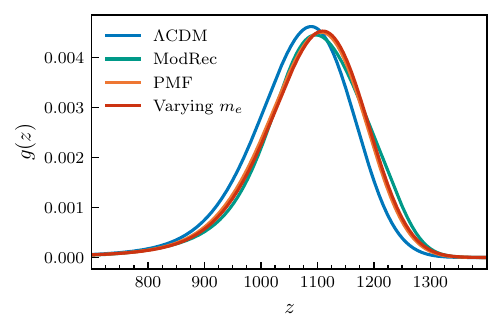}
    \caption{\label{fig:viz_comparison} The \lcdm\ visibility function computed using the \planck\ mean parameters, the mean visibility function from the \modrec\ \planck+BAO+\shoes\ chain, a PMF model visibility function from \cite{Galli:2021mxk} with clumping factor $b=0.5$, and a varying $m_e$ model visibility function using the mean parameters from the \planck+BAO+SN chain of \citep{Hart2020H0}. The PMF curve is almost entirely obscured by the varying $m_{\rm e}$ curve.}.
\end{figure}

\begin{table}[t]
\caption{\label{tab:visibility_moments}
Moments of the visibility functions compared in Fig.~\ref{fig:viz_comparison} }
\begin{ruledtabular}
\begin{tabular}{lcccc}
\textrm{Model} & Mean & Variance & Skewness & Kurtosis \\
\colrule

\lcdm\ & $1064.1$ & $9592$ & $-864\times 10^3$ & $4.700 \times 10^{8}$  \\
\modrec\  & $1088.0$ & $10144$ & $-913\times 10^3$ & $5.273 \times 10^{8}$  \\
PMF     & $1080.8$ & $9967$ & $-911\times 10^3$ & $5.058 \times 10^{8}$ \\
Varying $m_e$ & $1084.3$ & $10018$ & $-938\times 10^3$ & $5.217 \times 10^{8}$  \\
\end{tabular}
\end{ruledtabular}
\end{table}

The mean visibility function from our \planck+BAO+\shoes chain peaks at a slightly higher redshift relative to the \lcdm\ visibility function (computed using the \planck\ mean parameters), and is also skewed towards higher redshifts. As examples of physical models which recover similar features, in Fig.~\ref{fig:viz_comparison}, we show the mean $g(z)$ from the \shoes chain, the \lcdm\ $g(z)$, as well as $g(z)$ arising from a model with primordial magnetic fields and varying electron mass. It is apparent that the latter exhibit a similar level of skewness and even kurtosis (see Table~\ref{tab:visibility_moments}). However, for the PMF scenario, a number of approximations are made in the calculation of the related visibility function, which is based on computing the averaged recombination history including only second order effects of baryon-density perturbations on the recombination dynamics \citep[e.g.,][for more details of the model]{Jedamzik2020Relieving, Thiele2021}. Indeed the non-linear nature of the ionization history responses to baryon density fluctuations suggests that this may be omitting additional effects, which in a more rigorous treatment could manifest in enhanced higher order moments of the Thomson visibility function, an aspect that may be worth exploring in the future.

\begin{figure}[t]
    \centering
    \includegraphics{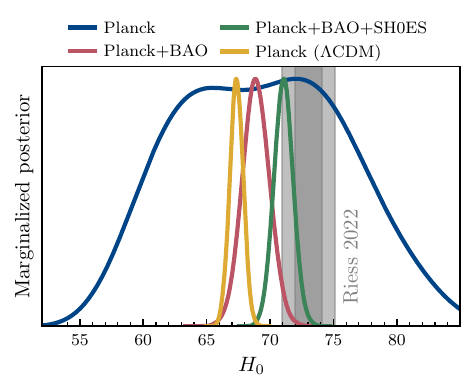}
    \caption{\label{fig:H0_posteriors} Marginalized 1D posterior distributions for $H_0$ in each of the three chains generated in our analysis.}
\end{figure}

The status of the $H_0$ tension in the modified recombination scenario presented here is summarized by the marginalized $H_0$ posteriors presented in Fig.~\ref{fig:H0_posteriors}. \planck\ data are able to accommodate SH0ES-consistent values of $H_0$, and there is a broad region of roughly equal posterior probability spanning low and high values of $H_0$. This increased uncertainty highlights the crucial role of the recombination model in the cosmological parameter inference. Once BAO data are included, the \shoes central value of $H_0 = 73.04$ km/s/Mpc cannot be accommodated through modifications to recombination alone. Including \shoes data does shift the mean $H_0$, but prefers a recombination history that looks very different from the standard picture. The crucial ingredients are a shift in the visibility maximum, and an increased width and skewness towards redshifts $z>z_*$.

\section{Forecasts for SPT-3G}\label{sec:Forecasts}

The recombination histories identified above lead to CMB power spectra which are consistent with current data, but which can differ significantly from \lcdm\ predictions at smaller scales which are unprobed by \planck. Upcoming data from ground-based CMB experiments, such as \sptthreeg, will have an impact on the constraints presented here. In this section we present forecasted constraints from forthcoming \sptthreeg\ observations.

\begin{figure}[t]
    \centering
    \includegraphics{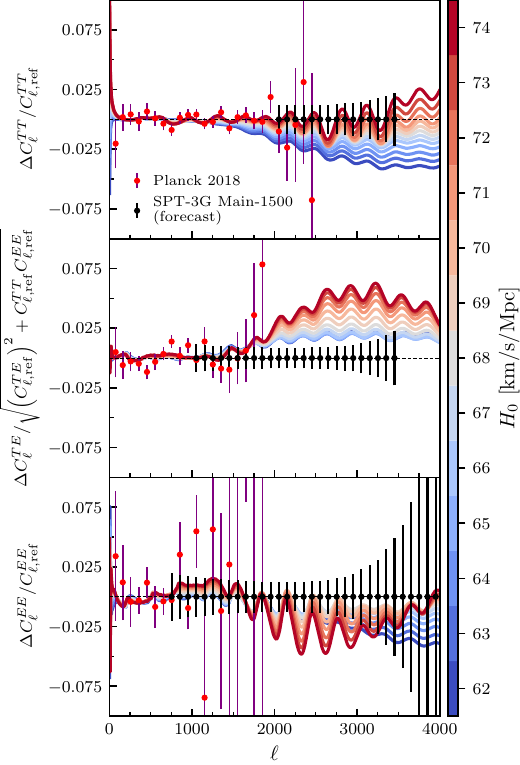}
    \caption{\label{fig:power_spectra_predictions} Power spectra residuals between predictions from the \modrec\ model and the \planck\ best-fit cosmology, taken as the reference cosmology. Also shown are binned data from \planck, as well as the forecasted bandpower errors for \sptthreeg{} in the Main-1500 configuration. Each have been binned with \changed{$\Delta \ell = 100$}. The power spectra correspond to models having the labeled values of $H_0$, in the same manner as Fig.~\ref{fig:bao_tension}.}
\end{figure}

\begin{figure*}[t]
    \centering
    \includegraphics{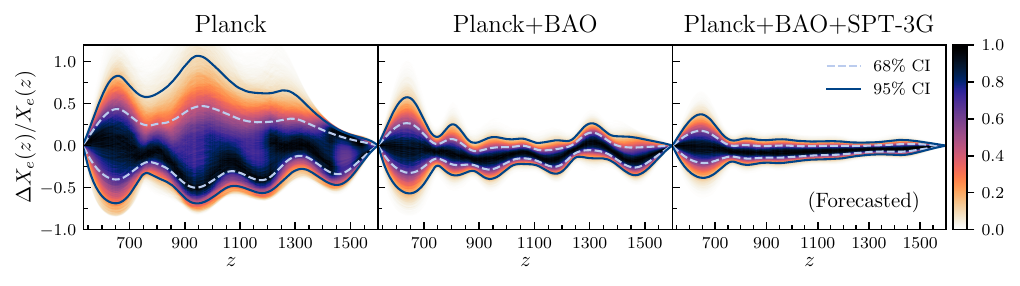}
    \caption{\label{fig:fractional_reconstruction} Fractional uncertainties in our recombination reconstructions for the \planck, \planck+BAO, and \planck+BAO+SPT data combinations. Dashed lines indicated the $68\%$ and $95\%$ confidence intervals. \changed{The color indicates the density of trajectories through a given point, as in Fig.~\ref{fig:reconstructions}}. SPT data will decrease the our reconstruction uncertainties by roughly a factor of 2, relatively uniformly across the redshift range in question. The fiducial cosmology for the SPT mock likelihood was taken to be the \planck\ mean cosmology. \changed{Our prior forces deviations from the standard ionization history to vanish at $z_{\rm min} = 533$ and $z_{\rm max} = 1600$, leading to artificially tight constraints at these boundaries.}}
\end{figure*}

We have implemented a mock Gaussian likelihood for \sptthreeg{} in \texttt{COBAYA}, 
\begin{equation}
    \log{\mathcal{L}} = \left( \vb{C}_\ell - \vb{C}_\ell^{\rm fid} \right)^\intercal \Sigma_\ell^{-1} \left( \vb{C}_\ell - \vb{C}_\ell^{\rm fid} \right)
\end{equation}
where $\vb{C}_\ell = \left(C_\ell^{TT}, C_\ell^{EE}, C_\ell^{TE}, C_\ell^{\phi \phi} \right)$ and the \planck\ 2018 best-fit cosmology is taken as the fiducial. The covariance matrix $\Sigma_\ell$ incorporates the effects of both instrumental noise and foregrounds and is constructed in the manner described in \citep{Prabhu:2024qix}. We use noise levels corresponding to the Main-1500 configuration of \sptthreeg{}, representing 5 years of observations on a sky area of 1500 $\rm deg^2$. This corresponds to the data that has already been collected since \sptthreeg{} began observations in 2019. 

As the observation regions for \planck\ and \sptthreeg\ are not independent, we remove large scale modes from the mock \sptthreeg{} data to avoid double counting information. The $\ell_{\rm min}$ threshold was determined as roughly the multipole at which the signal-to-noise for \sptthreeg\ is twice that of \planck. This approach was validated in \citet{Prabhu:2024qix}, and was found to give results less than 8\% worse than a full Fisher accounting for the overlapping sky regions. The multipole ranges we used were  $\ell \in [2000,3500], [1000, 3500], [750, 4000], [30,4000]$ for TT, TE, EE, and $\phi \phi$ respectively. \changed{We bin the data with $\Delta \ell = 100$, and discard bins which are centered outside of these ranges. This mock data is used along with \planck\ CMB and BOSS/eBOSS BAO data in an MCMC.}

We find that these data will significantly further reduce the remaining freedom in $\Xe(z)$. As shown in Fig.~\ref{fig:fractional_reconstruction}, assuming that \sptthreeg{} continues to observe power spectra consistent with \lcdm\ conditioned on \planck\ data, uncertainties for $\Xe(z)$ will be reduced to an approximately 15\% level at 95\% confidence around the standard prediction. Additionally, we forecast that the resulting error on the $H_0$ from Planck + BAO + SPT-3G will be 67\% larger than from Planck alone assuming \lcdm; i.e., the increased uncertainty in $H_0$ that one gets from freeing up recombination, will be less than a $1\sigma$ increase. The marginalized 1D posterior in the \planck+BAO+SPT-3G chain is $H_0 = 68.36 \pm 0.90$ km/s/Mpc.

\changed{In Fig.~\ref{fig:power_spectra_predictions}, we show predictions for the same models presented as dashed lines in in Figs.~\ref{fig:xe_colorcoded_H0} and ~\ref{fig:g_colorcoded_H0}. We first bin models from the \planck-only chain according to their value of $H_0$ and select only models which fit better than a \lcdm\ best fit, plotting the mean from each bin. We see that even for $H_0 \approx 68$ km/s/Mpc, the \modrec\ model power spectrum predictions differ from those of \lcdm. This indicates that the new freedom introduced in recombination improves the fit relative to the case of standard recombination --- in the \modrec\ model, the standard recombination history does not provide the best fit even for models near the \planck\ best fit assuming \lcdm. This is also reflected by the fact that the mean ionization fraction for the \planck\ chain in Fig.~\ref{fig:reconstructions} differs from the standard prediction. There are deviations from the \lcdm\ best fit (constrained by \planck) at the level of a few percent in the damping tail for the TT and EE spectra. The largest deviation from the \lcdm\ reference is seen in the cross-spectrum, where all of the predictions deviate from the \planck\ prediction.}

\changed{Taking the $H_0 = 71$ km/s/Mpc bin mean as a representative example of a high-$H_0$ recombination history, we find that $\chi^2_{\rm SPT} = 205$. The binned mock SPT data, after large scale modes are removed, consists of 113 data points which are constraining the 13 parameters of the \modrec\ model. The probability for $\chi^2_{\rm SPT}$ to exceed 205 for this mock data is approximately $2 \times 10^{-9}$, indicating that forthcoming SPT data will be able to rule out (or confirm) such a deviation at extremely high confidence \footnote{\changed{For a Gaussian distribution, this would correspond to a $5.9\sigma$ detection, although we note that the \modrec\ posteriors are highly non-Gaussian.}}. Precise measurements of the temperature and polarization power spectra at small scales will therefore be important for constraining alternative models of the recombination era.} 

\section{Summary and conclusions}\label{sec:conclusion}

We introduced a phenomenological model of modified recombination which allowed us to study possible departures that $\Xe(z)$, the ionization fraction, might have from its \lcdm\ values during hydrogen recombination, $533 < z < 1600$. Using this model, we reconstructed the range of possible recombination histories conditioned on combinations of: CMB data from \planck, BAO data from eBOSS and SDSS DR7, and the distance ladder-based $H_0$ determination from the \shoes team.  We did not employ a linear response approximation, which allowed us to explore a functional space containing good approximations to ionization histories arising from a variety of physical models. These reconstructions are possible due to the sensitivity of the photon scattering rate to the ionization history. This analysis relied on the use of a neural network emulator to accelerate the inference process.

We found no preference for non-standard recombination from \planck\ data alone, and from \planck+BAO data in combination. However, using only \planck\ data, we found that uncertainties in cosmological parameters are significantly increased when varying the recombination model, in contrast to the case when a linear response approximation is employed. The dual sensitivity of BAO measurements to the cosmic sound horizon at recombination and to low redshift distances makes it very helpful in our reconstructions and determination of cosmological parameters in the presence of modified recombination, highlighting its continued importance in constraining alternative models for the recombination era. We also forecasted the impact of forthcoming data from \sptthreeg, and found that these data will reduce the ionization fraction reconstruction uncertainty to approximately 10\% at 95\% confidence.

Our work has implications for solutions to the $H_0$ tension which rely on modifications to the recombination process. We have identified recombination histories consistent with current data which can accommodate high values of $H_0$, providing possible targets for model builders. However, the aforementioned data combination did not prefer these recombination histories over the standard one. Only when including the \shoes determination in our likelihood does a non-standard recombination history become preferred. Including \shoes data prefers a visibility function which peaks early, has increased width, and is skewed towards higher redshifts. The latter aspect is also present in models with varying electron mass or early structure formation. Throughout, we have emphasized the role of high dimensional information available from damping and polarization generation, which enables the reconstruction of these free functions.

Because of the tight constraints placed by BAO data on modified recombination, and the impact this has on such solutions to the $H_0$ tension, in Sec.~\ref{sec:Hubble} we have discussed how these constraints may be evaded by using the FFAT scaling symmetry of \citet{Cyr-Racine:2021oal}. We conclude that any model resolving the $H_0$ tension must have a mechanism to maintain BAO-consistent values of $D_A(z)$ and $D_H(z)$ at low redshifts. This cannot be accomplished through changes to recombination alone. 

Our analysis highlights a number of future directions. Firstly, it may be interesting to directly phenomenologically model the skewness and higher order visibility moments, going beyond a Gaussian with mean and variance. This might reveal another independent perspective on the origin of the Hubble tension in an efficient way. The addition of new data from SPT-3G may be vital in this.

Secondly, we demonstrated that the cosmological data can in principle be used to directly reconstruct the ionization history. However, once including a modeling of the freeze-out tail or reionization history one can expect a significant reconstruction penalty, highlighting how the constraints may still be driven by prior choices. Ultimately, it will be important to find more direct ways of {\it measuring} the recombination history. The most direct constraints could be deduced from the cosmological recombination radiation \citep{Sunyaev2009}, which for \lcdm\ can be computed accurately using {\tt CosmoSpec} \citep{Chluba2016Cosmospec}. 
However, non-standard recombination processes and parameter variations lead to changes of the recombination radiation \citep{Chluba:2007zz, Chluba:2008aw, Chluba2010dm}.
In this context, it was recently demonstrated that models relating to varying electron mass, early structure formation and even early-dark energy may be distinguishable \citep{Hart2023CRR, Lucca:2023cdl}. However, futuristic CMB spectrometers will be required for this endeavour \citep{Mayuri2015Detection, Vincent2015, Chluba2019Spectral, Hart2020Sensitivity, Chluba2019Voyage}, calling for a significant investment from the community.

Finally, we note that days before finishing this paper new BAO results were released by the DESI Collaboration \citep{DESI:2024uvr, DESI:2024lzq}. These data are in some degree of tension with the \lcdm\ model conditioned on \planck\ data \citep{DESI:2024mwx}. \changed{We explore the implications of the DESI data for modified recombination and the Hubble tension in a separate work \citep{Lynch:2024hzh}}. It can already be anticipated that from the observational point of view extra freedom in the ionization history will play a crucial role in quantifying the significance of any departure from the \lcdm\ model. In addition, the assumption of the standard recombination history implies a significant theoretical prior that has to be carefully evaluated. With the present work, we now have the tools to explore these important questions in upcoming analyses.

{\small
\section*{Acknowledgements}
GL would like to thank Emil Holm and Andreas Nygaard for many useful discussions regarding {\tt CONNECT}. LK and JC would like to thank the Aspen Center for Physics for their hospitality during the Summer of 2021, where this projects started.
LK and GL were supported in part by DOE Office of Science award DE-SC0009999.
JC was supported by the ERC Consolidator Grant {\it CMBSPEC} (No.~725456) and by the Royal Society as a Royal Society University Research Fellow at the University of Manchester, UK (No.~URF/R/191023).}

\bibliography{apssamp}

\providecommand{\noopsort}[1]{}\providecommand{\singleletter}[1]{#1}%
\begin{thebibliography}{102}%
\makeatletter
\providecommand \@ifxundefined [1]{%
 \@ifx{#1\undefined}
}%
\providecommand \@ifnum [1]{%
 \ifnum #1\expandafter \@firstoftwo
 \else \expandafter \@secondoftwo
 \fi
}%
\providecommand \@ifx [1]{%
 \ifx #1\expandafter \@firstoftwo
 \else \expandafter \@secondoftwo
 \fi
}%
\providecommand \natexlab [1]{#1}%
\providecommand \enquote  [1]{``#1''}%
\providecommand \bibnamefont  [1]{#1}%
\providecommand \bibfnamefont [1]{#1}%
\providecommand \citenamefont [1]{#1}%
\providecommand \href@noop [0]{\@secondoftwo}%
\providecommand \href [0]{\begingroup \@sanitize@url \@href}%
\providecommand \@href[1]{\@@startlink{#1}\@@href}%
\providecommand \@@href[1]{\endgroup#1\@@endlink}%
\providecommand \@sanitize@url [0]{\catcode `\\12\catcode `\$12\catcode
  `\&12\catcode `\#12\catcode `\^12\catcode `\_12\catcode `\%12\relax}%
\providecommand \@@startlink[1]{}%
\providecommand \@@endlink[0]{}%
\providecommand \url  [0]{\begingroup\@sanitize@url \@url }%
\providecommand \@url [1]{\endgroup\@href {#1}{\urlprefix }}%
\providecommand \urlprefix  [0]{URL }%
\providecommand \Eprint [0]{\href }%
\providecommand \doibase [0]{https://doi.org/}%
\providecommand \selectlanguage [0]{\@gobble}%
\providecommand \bibinfo  [0]{\@secondoftwo}%
\providecommand \bibfield  [0]{\@secondoftwo}%
\providecommand \translation [1]{[#1]}%
\providecommand \BibitemOpen [0]{}%
\providecommand \bibitemStop [0]{}%
\providecommand \bibitemNoStop [0]{.\EOS\space}%
\providecommand \EOS [0]{\spacefactor3000\relax}%
\providecommand \BibitemShut  [1]{\csname bibitem#1\endcsname}%
\let\auto@bib@innerbib\@empty
\bibitem [{\citenamefont {Aghanim}\ \emph
  {et~al.}(2020{\natexlab{a}})\citenamefont {Aghanim} \emph
  {et~al.}}]{Planck:2018vyg}%
  \BibitemOpen
  \bibfield  {author} {\bibinfo {author} {\bibfnamefont {N.}~\bibnamefont
  {Aghanim}} \emph {et~al.} (\bibinfo {collaboration} {Planck}),\ }\bibfield
  {title} {\bibinfo {title} {{Planck 2018 results. VI. Cosmological
  parameters}},\ }\href {https://doi.org/10.1051/0004-6361/201833910}
  {\bibfield  {journal} {\bibinfo  {journal} {Astron. Astrophys.}\ }\textbf
  {\bibinfo {volume} {641}},\ \bibinfo {pages} {A6} (\bibinfo {year}
  {2020}{\natexlab{a}})},\ \bibinfo {note} {[Erratum: Astron.Astrophys. 652, C4
  (2021)]},\ \Eprint {https://arxiv.org/abs/1807.06209} {arXiv:1807.06209
  [astro-ph.CO]} \BibitemShut {NoStop}%
\bibitem [{\citenamefont {{Hu}}\ \emph {et~al.}(1995)\citenamefont {{Hu}},
  \citenamefont {{Scott}}, \citenamefont {{Sugiyama}},\ and\ \citenamefont
  {{White}}}]{Hu1995}%
  \BibitemOpen
  \bibfield  {author} {\bibinfo {author} {\bibfnamefont {W.}~\bibnamefont
  {{Hu}}}, \bibinfo {author} {\bibfnamefont {D.}~\bibnamefont {{Scott}}},
  \bibinfo {author} {\bibfnamefont {N.}~\bibnamefont {{Sugiyama}}},\ and\
  \bibinfo {author} {\bibfnamefont {M.}~\bibnamefont {{White}}},\ }\bibfield
  {title} {\bibinfo {title} {{Effect of physical assumptions on the calculation
  of microwave background anisotropies}},\ }\href@noop {} {\bibfield  {journal}
  {\bibinfo  {journal} {\prd}\ }\textbf {\bibinfo {volume} {52}},\ \bibinfo
  {pages} {5498} (\bibinfo {year} {1995})},\ \Eprint
  {https://arxiv.org/abs/arXiv:astro-ph/9505043} {arXiv:astro-ph/9505043}
  \BibitemShut {NoStop}%
\bibitem [{\citenamefont {Chluba}\ and\ \citenamefont
  {Sunyaev}(2006)}]{Chluba:2005uz}%
  \BibitemOpen
  \bibfield  {author} {\bibinfo {author} {\bibfnamefont {J.}~\bibnamefont
  {Chluba}}\ and\ \bibinfo {author} {\bibfnamefont {R.~A.}\ \bibnamefont
  {Sunyaev}},\ }\bibfield  {title} {\bibinfo {title} {{Induced two-photon decay
  of the 2s level and the rate of cosmological hydrogen recombination}},\
  }\href {https://doi.org/10.1051/0004-6361:20053988} {\bibfield  {journal}
  {\bibinfo  {journal} {Astron. Astrophys.}\ }\textbf {\bibinfo {volume}
  {446}},\ \bibinfo {pages} {39} (\bibinfo {year} {2006})},\ \Eprint
  {https://arxiv.org/abs/astro-ph/0508144} {arXiv:astro-ph/0508144}
  \BibitemShut {NoStop}%
\bibitem [{\citenamefont {{Lewis}}\ \emph {et~al.}(2006)\citenamefont
  {{Lewis}}, \citenamefont {{Weller}},\ and\ \citenamefont
  {{Battye}}}]{Lewis2006}%
  \BibitemOpen
  \bibfield  {author} {\bibinfo {author} {\bibfnamefont {A.}~\bibnamefont
  {{Lewis}}}, \bibinfo {author} {\bibfnamefont {J.}~\bibnamefont {{Weller}}},\
  and\ \bibinfo {author} {\bibfnamefont {R.}~\bibnamefont {{Battye}}},\
  }\bibfield  {title} {\bibinfo {title} {{The cosmic microwave background and
  the ionization history of the Universe}},\ }\href
  {https://doi.org/10.1111/j.1365-2966.2006.10983.x} {\bibfield  {journal}
  {\bibinfo  {journal} {\mnras}\ }\textbf {\bibinfo {volume} {373}},\ \bibinfo
  {pages} {561} (\bibinfo {year} {2006})},\ \Eprint
  {https://arxiv.org/abs/astro-ph/0606552} {astro-ph/0606552} \BibitemShut
  {NoStop}%
\bibitem [{\citenamefont {{Rubi{\~n}o-Mart{\'\i}n}}\ \emph
  {et~al.}(2010)\citenamefont {{Rubi{\~n}o-Mart{\'\i}n}}, \citenamefont
  {{Chluba}}, \citenamefont {{Fendt}},\ and\ \citenamefont
  {{Wandelt}}}]{Jose2010}%
  \BibitemOpen
  \bibfield  {author} {\bibinfo {author} {\bibfnamefont {J.~A.}\ \bibnamefont
  {{Rubi{\~n}o-Mart{\'\i}n}}}, \bibinfo {author} {\bibfnamefont
  {J.}~\bibnamefont {{Chluba}}}, \bibinfo {author} {\bibfnamefont {W.~A.}\
  \bibnamefont {{Fendt}}},\ and\ \bibinfo {author} {\bibfnamefont {B.~D.}\
  \bibnamefont {{Wandelt}}},\ }\bibfield  {title} {\bibinfo {title}
  {{Estimating the impact of recombination uncertainties on the cosmological
  parameter constraints from cosmic microwave background experiments}},\ }\href
  {https://doi.org/10.1111/j.1365-2966.2009.16136.x} {\bibfield  {journal}
  {\bibinfo  {journal} {\mnras}\ }\textbf {\bibinfo {volume} {403}},\ \bibinfo
  {pages} {439} (\bibinfo {year} {2010})},\ \Eprint
  {https://arxiv.org/abs/0910.4383} {arXiv:0910.4383 [astro-ph.CO]}
  \BibitemShut {NoStop}%
\bibitem [{\citenamefont {Ade}\ \emph {et~al.}(2016)\citenamefont {Ade} \emph
  {et~al.}}]{Planck:2015fie}%
  \BibitemOpen
  \bibfield  {author} {\bibinfo {author} {\bibfnamefont {P.~A.~R.}\
  \bibnamefont {Ade}} \emph {et~al.} (\bibinfo {collaboration} {Planck}),\
  }\bibfield  {title} {\bibinfo {title} {{Planck 2015 results. XIII.
  Cosmological parameters}},\ }\href
  {https://doi.org/10.1051/0004-6361/201525830} {\bibfield  {journal} {\bibinfo
   {journal} {Astron. Astrophys.}\ }\textbf {\bibinfo {volume} {594}},\
  \bibinfo {pages} {A13} (\bibinfo {year} {2016})},\ \Eprint
  {https://arxiv.org/abs/1502.01589} {arXiv:1502.01589 [astro-ph.CO]}
  \BibitemShut {NoStop}%
\bibitem [{\citenamefont {Farhang}\ \emph {et~al.}(2012)\citenamefont
  {Farhang}, \citenamefont {Bond},\ and\ \citenamefont
  {Chluba}}]{Farhang:2011pt}%
  \BibitemOpen
  \bibfield  {author} {\bibinfo {author} {\bibfnamefont {M.}~\bibnamefont
  {Farhang}}, \bibinfo {author} {\bibfnamefont {J.~R.}\ \bibnamefont {Bond}},\
  and\ \bibinfo {author} {\bibfnamefont {J.}~\bibnamefont {Chluba}},\
  }\bibfield  {title} {\bibinfo {title} {{Semi-blind Eigen-analyses of
  Recombination Histories Using CMB Data}},\ }\href
  {https://doi.org/10.1088/0004-637X/752/2/88} {\bibfield  {journal} {\bibinfo
  {journal} {Astrophys. J.}\ }\textbf {\bibinfo {volume} {752}},\ \bibinfo
  {pages} {88} (\bibinfo {year} {2012})},\ \Eprint
  {https://arxiv.org/abs/1110.4608} {arXiv:1110.4608 [astro-ph.CO]}
  \BibitemShut {NoStop}%
\bibitem [{\citenamefont {{Hart}}\ and\ \citenamefont
  {{Chluba}}(2020{\natexlab{a}})}]{Hart:2019gvj}%
  \BibitemOpen
  \bibfield  {author} {\bibinfo {author} {\bibfnamefont {L.}~\bibnamefont
  {{Hart}}}\ and\ \bibinfo {author} {\bibfnamefont {J.}~\bibnamefont
  {{Chluba}}},\ }\bibfield  {title} {\bibinfo {title} {{Improved
  model-independent constraints on the recombination era and development of a
  direct projection method}},\ }\href {https://doi.org/10.1093/mnras/staa1426}
  {\bibfield  {journal} {\bibinfo  {journal} {\mnras}\ }\textbf {\bibinfo
  {volume} {495}},\ \bibinfo {pages} {4210} (\bibinfo {year}
  {2020}{\natexlab{a}})},\ \Eprint {https://arxiv.org/abs/1912.04682}
  {arXiv:1912.04682 [astro-ph.CO]} \BibitemShut {NoStop}%
\bibitem [{\citenamefont {Adams}\ \emph {et~al.}(1998)\citenamefont {Adams},
  \citenamefont {Sarkar},\ and\ \citenamefont {Sciama}}]{Adams:1998nr}%
  \BibitemOpen
  \bibfield  {author} {\bibinfo {author} {\bibfnamefont {J.~A.}\ \bibnamefont
  {Adams}}, \bibinfo {author} {\bibfnamefont {S.}~\bibnamefont {Sarkar}},\ and\
  \bibinfo {author} {\bibfnamefont {D.~W.}\ \bibnamefont {Sciama}},\ }\bibfield
   {title} {\bibinfo {title} {{CMB anisotropy in the decaying neutrino
  cosmology}},\ }\href {https://doi.org/10.1046/j.1365-8711.1998.02017.x}
  {\bibfield  {journal} {\bibinfo  {journal} {Mon. Not. Roy. Astron. Soc.}\
  }\textbf {\bibinfo {volume} {301}},\ \bibinfo {pages} {210} (\bibinfo {year}
  {1998})},\ \Eprint {https://arxiv.org/abs/astro-ph/9805108}
  {arXiv:astro-ph/9805108} \BibitemShut {NoStop}%
\bibitem [{\citenamefont {{Chen}}\ and\ \citenamefont
  {{Kamionkowski}}(2004)}]{Chen2004}%
  \BibitemOpen
  \bibfield  {author} {\bibinfo {author} {\bibfnamefont {X.}~\bibnamefont
  {{Chen}}}\ and\ \bibinfo {author} {\bibfnamefont {M.}~\bibnamefont
  {{Kamionkowski}}},\ }\bibfield  {title} {\bibinfo {title} {{Particle decays
  during the cosmic dark ages}},\ }\href
  {https://doi.org/10.1103/PhysRevD.70.043502} {\bibfield  {journal} {\bibinfo
  {journal} {\prd}\ }\textbf {\bibinfo {volume} {70}},\ \bibinfo {pages}
  {043502} (\bibinfo {year} {2004})},\ \Eprint
  {https://arxiv.org/abs/arXiv:astro-ph/0310473} {arXiv:astro-ph/0310473}
  \BibitemShut {NoStop}%
\bibitem [{\citenamefont {Galli}\ \emph {et~al.}(2013)\citenamefont {Galli},
  \citenamefont {Slatyer}, \citenamefont {Valdes},\ and\ \citenamefont
  {Iocco}}]{Galli:2013dna}%
  \BibitemOpen
  \bibfield  {author} {\bibinfo {author} {\bibfnamefont {S.}~\bibnamefont
  {Galli}}, \bibinfo {author} {\bibfnamefont {T.~R.}\ \bibnamefont {Slatyer}},
  \bibinfo {author} {\bibfnamefont {M.}~\bibnamefont {Valdes}},\ and\ \bibinfo
  {author} {\bibfnamefont {F.}~\bibnamefont {Iocco}},\ }\bibfield  {title}
  {\bibinfo {title} {{Systematic Uncertainties In Constraining Dark Matter
  Annihilation From The Cosmic Microwave Background}},\ }\href
  {https://doi.org/10.1103/PhysRevD.88.063502} {\bibfield  {journal} {\bibinfo
  {journal} {Phys. Rev. D}\ }\textbf {\bibinfo {volume} {88}},\ \bibinfo
  {pages} {063502} (\bibinfo {year} {2013})},\ \Eprint
  {https://arxiv.org/abs/1306.0563} {arXiv:1306.0563 [astro-ph.CO]}
  \BibitemShut {NoStop}%
\bibitem [{\citenamefont {Slatyer}\ and\ \citenamefont
  {Wu}(2017)}]{Slatyer:2016qyl}%
  \BibitemOpen
  \bibfield  {author} {\bibinfo {author} {\bibfnamefont {T.~R.}\ \bibnamefont
  {Slatyer}}\ and\ \bibinfo {author} {\bibfnamefont {C.-L.}\ \bibnamefont
  {Wu}},\ }\bibfield  {title} {\bibinfo {title} {{General Constraints on Dark
  Matter Decay from the Cosmic Microwave Background}},\ }\href
  {https://doi.org/10.1103/PhysRevD.95.023010} {\bibfield  {journal} {\bibinfo
  {journal} {Phys. Rev. D}\ }\textbf {\bibinfo {volume} {95}},\ \bibinfo
  {pages} {023010} (\bibinfo {year} {2017})},\ \Eprint
  {https://arxiv.org/abs/1610.06933} {arXiv:1610.06933 [astro-ph.CO]}
  \BibitemShut {NoStop}%
\bibitem [{\citenamefont {Finkbeiner}\ \emph {et~al.}(2012)\citenamefont
  {Finkbeiner}, \citenamefont {Galli}, \citenamefont {Lin},\ and\ \citenamefont
  {Slatyer}}]{Finkbeiner:2011dx}%
  \BibitemOpen
  \bibfield  {author} {\bibinfo {author} {\bibfnamefont {D.~P.}\ \bibnamefont
  {Finkbeiner}}, \bibinfo {author} {\bibfnamefont {S.}~\bibnamefont {Galli}},
  \bibinfo {author} {\bibfnamefont {T.}~\bibnamefont {Lin}},\ and\ \bibinfo
  {author} {\bibfnamefont {T.~R.}\ \bibnamefont {Slatyer}},\ }\bibfield
  {title} {\bibinfo {title} {{Searching for Dark Matter in the CMB: A Compact
  Parameterization of Energy Injection from New Physics}},\ }\href
  {https://doi.org/10.1103/PhysRevD.85.043522} {\bibfield  {journal} {\bibinfo
  {journal} {Phys. Rev. D}\ }\textbf {\bibinfo {volume} {85}},\ \bibinfo
  {pages} {043522} (\bibinfo {year} {2012})},\ \Eprint
  {https://arxiv.org/abs/1109.6322} {arXiv:1109.6322 [astro-ph.CO]}
  \BibitemShut {NoStop}%
\bibitem [{\citenamefont {Ade}\ \emph {et~al.}(2015)\citenamefont {Ade} \emph
  {et~al.}}]{Planck:2014ylh}%
  \BibitemOpen
  \bibfield  {author} {\bibinfo {author} {\bibfnamefont {P.~A.~R.}\
  \bibnamefont {Ade}} \emph {et~al.} (\bibinfo {collaboration} {Planck}),\
  }\bibfield  {title} {\bibinfo {title} {{Planck intermediate results - XXIV.
  Constraints on variations in fundamental constants}},\ }\href
  {https://doi.org/10.1051/0004-6361/201424496} {\bibfield  {journal} {\bibinfo
   {journal} {Astron. Astrophys.}\ }\textbf {\bibinfo {volume} {580}},\
  \bibinfo {pages} {A22} (\bibinfo {year} {2015})},\ \Eprint
  {https://arxiv.org/abs/1406.7482} {arXiv:1406.7482 [astro-ph.CO]}
  \BibitemShut {NoStop}%
\bibitem [{\citenamefont {Hart}\ and\ \citenamefont
  {Chluba}(2018)}]{Hart:2017ndk}%
  \BibitemOpen
  \bibfield  {author} {\bibinfo {author} {\bibfnamefont {L.}~\bibnamefont
  {Hart}}\ and\ \bibinfo {author} {\bibfnamefont {J.}~\bibnamefont {Chluba}},\
  }\bibfield  {title} {\bibinfo {title} {{New constraints on time-dependent
  variations of fundamental constants using Planck data}},\ }\href
  {https://doi.org/10.1093/mnras/stx2783} {\bibfield  {journal} {\bibinfo
  {journal} {Mon. Not. Roy. Astron. Soc.}\ }\textbf {\bibinfo {volume} {474}},\
  \bibinfo {pages} {1850} (\bibinfo {year} {2018})},\ \Eprint
  {https://arxiv.org/abs/1705.03925} {arXiv:1705.03925 [astro-ph.CO]}
  \BibitemShut {NoStop}%
\bibitem [{\citenamefont {{Hart}}\ and\ \citenamefont
  {{Chluba}}(2020{\natexlab{b}})}]{Hart2020H0}%
  \BibitemOpen
  \bibfield  {author} {\bibinfo {author} {\bibfnamefont {L.}~\bibnamefont
  {{Hart}}}\ and\ \bibinfo {author} {\bibfnamefont {J.}~\bibnamefont
  {{Chluba}}},\ }\bibfield  {title} {\bibinfo {title} {{Updated fundamental
  constant constraints from Planck 2018 data and possible relations to the
  Hubble tension}},\ }\href {https://doi.org/10.1093/mnras/staa412} {\bibfield
  {journal} {\bibinfo  {journal} {\mnras}\ }\textbf {\bibinfo {volume} {493}},\
  \bibinfo {pages} {3255} (\bibinfo {year} {2020}{\natexlab{b}})},\ \Eprint
  {https://arxiv.org/abs/1912.03986} {arXiv:1912.03986 [astro-ph.CO]}
  \BibitemShut {NoStop}%
\bibitem [{\citenamefont {Sekiguchi}\ and\ \citenamefont
  {Takahashi}(2021)}]{Sekiguchi:2020teg}%
  \BibitemOpen
  \bibfield  {author} {\bibinfo {author} {\bibfnamefont {T.}~\bibnamefont
  {Sekiguchi}}\ and\ \bibinfo {author} {\bibfnamefont {T.}~\bibnamefont
  {Takahashi}},\ }\bibfield  {title} {\bibinfo {title} {{Early recombination as
  a solution to the $H_0$ tension}},\ }\href
  {https://doi.org/10.1103/PhysRevD.103.083507} {\bibfield  {journal} {\bibinfo
   {journal} {Phys. Rev. D}\ }\textbf {\bibinfo {volume} {103}},\ \bibinfo
  {pages} {083507} (\bibinfo {year} {2021})},\ \Eprint
  {https://arxiv.org/abs/2007.03381} {arXiv:2007.03381 [astro-ph.CO]}
  \BibitemShut {NoStop}%
\bibitem [{\citenamefont {Jedamzik}\ and\ \citenamefont
  {Pogosian}(2020)}]{Jedamzik2020Relieving}%
  \BibitemOpen
  \bibfield  {author} {\bibinfo {author} {\bibfnamefont {K.}~\bibnamefont
  {Jedamzik}}\ and\ \bibinfo {author} {\bibfnamefont {L.}~\bibnamefont
  {Pogosian}},\ }\bibfield  {title} {\bibinfo {title} {{Relieving the Hubble
  tension with primordial magnetic fields}},\ }\href
  {https://doi.org/10.1103/PhysRevLett.125.181302} {\bibfield  {journal}
  {\bibinfo  {journal} {Phys. Rev. Lett.}\ }\textbf {\bibinfo {volume} {125}},\
  \bibinfo {pages} {181302} (\bibinfo {year} {2020})},\ \Eprint
  {https://arxiv.org/abs/2004.09487} {arXiv:2004.09487 [astro-ph.CO]}
  \BibitemShut {NoStop}%
\bibitem [{\citenamefont {{Thiele}}\ \emph {et~al.}(2021)\citenamefont
  {{Thiele}}, \citenamefont {{Guan}}, \citenamefont {{Hill}}, \citenamefont
  {{Kosowsky}},\ and\ \citenamefont {{Spergel}}}]{Thiele2021}%
  \BibitemOpen
  \bibfield  {author} {\bibinfo {author} {\bibfnamefont {L.}~\bibnamefont
  {{Thiele}}}, \bibinfo {author} {\bibfnamefont {Y.}~\bibnamefont {{Guan}}},
  \bibinfo {author} {\bibfnamefont {J.~C.}\ \bibnamefont {{Hill}}}, \bibinfo
  {author} {\bibfnamefont {A.}~\bibnamefont {{Kosowsky}}},\ and\ \bibinfo
  {author} {\bibfnamefont {D.~N.}\ \bibnamefont {{Spergel}}},\ }\bibfield
  {title} {\bibinfo {title} {{Can small-scale baryon inhomogeneities resolve
  the Hubble tension? An investigation with ACT DR4}},\ }\href
  {https://doi.org/10.1103/PhysRevD.104.063535} {\bibfield  {journal} {\bibinfo
   {journal} {\prd}\ }\textbf {\bibinfo {volume} {104}},\ \bibinfo {eid}
  {063535} (\bibinfo {year} {2021})},\ \Eprint
  {https://arxiv.org/abs/2105.03003} {arXiv:2105.03003 [astro-ph.CO]}
  \BibitemShut {NoStop}%
\bibitem [{\citenamefont {Rashkovetskyi}\ \emph {et~al.}(2021)\citenamefont
  {Rashkovetskyi}, \citenamefont {Mu\~noz}, \citenamefont {Eisenstein},\ and\
  \citenamefont {Dvorkin}}]{Rashkovetskyi:2021rwg}%
  \BibitemOpen
  \bibfield  {author} {\bibinfo {author} {\bibfnamefont {M.}~\bibnamefont
  {Rashkovetskyi}}, \bibinfo {author} {\bibfnamefont {J.~B.}\ \bibnamefont
  {Mu\~noz}}, \bibinfo {author} {\bibfnamefont {D.~J.}\ \bibnamefont
  {Eisenstein}},\ and\ \bibinfo {author} {\bibfnamefont {C.}~\bibnamefont
  {Dvorkin}},\ }\bibfield  {title} {\bibinfo {title} {{Small-scale clumping at
  recombination and the Hubble tension}},\ }\href
  {https://doi.org/10.1103/PhysRevD.104.103517} {\bibfield  {journal} {\bibinfo
   {journal} {Phys. Rev. D}\ }\textbf {\bibinfo {volume} {104}},\ \bibinfo
  {pages} {103517} (\bibinfo {year} {2021})},\ \Eprint
  {https://arxiv.org/abs/2108.02747} {arXiv:2108.02747 [astro-ph.CO]}
  \BibitemShut {NoStop}%
\bibitem [{\citenamefont {Galli}\ \emph {et~al.}(2022)\citenamefont {Galli},
  \citenamefont {Pogosian}, \citenamefont {Jedamzik},\ and\ \citenamefont
  {Balkenhol}}]{Galli:2021mxk}%
  \BibitemOpen
  \bibfield  {author} {\bibinfo {author} {\bibfnamefont {S.}~\bibnamefont
  {Galli}}, \bibinfo {author} {\bibfnamefont {L.}~\bibnamefont {Pogosian}},
  \bibinfo {author} {\bibfnamefont {K.}~\bibnamefont {Jedamzik}},\ and\
  \bibinfo {author} {\bibfnamefont {L.}~\bibnamefont {Balkenhol}},\ }\bibfield
  {title} {\bibinfo {title} {{Consistency of Planck, ACT, and SPT constraints
  on magnetically assisted recombination and forecasts for future
  experiments}},\ }\href {https://doi.org/10.1103/PhysRevD.105.023513}
  {\bibfield  {journal} {\bibinfo  {journal} {Phys. Rev. D}\ }\textbf {\bibinfo
  {volume} {105}},\ \bibinfo {pages} {023513} (\bibinfo {year} {2022})},\
  \Eprint {https://arxiv.org/abs/2109.03816} {arXiv:2109.03816 [astro-ph.CO]}
  \BibitemShut {NoStop}%
\bibitem [{\citenamefont {Farhang}\ \emph {et~al.}(2013)\citenamefont
  {Farhang}, \citenamefont {Bond}, \citenamefont {Chluba},\ and\ \citenamefont
  {Switzer}}]{Farhang:2012jz}%
  \BibitemOpen
  \bibfield  {author} {\bibinfo {author} {\bibfnamefont {M.}~\bibnamefont
  {Farhang}}, \bibinfo {author} {\bibfnamefont {J.~R.}\ \bibnamefont {Bond}},
  \bibinfo {author} {\bibfnamefont {J.}~\bibnamefont {Chluba}},\ and\ \bibinfo
  {author} {\bibfnamefont {E.~R.}\ \bibnamefont {Switzer}},\ }\bibfield
  {title} {\bibinfo {title} {{Constraints on perturbations to the recombination
  history from measurements of the CMB damping tail}},\ }\href
  {https://doi.org/10.1088/0004-637X/764/2/137} {\bibfield  {journal} {\bibinfo
   {journal} {Astrophys. J.}\ }\textbf {\bibinfo {volume} {764}},\ \bibinfo
  {pages} {137} (\bibinfo {year} {2013})},\ \Eprint
  {https://arxiv.org/abs/1211.4634} {arXiv:1211.4634 [astro-ph.CO]}
  \BibitemShut {NoStop}%
\bibitem [{\citenamefont {Lee}\ \emph {et~al.}(2023)\citenamefont {Lee},
  \citenamefont {Ali-Ha\"\i{}moud}, \citenamefont {Sch\"oneberg},\ and\
  \citenamefont {Poulin}}]{Lee:2022gzh}%
  \BibitemOpen
  \bibfield  {author} {\bibinfo {author} {\bibfnamefont {N.}~\bibnamefont
  {Lee}}, \bibinfo {author} {\bibfnamefont {Y.}~\bibnamefont
  {Ali-Ha\"\i{}moud}}, \bibinfo {author} {\bibfnamefont {N.}~\bibnamefont
  {Sch\"oneberg}},\ and\ \bibinfo {author} {\bibfnamefont {V.}~\bibnamefont
  {Poulin}},\ }\bibfield  {title} {\bibinfo {title} {{What It Takes to Solve
  the Hubble Tension through Modifications of Cosmological Recombination}},\
  }\href {https://doi.org/10.1103/PhysRevLett.130.161003} {\bibfield  {journal}
  {\bibinfo  {journal} {Phys. Rev. Lett.}\ }\textbf {\bibinfo {volume} {130}},\
  \bibinfo {pages} {161003} (\bibinfo {year} {2023})},\ \Eprint
  {https://arxiv.org/abs/2212.04494} {arXiv:2212.04494 [astro-ph.CO]}
  \BibitemShut {NoStop}%
\bibitem [{\citenamefont {Nygaard}\ \emph {et~al.}(2023)\citenamefont
  {Nygaard}, \citenamefont {Holm}, \citenamefont {Hannestad},\ and\
  \citenamefont {Tram}}]{Nygaard:2022wri}%
  \BibitemOpen
  \bibfield  {author} {\bibinfo {author} {\bibfnamefont {A.}~\bibnamefont
  {Nygaard}}, \bibinfo {author} {\bibfnamefont {E.~B.}\ \bibnamefont {Holm}},
  \bibinfo {author} {\bibfnamefont {S.}~\bibnamefont {Hannestad}},\ and\
  \bibinfo {author} {\bibfnamefont {T.}~\bibnamefont {Tram}},\ }\bibfield
  {title} {\bibinfo {title} {{CONNECT: a neural network based framework for
  emulating cosmological observables and cosmological parameter inference}},\
  }\href {https://doi.org/10.1088/1475-7516/2023/05/025} {\bibfield  {journal}
  {\bibinfo  {journal} {JCAP}\ }\textbf {\bibinfo {volume} {05}},\ \bibinfo
  {pages} {025}},\ \Eprint {https://arxiv.org/abs/2205.15726} {arXiv:2205.15726
  [astro-ph.IM]} \BibitemShut {NoStop}%
\bibitem [{\citenamefont {Chiang}\ and\ \citenamefont
  {Slosar}(2018)}]{Chiang:2018xpn}%
  \BibitemOpen
  \bibfield  {author} {\bibinfo {author} {\bibfnamefont {C.-T.}\ \bibnamefont
  {Chiang}}\ and\ \bibinfo {author} {\bibfnamefont {A.}~\bibnamefont
  {Slosar}},\ }\bibfield  {title} {\bibinfo {title} {{Inferences of $H_0$ in
  presence of a non-standard recombination}},\ }\href@noop {} {\  (\bibinfo
  {year} {2018})},\ \Eprint {https://arxiv.org/abs/1811.03624}
  {arXiv:1811.03624 [astro-ph.CO]} \BibitemShut {NoStop}%
\bibitem [{\citenamefont {Verde}\ \emph {et~al.}(2023)\citenamefont {Verde},
  \citenamefont {Sch\"oneberg},\ and\ \citenamefont
  {Gil-Mar\'\i{}n}}]{Verde:2023lmm}%
  \BibitemOpen
  \bibfield  {author} {\bibinfo {author} {\bibfnamefont {L.}~\bibnamefont
  {Verde}}, \bibinfo {author} {\bibfnamefont {N.}~\bibnamefont
  {Sch\"oneberg}},\ and\ \bibinfo {author} {\bibfnamefont {H.}~\bibnamefont
  {Gil-Mar\'\i{}n}},\ }\bibfield  {title} {\bibinfo {title} {{A tale of many
  $H_0$}},\ }\href@noop {} {\  (\bibinfo {year} {2023})},\ \Eprint
  {https://arxiv.org/abs/2311.13305} {arXiv:2311.13305 [astro-ph.CO]}
  \BibitemShut {NoStop}%
\bibitem [{\citenamefont {Lucca}\ \emph {et~al.}(2023)\citenamefont {Lucca},
  \citenamefont {Chluba},\ and\ \citenamefont {Rotti}}]{Lucca:2023cdl}%
  \BibitemOpen
  \bibfield  {author} {\bibinfo {author} {\bibfnamefont {M.}~\bibnamefont
  {Lucca}}, \bibinfo {author} {\bibfnamefont {J.}~\bibnamefont {Chluba}},\ and\
  \bibinfo {author} {\bibfnamefont {A.}~\bibnamefont {Rotti}},\ }\bibfield
  {title} {\bibinfo {title} {{CRRfast: An emulator for the Cosmological
  Recombination Radiation with effects from inhomogeneous recombination}},\
  }\href@noop {} {\  (\bibinfo {year} {2023})},\ \Eprint
  {https://arxiv.org/abs/2306.08085} {arXiv:2306.08085 [astro-ph.CO]}
  \BibitemShut {NoStop}%
\bibitem [{\citenamefont {Cyr-Racine}\ \emph {et~al.}(2022)\citenamefont
  {Cyr-Racine}, \citenamefont {Ge},\ and\ \citenamefont
  {Knox}}]{Cyr-Racine:2021oal}%
  \BibitemOpen
  \bibfield  {author} {\bibinfo {author} {\bibfnamefont {F.-Y.}\ \bibnamefont
  {Cyr-Racine}}, \bibinfo {author} {\bibfnamefont {F.}~\bibnamefont {Ge}},\
  and\ \bibinfo {author} {\bibfnamefont {L.}~\bibnamefont {Knox}},\ }\bibfield
  {title} {\bibinfo {title} {{Symmetry of Cosmological Observables, a Mirror
  World Dark Sector, and the Hubble Constant}},\ }\href
  {https://doi.org/10.1103/PhysRevLett.128.201301} {\bibfield  {journal}
  {\bibinfo  {journal} {Phys. Rev. Lett.}\ }\textbf {\bibinfo {volume} {128}},\
  \bibinfo {pages} {201301} (\bibinfo {year} {2022})},\ \Eprint
  {https://arxiv.org/abs/2107.13000} {arXiv:2107.13000 [astro-ph.CO]}
  \BibitemShut {NoStop}%
\bibitem [{\citenamefont {Chluba}\ and\ \citenamefont
  {Thomas}(2011)}]{Chluba:2010ca}%
  \BibitemOpen
  \bibfield  {author} {\bibinfo {author} {\bibfnamefont {J.}~\bibnamefont
  {Chluba}}\ and\ \bibinfo {author} {\bibfnamefont {R.~M.}\ \bibnamefont
  {Thomas}},\ }\bibfield  {title} {\bibinfo {title} {{Towards a complete
  treatment of the cosmological recombination problem}},\ }\href
  {https://doi.org/10.1111/j.1365-2966.2010.17940.x} {\bibfield  {journal}
  {\bibinfo  {journal} {Mon. Not. Roy. Astron. Soc.}\ }\textbf {\bibinfo
  {volume} {412}},\ \bibinfo {pages} {748} (\bibinfo {year} {2011})},\ \Eprint
  {https://arxiv.org/abs/1010.3631} {arXiv:1010.3631 [astro-ph.CO]}
  \BibitemShut {NoStop}%
\bibitem [{\citenamefont {{Ali-Ha{\"\i}moud}}\ and\ \citenamefont
  {{Hirata}}(2011)}]{2011PhRvD..83d3513A}%
  \BibitemOpen
  \bibfield  {author} {\bibinfo {author} {\bibfnamefont {Y.}~\bibnamefont
  {{Ali-Ha{\"\i}moud}}}\ and\ \bibinfo {author} {\bibfnamefont {C.~M.}\
  \bibnamefont {{Hirata}}},\ }\bibfield  {title} {\bibinfo {title} {{HyRec: A
  fast and highly accurate primordial hydrogen and helium recombination
  code}},\ }\href {https://doi.org/10.1103/PhysRevD.83.043513} {\bibfield
  {journal} {\bibinfo  {journal} {\prd}\ }\textbf {\bibinfo {volume} {83}},\
  \bibinfo {eid} {043513} (\bibinfo {year} {2011})},\ \Eprint
  {https://arxiv.org/abs/1011.3758} {arXiv:1011.3758 [astro-ph.CO]}
  \BibitemShut {NoStop}%
\bibitem [{\citenamefont {{Switzer}}\ and\ \citenamefont
  {{Hirata}}(2008)}]{Switzer2007I}%
  \BibitemOpen
  \bibfield  {author} {\bibinfo {author} {\bibfnamefont {E.~R.}\ \bibnamefont
  {{Switzer}}}\ and\ \bibinfo {author} {\bibfnamefont {C.~M.}\ \bibnamefont
  {{Hirata}}},\ }\bibfield  {title} {\bibinfo {title} {{Primordial helium
  recombination. I. Feedback, line transfer, and continuum opacity}},\ }\href
  {https://doi.org/10.1103/PhysRevD.77.083006} {\bibfield  {journal} {\bibinfo
  {journal} {\prd}\ }\textbf {\bibinfo {volume} {77}},\ \bibinfo {pages}
  {083006} (\bibinfo {year} {2008})},\ \Eprint
  {https://arxiv.org/abs/arXiv:astro-ph/0702143} {arXiv:astro-ph/0702143}
  \BibitemShut {NoStop}%
\bibitem [{\citenamefont {{Kholupenko}}\ \emph {et~al.}(2007)\citenamefont
  {{Kholupenko}}, \citenamefont {{Ivanchik}},\ and\ \citenamefont
  {{Varshalovich}}}]{Kholupenko2007}%
  \BibitemOpen
  \bibfield  {author} {\bibinfo {author} {\bibfnamefont {E.~E.}\ \bibnamefont
  {{Kholupenko}}}, \bibinfo {author} {\bibfnamefont {A.~V.}\ \bibnamefont
  {{Ivanchik}}},\ and\ \bibinfo {author} {\bibfnamefont {D.~A.}\ \bibnamefont
  {{Varshalovich}}},\ }\bibfield  {title} {\bibinfo {title} {{Rapid HeII->HeI
  recombination and radiation arising from this process}},\ }\href
  {https://doi.org/10.1111/j.1745-3933.2007.00316.x} {\bibfield  {journal}
  {\bibinfo  {journal} {\mnras}\ }\textbf {\bibinfo {volume} {378}},\ \bibinfo
  {pages} {L39} (\bibinfo {year} {2007})},\ \Eprint
  {https://arxiv.org/abs/arXiv:astro-ph/0703438} {arXiv:astro-ph/0703438}
  \BibitemShut {NoStop}%
\bibitem [{\citenamefont {Rubino-Martin}\ \emph {et~al.}(2008)\citenamefont
  {Rubino-Martin}, \citenamefont {Chluba},\ and\ \citenamefont
  {Sunyaev}}]{Rubino-Martin:2007tua}%
  \BibitemOpen
  \bibfield  {author} {\bibinfo {author} {\bibfnamefont {J.~A.}\ \bibnamefont
  {Rubino-Martin}}, \bibinfo {author} {\bibfnamefont {J.}~\bibnamefont
  {Chluba}},\ and\ \bibinfo {author} {\bibfnamefont {R.~A.}\ \bibnamefont
  {Sunyaev}},\ }\bibfield  {title} {\bibinfo {title} {{Lines in the cosmic
  microwave background spectrum from the epoch of cosmological helium
  recombination}},\ }\href {https://doi.org/10.1051/0004-6361:20078993}
  {\bibfield  {journal} {\bibinfo  {journal} {Astron. Astrophys.}\ }\textbf
  {\bibinfo {volume} {485}},\ \bibinfo {pages} {377} (\bibinfo {year}
  {2008})},\ \Eprint {https://arxiv.org/abs/0711.0594} {arXiv:0711.0594
  [astro-ph]} \BibitemShut {NoStop}%
\bibitem [{\citenamefont {Blas}\ \emph {et~al.}(2011)\citenamefont {Blas},
  \citenamefont {Lesgourgues},\ and\ \citenamefont {Tram}}]{Blas:2011rf}%
  \BibitemOpen
  \bibfield  {author} {\bibinfo {author} {\bibfnamefont {D.}~\bibnamefont
  {Blas}}, \bibinfo {author} {\bibfnamefont {J.}~\bibnamefont {Lesgourgues}},\
  and\ \bibinfo {author} {\bibfnamefont {T.}~\bibnamefont {Tram}},\ }\bibfield
  {title} {\bibinfo {title} {{The Cosmic Linear Anisotropy Solving System
  (CLASS) II: Approximation schemes}},\ }\href
  {https://doi.org/10.1088/1475-7516/2011/07/034} {\bibfield  {journal}
  {\bibinfo  {journal} {JCAP}\ }\textbf {\bibinfo {volume} {07}},\ \bibinfo
  {pages} {034}},\ \Eprint {https://arxiv.org/abs/1104.2933} {arXiv:1104.2933
  [astro-ph.CO]} \BibitemShut {NoStop}%
\bibitem [{\citenamefont {Lucca}\ \emph {et~al.}(2020)\citenamefont {Lucca},
  \citenamefont {Sch\"oneberg}, \citenamefont {Hooper}, \citenamefont
  {Lesgourgues},\ and\ \citenamefont {Chluba}}]{Lucca:2019rxf}%
  \BibitemOpen
  \bibfield  {author} {\bibinfo {author} {\bibfnamefont {M.}~\bibnamefont
  {Lucca}}, \bibinfo {author} {\bibfnamefont {N.}~\bibnamefont {Sch\"oneberg}},
  \bibinfo {author} {\bibfnamefont {D.~C.}\ \bibnamefont {Hooper}}, \bibinfo
  {author} {\bibfnamefont {J.}~\bibnamefont {Lesgourgues}},\ and\ \bibinfo
  {author} {\bibfnamefont {J.}~\bibnamefont {Chluba}},\ }\bibfield  {title}
  {\bibinfo {title} {{The synergy between CMB spectral distortions and
  anisotropies}},\ }\href {https://doi.org/10.1088/1475-7516/2020/02/026}
  {\bibfield  {journal} {\bibinfo  {journal} {JCAP}\ }\textbf {\bibinfo
  {volume} {02}},\ \bibinfo {pages} {026}},\ \Eprint
  {https://arxiv.org/abs/1910.04619} {arXiv:1910.04619 [astro-ph.CO]}
  \BibitemShut {NoStop}%
\bibitem [{\citenamefont {Bolliet}\ \emph {et~al.}(2023)\citenamefont
  {Bolliet}, \citenamefont {Spurio~Mancini}, \citenamefont {Hill},
  \citenamefont {Madhavacheril}, \citenamefont {Jense}, \citenamefont
  {Calabrese},\ and\ \citenamefont {Dunkley}}]{Bolliet:2023sst}%
  \BibitemOpen
  \bibfield  {author} {\bibinfo {author} {\bibfnamefont {B.}~\bibnamefont
  {Bolliet}}, \bibinfo {author} {\bibfnamefont {A.}~\bibnamefont
  {Spurio~Mancini}}, \bibinfo {author} {\bibfnamefont {J.~C.}\ \bibnamefont
  {Hill}}, \bibinfo {author} {\bibfnamefont {M.}~\bibnamefont {Madhavacheril}},
  \bibinfo {author} {\bibfnamefont {H.~T.}\ \bibnamefont {Jense}}, \bibinfo
  {author} {\bibfnamefont {E.}~\bibnamefont {Calabrese}},\ and\ \bibinfo
  {author} {\bibfnamefont {J.}~\bibnamefont {Dunkley}},\ }\bibfield  {title}
  {\bibinfo {title} {{High-accuracy emulators for observables in $\Lambda$CDM,
  $N_\mathrm{eff}$, $\Sigma m_\nu$, and $w$ cosmologies}},\ }\href@noop {} {\
  (\bibinfo {year} {2023})},\ \Eprint {https://arxiv.org/abs/2303.01591}
  {arXiv:2303.01591 [astro-ph.CO]} \BibitemShut {NoStop}%
\bibitem [{\citenamefont {Mootoovaloo}\ \emph {et~al.}(2022)\citenamefont
  {Mootoovaloo}, \citenamefont {Jaffe}, \citenamefont {Heavens},\ and\
  \citenamefont {Leclercq}}]{Mootoovaloo:2021rot}%
  \BibitemOpen
  \bibfield  {author} {\bibinfo {author} {\bibfnamefont {A.}~\bibnamefont
  {Mootoovaloo}}, \bibinfo {author} {\bibfnamefont {A.~H.}\ \bibnamefont
  {Jaffe}}, \bibinfo {author} {\bibfnamefont {A.~F.}\ \bibnamefont {Heavens}},\
  and\ \bibinfo {author} {\bibfnamefont {F.}~\bibnamefont {Leclercq}},\
  }\bibfield  {title} {\bibinfo {title} {{Kernel-based emulator for the 3D
  matter power spectrum from CLASS}},\ }\href
  {https://doi.org/10.1016/j.ascom.2021.100508} {\bibfield  {journal} {\bibinfo
   {journal} {Astron. Comput.}\ }\textbf {\bibinfo {volume} {38}},\ \bibinfo
  {pages} {100508} (\bibinfo {year} {2022})},\ \Eprint
  {https://arxiv.org/abs/2105.02256} {arXiv:2105.02256 [astro-ph.CO]}
  \BibitemShut {NoStop}%
\bibitem [{\citenamefont {Aric\`o}\ \emph {et~al.}(2021)\citenamefont
  {Aric\`o}, \citenamefont {Angulo},\ and\ \citenamefont
  {Zennaro}}]{Arico:2021izc}%
  \BibitemOpen
  \bibfield  {author} {\bibinfo {author} {\bibfnamefont {G.}~\bibnamefont
  {Aric\`o}}, \bibinfo {author} {\bibfnamefont {R.~E.}\ \bibnamefont
  {Angulo}},\ and\ \bibinfo {author} {\bibfnamefont {M.}~\bibnamefont
  {Zennaro}},\ }\bibfield  {title} {\bibinfo {title} {{Accelerating
  Large-Scale-Structure data analyses by emulating Boltzmann solvers and
  Lagrangian Perturbation Theory}}\ }\href
  {https://doi.org/10.12688/openreseurope.14310.2}
  {10.12688/openreseurope.14310.2} (\bibinfo {year} {2021}),\ \Eprint
  {https://arxiv.org/abs/2104.14568} {arXiv:2104.14568 [astro-ph.CO]}
  \BibitemShut {NoStop}%
\bibitem [{\citenamefont {Albers}\ \emph {et~al.}(2019)\citenamefont {Albers},
  \citenamefont {Fidler}, \citenamefont {Lesgourgues}, \citenamefont
  {Sch\"oneberg},\ and\ \citenamefont {Torrado}}]{Albers:2019rzt}%
  \BibitemOpen
  \bibfield  {author} {\bibinfo {author} {\bibfnamefont {J.}~\bibnamefont
  {Albers}}, \bibinfo {author} {\bibfnamefont {C.}~\bibnamefont {Fidler}},
  \bibinfo {author} {\bibfnamefont {J.}~\bibnamefont {Lesgourgues}}, \bibinfo
  {author} {\bibfnamefont {N.}~\bibnamefont {Sch\"oneberg}},\ and\ \bibinfo
  {author} {\bibfnamefont {J.}~\bibnamefont {Torrado}},\ }\bibfield  {title}
  {\bibinfo {title} {{CosmicNet. Part I. Physics-driven implementation of
  neural networks within Einstein-Boltzmann Solvers}},\ }\href
  {https://doi.org/10.1088/1475-7516/2019/09/028} {\bibfield  {journal}
  {\bibinfo  {journal} {JCAP}\ }\textbf {\bibinfo {volume} {09}},\ \bibinfo
  {pages} {028}},\ \Eprint {https://arxiv.org/abs/1907.05764} {arXiv:1907.05764
  [astro-ph.CO]} \BibitemShut {NoStop}%
\bibitem [{\citenamefont {G\"unther}\ \emph {et~al.}(2022)\citenamefont
  {G\"unther}, \citenamefont {Lesgourgues}, \citenamefont {Samaras},
  \citenamefont {Sch\"oneberg}, \citenamefont {Stadtmann}, \citenamefont
  {Fidler},\ and\ \citenamefont {Torrado}}]{Gunther:2022pto}%
  \BibitemOpen
  \bibfield  {author} {\bibinfo {author} {\bibfnamefont {S.}~\bibnamefont
  {G\"unther}}, \bibinfo {author} {\bibfnamefont {J.}~\bibnamefont
  {Lesgourgues}}, \bibinfo {author} {\bibfnamefont {G.}~\bibnamefont
  {Samaras}}, \bibinfo {author} {\bibfnamefont {N.}~\bibnamefont
  {Sch\"oneberg}}, \bibinfo {author} {\bibfnamefont {F.}~\bibnamefont
  {Stadtmann}}, \bibinfo {author} {\bibfnamefont {C.}~\bibnamefont {Fidler}},\
  and\ \bibinfo {author} {\bibfnamefont {J.}~\bibnamefont {Torrado}},\
  }\bibfield  {title} {\bibinfo {title} {{CosmicNet II: emulating extended
  cosmologies with efficient and accurate neural networks}},\ }\href
  {https://doi.org/10.1088/1475-7516/2022/11/035} {\bibfield  {journal}
  {\bibinfo  {journal} {JCAP}\ }\textbf {\bibinfo {volume} {11}},\ \bibinfo
  {pages} {035}},\ \Eprint {https://arxiv.org/abs/2207.05707} {arXiv:2207.05707
  [astro-ph.CO]} \BibitemShut {NoStop}%
\bibitem [{\citenamefont {Spurio~Mancini}\ \emph {et~al.}(2022)\citenamefont
  {Spurio~Mancini}, \citenamefont {Piras}, \citenamefont {Alsing},
  \citenamefont {Joachimi},\ and\ \citenamefont
  {Hobson}}]{SpurioMancini:2021ppk}%
  \BibitemOpen
  \bibfield  {author} {\bibinfo {author} {\bibfnamefont {A.}~\bibnamefont
  {Spurio~Mancini}}, \bibinfo {author} {\bibfnamefont {D.}~\bibnamefont
  {Piras}}, \bibinfo {author} {\bibfnamefont {J.}~\bibnamefont {Alsing}},
  \bibinfo {author} {\bibfnamefont {B.}~\bibnamefont {Joachimi}},\ and\
  \bibinfo {author} {\bibfnamefont {M.~P.}\ \bibnamefont {Hobson}},\ }\bibfield
   {title} {\bibinfo {title} {{CosmoPower: emulating cosmological power spectra
  for accelerated Bayesian inference from next-generation surveys}},\ }\href
  {https://doi.org/10.1093/mnras/stac064} {\bibfield  {journal} {\bibinfo
  {journal} {Mon. Not. Roy. Astron. Soc.}\ }\textbf {\bibinfo {volume} {511}},\
  \bibinfo {pages} {1771} (\bibinfo {year} {2022})},\ \Eprint
  {https://arxiv.org/abs/2106.03846} {arXiv:2106.03846 [astro-ph.CO]}
  \BibitemShut {NoStop}%
\bibitem [{\citenamefont {Tang}(1993)}]{Tang:1993}%
  \BibitemOpen
  \bibfield  {author} {\bibinfo {author} {\bibfnamefont {B.}~\bibnamefont
  {Tang}},\ }\bibfield  {title} {\bibinfo {title} {Orthogonal array-based latin
  hypercubes},\ }\href {https://doi.org/10.1080/01621459.1993.10476423}
  {\bibfield  {journal} {\bibinfo  {journal} {Journal of the American
  Statistical Association}\ }\textbf {\bibinfo {volume} {88}},\ \bibinfo
  {pages} {1392} (\bibinfo {year} {1993})}\BibitemShut {NoStop}%
\bibitem [{\citenamefont {{Schneider}}\ \emph {et~al.}(2011)\citenamefont
  {{Schneider}}, \citenamefont {{Holm}},\ and\ \citenamefont
  {{Knox}}}]{Schneider:2011}%
  \BibitemOpen
  \bibfield  {author} {\bibinfo {author} {\bibfnamefont {M.~D.}\ \bibnamefont
  {{Schneider}}}, \bibinfo {author} {\bibfnamefont {{\'O}.}~\bibnamefont
  {{Holm}}},\ and\ \bibinfo {author} {\bibfnamefont {L.}~\bibnamefont
  {{Knox}}},\ }\bibfield  {title} {\bibinfo {title} {{Intelligent Design: On
  the Emulation of Cosmological Simulations}},\ }\href
  {https://doi.org/10.1088/0004-637X/728/2/137} {\bibfield  {journal} {\bibinfo
   {journal} {\apj}\ }\textbf {\bibinfo {volume} {728}},\ \bibinfo {eid} {137}
  (\bibinfo {year} {2011})},\ \Eprint {https://arxiv.org/abs/1002.1752}
  {arXiv:1002.1752 [astro-ph.CO]} \BibitemShut {NoStop}%
\bibitem [{\citenamefont {Aghanim}\ \emph
  {et~al.}(2020{\natexlab{b}})\citenamefont {Aghanim} \emph
  {et~al.}}]{Planck:2019nip}%
  \BibitemOpen
  \bibfield  {author} {\bibinfo {author} {\bibfnamefont {N.}~\bibnamefont
  {Aghanim}} \emph {et~al.} (\bibinfo {collaboration} {Planck}),\ }\bibfield
  {title} {\bibinfo {title} {{Planck 2018 results. V. CMB power spectra and
  likelihoods}},\ }\href {https://doi.org/10.1051/0004-6361/201936386}
  {\bibfield  {journal} {\bibinfo  {journal} {Astron. Astrophys.}\ }\textbf
  {\bibinfo {volume} {641}},\ \bibinfo {pages} {A5} (\bibinfo {year}
  {2020}{\natexlab{b}})},\ \Eprint {https://arxiv.org/abs/1907.12875}
  {arXiv:1907.12875 [astro-ph.CO]} \BibitemShut {NoStop}%
\bibitem [{\citenamefont {Prince}\ and\ \citenamefont
  {Dunkley}(2019)}]{Prince:2019hse}%
  \BibitemOpen
  \bibfield  {author} {\bibinfo {author} {\bibfnamefont {H.}~\bibnamefont
  {Prince}}\ and\ \bibinfo {author} {\bibfnamefont {J.}~\bibnamefont
  {Dunkley}},\ }\bibfield  {title} {\bibinfo {title} {{Data compression in
  cosmology: A compressed likelihood for Planck data}},\ }\href
  {https://doi.org/10.1103/PhysRevD.100.083502} {\bibfield  {journal} {\bibinfo
   {journal} {Phys. Rev. D}\ }\textbf {\bibinfo {volume} {100}},\ \bibinfo
  {pages} {083502} (\bibinfo {year} {2019})},\ \Eprint
  {https://arxiv.org/abs/1909.05869} {arXiv:1909.05869 [astro-ph.CO]}
  \BibitemShut {NoStop}%
\bibitem [{\citenamefont {Torrado}\ and\ \citenamefont
  {Lewis}(2021)}]{Torrado:2020dgo}%
  \BibitemOpen
  \bibfield  {author} {\bibinfo {author} {\bibfnamefont {J.}~\bibnamefont
  {Torrado}}\ and\ \bibinfo {author} {\bibfnamefont {A.}~\bibnamefont
  {Lewis}},\ }\bibfield  {title} {\bibinfo {title} {{Cobaya: Code for Bayesian
  Analysis of hierarchical physical models}},\ }\href
  {https://doi.org/10.1088/1475-7516/2021/05/057} {\bibfield  {journal}
  {\bibinfo  {journal} {JCAP}\ }\textbf {\bibinfo {volume} {05}},\ \bibinfo
  {pages} {057}},\ \Eprint {https://arxiv.org/abs/2005.05290} {arXiv:2005.05290
  [astro-ph.IM]} \BibitemShut {NoStop}%
\bibitem [{\citenamefont {Lewis}(2019)}]{Lewis:2019xzd}%
  \BibitemOpen
  \bibfield  {author} {\bibinfo {author} {\bibfnamefont {A.}~\bibnamefont
  {Lewis}},\ }\bibfield  {title} {\bibinfo {title} {{GetDist: a Python package
  for analysing Monte Carlo samples}},\ }\href {https://getdist.readthedocs.io}
  {\  (\bibinfo {year} {2019})},\ \Eprint {https://arxiv.org/abs/1910.13970}
  {arXiv:1910.13970 [astro-ph.IM]} \BibitemShut {NoStop}%
\bibitem [{\citenamefont {Alam}\ \emph {et~al.}(2021)\citenamefont {Alam} \emph
  {et~al.}}]{eBOSS:2020yzd}%
  \BibitemOpen
  \bibfield  {author} {\bibinfo {author} {\bibfnamefont {S.}~\bibnamefont
  {Alam}} \emph {et~al.} (\bibinfo {collaboration} {eBOSS}),\ }\bibfield
  {title} {\bibinfo {title} {{Completed SDSS-IV extended Baryon Oscillation
  Spectroscopic Survey: Cosmological implications from two decades of
  spectroscopic surveys at the Apache Point Observatory}},\ }\href
  {https://doi.org/10.1103/PhysRevD.103.083533} {\bibfield  {journal} {\bibinfo
   {journal} {Phys. Rev. D}\ }\textbf {\bibinfo {volume} {103}},\ \bibinfo
  {pages} {083533} (\bibinfo {year} {2021})},\ \bibinfo {note} {see Appendix A
  for a discussion of observational considerations.},\ \Eprint
  {https://arxiv.org/abs/2007.08991} {arXiv:2007.08991 [astro-ph.CO]}
  \BibitemShut {NoStop}%
\bibitem [{\citenamefont {Ross}\ \emph {et~al.}(2015)\citenamefont {Ross},
  \citenamefont {Samushia}, \citenamefont {Howlett}, \citenamefont {Percival},
  \citenamefont {Burden},\ and\ \citenamefont {Manera}}]{Ross:2014qpa}%
  \BibitemOpen
  \bibfield  {author} {\bibinfo {author} {\bibfnamefont {A.~J.}\ \bibnamefont
  {Ross}}, \bibinfo {author} {\bibfnamefont {L.}~\bibnamefont {Samushia}},
  \bibinfo {author} {\bibfnamefont {C.}~\bibnamefont {Howlett}}, \bibinfo
  {author} {\bibfnamefont {W.~J.}\ \bibnamefont {Percival}}, \bibinfo {author}
  {\bibfnamefont {A.}~\bibnamefont {Burden}},\ and\ \bibinfo {author}
  {\bibfnamefont {M.}~\bibnamefont {Manera}},\ }\bibfield  {title} {\bibinfo
  {title} {{The clustering of the SDSS DR7 main Galaxy sample \textendash{} I.
  A 4 per cent distance measure at $z = 0.15$}},\ }\href
  {https://doi.org/10.1093/mnras/stv154} {\bibfield  {journal} {\bibinfo
  {journal} {Mon. Not. Roy. Astron. Soc.}\ }\textbf {\bibinfo {volume} {449}},\
  \bibinfo {pages} {835} (\bibinfo {year} {2015})},\ \Eprint
  {https://arxiv.org/abs/1409.3242} {arXiv:1409.3242 [astro-ph.CO]}
  \BibitemShut {NoStop}%
\bibitem [{\citenamefont {Hu}\ and\ \citenamefont
  {Sugiyama}(1995)}]{Hu:1994uz}%
  \BibitemOpen
  \bibfield  {author} {\bibinfo {author} {\bibfnamefont {W.}~\bibnamefont
  {Hu}}\ and\ \bibinfo {author} {\bibfnamefont {N.}~\bibnamefont {Sugiyama}},\
  }\bibfield  {title} {\bibinfo {title} {{Anisotropies in the cosmic microwave
  background: An Analytic approach}},\ }\href {https://doi.org/10.1086/175624}
  {\bibfield  {journal} {\bibinfo  {journal} {Astrophys. J.}\ }\textbf
  {\bibinfo {volume} {444}},\ \bibinfo {pages} {489} (\bibinfo {year}
  {1995})},\ \Eprint {https://arxiv.org/abs/astro-ph/9407093}
  {arXiv:astro-ph/9407093} \BibitemShut {NoStop}%
\bibitem [{\citenamefont {{Bond}}\ and\ \citenamefont
  {{Efstathiou}}(1984)}]{Bond:1984}%
  \BibitemOpen
  \bibfield  {author} {\bibinfo {author} {\bibfnamefont {J.~R.}\ \bibnamefont
  {{Bond}}}\ and\ \bibinfo {author} {\bibfnamefont {G.}~\bibnamefont
  {{Efstathiou}}},\ }\bibfield  {title} {\bibinfo {title} {{Cosmic background
  radiation anisotropies in universes dominated by nonbaryonic dark matter}},\
  }\href {https://doi.org/10.1086/184362} {\bibfield  {journal} {\bibinfo
  {journal} {\apjl}\ }\textbf {\bibinfo {volume} {285}},\ \bibinfo {pages}
  {L45} (\bibinfo {year} {1984})}\BibitemShut {NoStop}%
\bibitem [{\citenamefont {Hu}\ and\ \citenamefont
  {Dodelson}(2002)}]{Hu:2001bc}%
  \BibitemOpen
  \bibfield  {author} {\bibinfo {author} {\bibfnamefont {W.}~\bibnamefont
  {Hu}}\ and\ \bibinfo {author} {\bibfnamefont {S.}~\bibnamefont {Dodelson}},\
  }\bibfield  {title} {\bibinfo {title} {{Cosmic Microwave Background
  Anisotropies}},\ }\href
  {https://doi.org/10.1146/annurev.astro.40.060401.093926} {\bibfield
  {journal} {\bibinfo  {journal} {Ann. Rev. Astron. Astrophys.}\ }\textbf
  {\bibinfo {volume} {40}},\ \bibinfo {pages} {171} (\bibinfo {year} {2002})},\
  \Eprint {https://arxiv.org/abs/astro-ph/0110414} {arXiv:astro-ph/0110414}
  \BibitemShut {NoStop}%
\bibitem [{\citenamefont {Hu}\ and\ \citenamefont {White}(1997)}]{Hu:1996mn}%
  \BibitemOpen
  \bibfield  {author} {\bibinfo {author} {\bibfnamefont {W.}~\bibnamefont
  {Hu}}\ and\ \bibinfo {author} {\bibfnamefont {M.~J.}\ \bibnamefont {White}},\
  }\bibfield  {title} {\bibinfo {title} {{The Damping tail of CMB
  anisotropies}},\ }\href {https://doi.org/10.1086/303928} {\bibfield
  {journal} {\bibinfo  {journal} {Astrophys. J.}\ }\textbf {\bibinfo {volume}
  {479}},\ \bibinfo {pages} {568} (\bibinfo {year} {1997})},\ \Eprint
  {https://arxiv.org/abs/astro-ph/9609079} {arXiv:astro-ph/9609079}
  \BibitemShut {NoStop}%
\bibitem [{\citenamefont {Zaldarriaga}\ and\ \citenamefont
  {Harari}(1995)}]{Zaldarriaga:1995gi}%
  \BibitemOpen
  \bibfield  {author} {\bibinfo {author} {\bibfnamefont {M.}~\bibnamefont
  {Zaldarriaga}}\ and\ \bibinfo {author} {\bibfnamefont {D.~D.}\ \bibnamefont
  {Harari}},\ }\bibfield  {title} {\bibinfo {title} {{Analytic approach to the
  polarization of the cosmic microwave background in flat and open
  universes}},\ }\href {https://doi.org/10.1103/PhysRevD.52.3276} {\bibfield
  {journal} {\bibinfo  {journal} {Phys. Rev. D}\ }\textbf {\bibinfo {volume}
  {52}},\ \bibinfo {pages} {3276} (\bibinfo {year} {1995})},\ \Eprint
  {https://arxiv.org/abs/astro-ph/9504085} {arXiv:astro-ph/9504085}
  \BibitemShut {NoStop}%
\bibitem [{\citenamefont {Pan}\ \emph {et~al.}(2016)\citenamefont {Pan},
  \citenamefont {Knox}, \citenamefont {Mulroe},\ and\ \citenamefont
  {Narimani}}]{Pan:2016zla}%
  \BibitemOpen
  \bibfield  {author} {\bibinfo {author} {\bibfnamefont {Z.}~\bibnamefont
  {Pan}}, \bibinfo {author} {\bibfnamefont {L.}~\bibnamefont {Knox}}, \bibinfo
  {author} {\bibfnamefont {B.}~\bibnamefont {Mulroe}},\ and\ \bibinfo {author}
  {\bibfnamefont {A.}~\bibnamefont {Narimani}},\ }\bibfield  {title} {\bibinfo
  {title} {{Cosmic Microwave Background Acoustic Peak Locations}},\ }\href
  {https://doi.org/10.1093/mnras/stw833} {\bibfield  {journal} {\bibinfo
  {journal} {Mon. Not. Roy. Astron. Soc.}\ }\textbf {\bibinfo {volume} {459}},\
  \bibinfo {pages} {2513} (\bibinfo {year} {2016})},\ \Eprint
  {https://arxiv.org/abs/1603.03091} {arXiv:1603.03091 [astro-ph.CO]}
  \BibitemShut {NoStop}%
\bibitem [{\citenamefont {Hadzhiyska}\ and\ \citenamefont
  {Spergel}(2019)}]{Hadzhiyska:2018mwh}%
  \BibitemOpen
  \bibfield  {author} {\bibinfo {author} {\bibfnamefont {B.}~\bibnamefont
  {Hadzhiyska}}\ and\ \bibinfo {author} {\bibfnamefont {D.~N.}\ \bibnamefont
  {Spergel}},\ }\bibfield  {title} {\bibinfo {title} {{Measuring the Duration
  of Last Scattering}},\ }\href {https://doi.org/10.1103/PhysRevD.99.043537}
  {\bibfield  {journal} {\bibinfo  {journal} {Phys. Rev. D}\ }\textbf {\bibinfo
  {volume} {99}},\ \bibinfo {pages} {043537} (\bibinfo {year} {2019})},\
  \Eprint {https://arxiv.org/abs/1808.04083} {arXiv:1808.04083 [astro-ph.CO]}
  \BibitemShut {NoStop}%
\bibitem [{\citenamefont {Hu}\ and\ \citenamefont
  {Sugiyama}(1996)}]{Hu:1995en}%
  \BibitemOpen
  \bibfield  {author} {\bibinfo {author} {\bibfnamefont {W.}~\bibnamefont
  {Hu}}\ and\ \bibinfo {author} {\bibfnamefont {N.}~\bibnamefont {Sugiyama}},\
  }\bibfield  {title} {\bibinfo {title} {{Small scale cosmological
  perturbations: An Analytic approach}},\ }\href
  {https://doi.org/10.1086/177989} {\bibfield  {journal} {\bibinfo  {journal}
  {Astrophys. J.}\ }\textbf {\bibinfo {volume} {471}},\ \bibinfo {pages} {542}
  (\bibinfo {year} {1996})},\ \Eprint {https://arxiv.org/abs/astro-ph/9510117}
  {arXiv:astro-ph/9510117} \BibitemShut {NoStop}%
\bibitem [{\citenamefont {Eisenstein}\ and\ \citenamefont
  {Hu}(1998)}]{Eisenstein:1997ik}%
  \BibitemOpen
  \bibfield  {author} {\bibinfo {author} {\bibfnamefont {D.~J.}\ \bibnamefont
  {Eisenstein}}\ and\ \bibinfo {author} {\bibfnamefont {W.}~\bibnamefont
  {Hu}},\ }\bibfield  {title} {\bibinfo {title} {{Baryonic features in the
  matter transfer function}},\ }\href {https://doi.org/10.1086/305424}
  {\bibfield  {journal} {\bibinfo  {journal} {Astrophys. J.}\ }\textbf
  {\bibinfo {volume} {496}},\ \bibinfo {pages} {605} (\bibinfo {year}
  {1998})},\ \Eprint {https://arxiv.org/abs/astro-ph/9709112}
  {arXiv:astro-ph/9709112} \BibitemShut {NoStop}%
\bibitem [{\citenamefont {Eisenstein}\ \emph {et~al.}(1998)\citenamefont
  {Eisenstein}, \citenamefont {Hu},\ and\ \citenamefont
  {Tegmark}}]{Eisenstein:1998tu}%
  \BibitemOpen
  \bibfield  {author} {\bibinfo {author} {\bibfnamefont {D.~J.}\ \bibnamefont
  {Eisenstein}}, \bibinfo {author} {\bibfnamefont {W.}~\bibnamefont {Hu}},\
  and\ \bibinfo {author} {\bibfnamefont {M.}~\bibnamefont {Tegmark}},\
  }\bibfield  {title} {\bibinfo {title} {{Cosmic complementarity: H(0) and
  Omega(m) from combining CMB experiments and redshift surveys}},\ }\href
  {https://doi.org/10.1086/311582} {\bibfield  {journal} {\bibinfo  {journal}
  {Astrophys. J. Lett.}\ }\textbf {\bibinfo {volume} {504}},\ \bibinfo {pages}
  {L57} (\bibinfo {year} {1998})},\ \Eprint
  {https://arxiv.org/abs/astro-ph/9805239} {arXiv:astro-ph/9805239}
  \BibitemShut {NoStop}%
\bibitem [{\citenamefont {{Weinberg}}\ \emph {et~al.}(2013)\citenamefont
  {{Weinberg}}, \citenamefont {{Mortonson}}, \citenamefont {{Eisenstein}},
  \citenamefont {{Hirata}}, \citenamefont {{Riess}},\ and\ \citenamefont
  {{Rozo}}}]{Weinberg:2013}%
  \BibitemOpen
  \bibfield  {author} {\bibinfo {author} {\bibfnamefont {D.~H.}\ \bibnamefont
  {{Weinberg}}}, \bibinfo {author} {\bibfnamefont {M.~J.}\ \bibnamefont
  {{Mortonson}}}, \bibinfo {author} {\bibfnamefont {D.~J.}\ \bibnamefont
  {{Eisenstein}}}, \bibinfo {author} {\bibfnamefont {C.}~\bibnamefont
  {{Hirata}}}, \bibinfo {author} {\bibfnamefont {A.~G.}\ \bibnamefont
  {{Riess}}},\ and\ \bibinfo {author} {\bibfnamefont {E.}~\bibnamefont
  {{Rozo}}},\ }\bibfield  {title} {\bibinfo {title} {{Observational probes of
  cosmic acceleration}},\ }\href
  {https://doi.org/10.1016/j.physrep.2013.05.001} {\bibfield  {journal}
  {\bibinfo  {journal} {\physrep}\ }\textbf {\bibinfo {volume} {530}},\
  \bibinfo {pages} {87} (\bibinfo {year} {2013})},\ \Eprint
  {https://arxiv.org/abs/1201.2434} {arXiv:1201.2434 [astro-ph.CO]}
  \BibitemShut {NoStop}%
\bibitem [{\citenamefont {Alam}\ \emph {et~al.}(2017)\citenamefont {Alam} \emph
  {et~al.}}]{BOSS:2016wmc}%
  \BibitemOpen
  \bibfield  {author} {\bibinfo {author} {\bibfnamefont {S.}~\bibnamefont
  {Alam}} \emph {et~al.} (\bibinfo {collaboration} {BOSS}),\ }\bibfield
  {title} {\bibinfo {title} {{The clustering of galaxies in the completed
  SDSS-III Baryon Oscillation Spectroscopic Survey: cosmological analysis of
  the DR12 galaxy sample}},\ }\href {https://doi.org/10.1093/mnras/stx721}
  {\bibfield  {journal} {\bibinfo  {journal} {Mon. Not. Roy. Astron. Soc.}\
  }\textbf {\bibinfo {volume} {470}},\ \bibinfo {pages} {2617} (\bibinfo {year}
  {2017})},\ \Eprint {https://arxiv.org/abs/1607.03155} {arXiv:1607.03155
  [astro-ph.CO]} \BibitemShut {NoStop}%
\bibitem [{\citenamefont {Riess}\ \emph {et~al.}(2022)\citenamefont {Riess}
  \emph {et~al.}}]{Riess:2021jrx}%
  \BibitemOpen
  \bibfield  {author} {\bibinfo {author} {\bibfnamefont {A.~G.}\ \bibnamefont
  {Riess}} \emph {et~al.},\ }\bibfield  {title} {\bibinfo {title} {{A
  Comprehensive Measurement of the Local Value of the Hubble Constant with 1 km
  s$^{−1}$ Mpc$^{−1}$ Uncertainty from the Hubble Space Telescope and the
  SH0ES Team}},\ }\href {https://doi.org/10.3847/2041-8213/ac5c5b} {\bibfield
  {journal} {\bibinfo  {journal} {Astrophys. J. Lett.}\ }\textbf {\bibinfo
  {volume} {934}},\ \bibinfo {pages} {L7} (\bibinfo {year} {2022})},\ \Eprint
  {https://arxiv.org/abs/2112.04510} {arXiv:2112.04510 [astro-ph.CO]}
  \BibitemShut {NoStop}%
\bibitem [{\citenamefont {Huang}\ \emph {et~al.}(2018)\citenamefont {Huang},
  \citenamefont {Addison}, \citenamefont {Weiland},\ and\ \citenamefont
  {Bennett}}]{Huang:2018xle}%
  \BibitemOpen
  \bibfield  {author} {\bibinfo {author} {\bibfnamefont {Y.}~\bibnamefont
  {Huang}}, \bibinfo {author} {\bibfnamefont {G.~E.}\ \bibnamefont {Addison}},
  \bibinfo {author} {\bibfnamefont {J.~L.}\ \bibnamefont {Weiland}},\ and\
  \bibinfo {author} {\bibfnamefont {C.~L.}\ \bibnamefont {Bennett}},\
  }\bibfield  {title} {\bibinfo {title} {{Assessing Consistency Between WMAP
  9-year and Planck 2015 Temperature Power Spectra}},\ }\href
  {https://doi.org/10.3847/1538-4357/aaeb1f} {\bibfield  {journal} {\bibinfo
  {journal} {Astrophys. J.}\ }\textbf {\bibinfo {volume} {869}},\ \bibinfo
  {pages} {38} (\bibinfo {year} {2018})},\ \Eprint
  {https://arxiv.org/abs/1804.05428} {arXiv:1804.05428 [astro-ph.CO]}
  \BibitemShut {NoStop}%
\bibitem [{\citenamefont {Aylor}\ \emph {et~al.}(2019)\citenamefont {Aylor},
  \citenamefont {Joy}, \citenamefont {Knox}, \citenamefont {Millea},
  \citenamefont {Raghunathan},\ and\ \citenamefont {Wu}}]{Aylor:2018drw}%
  \BibitemOpen
  \bibfield  {author} {\bibinfo {author} {\bibfnamefont {K.}~\bibnamefont
  {Aylor}}, \bibinfo {author} {\bibfnamefont {M.}~\bibnamefont {Joy}}, \bibinfo
  {author} {\bibfnamefont {L.}~\bibnamefont {Knox}}, \bibinfo {author}
  {\bibfnamefont {M.}~\bibnamefont {Millea}}, \bibinfo {author} {\bibfnamefont
  {S.}~\bibnamefont {Raghunathan}},\ and\ \bibinfo {author} {\bibfnamefont
  {W.~L.~K.}\ \bibnamefont {Wu}},\ }\bibfield  {title} {\bibinfo {title}
  {{Sounds Discordant: Classical Distance Ladder \& $\Lambda$CDM -based
  Determinations of the Cosmological Sound Horizon}},\ }\href
  {https://doi.org/10.3847/1538-4357/ab0898} {\bibfield  {journal} {\bibinfo
  {journal} {Astrophys. J.}\ }\textbf {\bibinfo {volume} {874}},\ \bibinfo
  {pages} {4} (\bibinfo {year} {2019})},\ \Eprint
  {https://arxiv.org/abs/1811.00537} {arXiv:1811.00537 [astro-ph.CO]}
  \BibitemShut {NoStop}%
\bibitem [{\citenamefont {Balkenhol}\ \emph {et~al.}(2023)\citenamefont
  {Balkenhol} \emph {et~al.}}]{SPT-3G:2022hvq}%
  \BibitemOpen
  \bibfield  {author} {\bibinfo {author} {\bibfnamefont {L.}~\bibnamefont
  {Balkenhol}} \emph {et~al.} (\bibinfo {collaboration} {SPT-3G}),\ }\bibfield
  {title} {\bibinfo {title} {{Measurement of the CMB temperature power spectrum
  and constraints on cosmology from the SPT-3G 2018 TT, TE, and EE dataset}},\
  }\href {https://doi.org/10.1103/PhysRevD.108.023510} {\bibfield  {journal}
  {\bibinfo  {journal} {Phys. Rev. D}\ }\textbf {\bibinfo {volume} {108}},\
  \bibinfo {pages} {023510} (\bibinfo {year} {2023})},\ \Eprint
  {https://arxiv.org/abs/2212.05642} {arXiv:2212.05642 [astro-ph.CO]}
  \BibitemShut {NoStop}%
\bibitem [{\citenamefont {Aiola}\ \emph {et~al.}(2020)\citenamefont {Aiola}
  \emph {et~al.}}]{ACT:2020gnv}%
  \BibitemOpen
  \bibfield  {author} {\bibinfo {author} {\bibfnamefont {S.}~\bibnamefont
  {Aiola}} \emph {et~al.} (\bibinfo {collaboration} {ACT}),\ }\bibfield
  {title} {\bibinfo {title} {{The Atacama Cosmology Telescope: DR4 Maps and
  Cosmological Parameters}},\ }\href
  {https://doi.org/10.1088/1475-7516/2020/12/047} {\bibfield  {journal}
  {\bibinfo  {journal} {JCAP}\ }\textbf {\bibinfo {volume} {12}},\ \bibinfo
  {pages} {047}},\ \Eprint {https://arxiv.org/abs/2007.07288} {arXiv:2007.07288
  [astro-ph.CO]} \BibitemShut {NoStop}%
\bibitem [{\citenamefont {Addison}\ \emph {et~al.}(2018)\citenamefont
  {Addison}, \citenamefont {Watts}, \citenamefont {Bennett}, \citenamefont
  {Halpern}, \citenamefont {Hinshaw},\ and\ \citenamefont
  {Weiland}}]{Addison:2017fdm}%
  \BibitemOpen
  \bibfield  {author} {\bibinfo {author} {\bibfnamefont {G.~E.}\ \bibnamefont
  {Addison}}, \bibinfo {author} {\bibfnamefont {D.~J.}\ \bibnamefont {Watts}},
  \bibinfo {author} {\bibfnamefont {C.~L.}\ \bibnamefont {Bennett}}, \bibinfo
  {author} {\bibfnamefont {M.}~\bibnamefont {Halpern}}, \bibinfo {author}
  {\bibfnamefont {G.}~\bibnamefont {Hinshaw}},\ and\ \bibinfo {author}
  {\bibfnamefont {J.~L.}\ \bibnamefont {Weiland}},\ }\bibfield  {title}
  {\bibinfo {title} {{Elucidating $\Lambda$CDM: Impact of Baryon Acoustic
  Oscillation Measurements on the Hubble Constant Discrepancy}},\ }\href
  {https://doi.org/10.3847/1538-4357/aaa1ed} {\bibfield  {journal} {\bibinfo
  {journal} {Astrophys. J.}\ }\textbf {\bibinfo {volume} {853}},\ \bibinfo
  {pages} {119} (\bibinfo {year} {2018})},\ \Eprint
  {https://arxiv.org/abs/1707.06547} {arXiv:1707.06547 [astro-ph.CO]}
  \BibitemShut {NoStop}%
\bibitem [{\citenamefont {{Di Valentino}}\ \emph {et~al.}(2021)\citenamefont
  {{Di Valentino}}, \citenamefont {{Anchordoqui}}, \citenamefont {{Akarsu}},
  \citenamefont {{Ali-Haimoud}}, \citenamefont {{Amendola}}, \citenamefont
  {{Arendse}} \emph {et~al.}}]{Valentino2021H0}%
  \BibitemOpen
  \bibfield  {author} {\bibinfo {author} {\bibfnamefont {E.}~\bibnamefont {{Di
  Valentino}}}, \bibinfo {author} {\bibfnamefont {L.~A.}\ \bibnamefont
  {{Anchordoqui}}}, \bibinfo {author} {\bibfnamefont {{\"O}.}~\bibnamefont
  {{Akarsu}}}, \bibinfo {author} {\bibfnamefont {Y.}~\bibnamefont
  {{Ali-Haimoud}}}, \bibinfo {author} {\bibfnamefont {L.}~\bibnamefont
  {{Amendola}}}, \bibinfo {author} {\bibfnamefont {N.}~\bibnamefont
  {{Arendse}}}, \emph {et~al.},\ }\bibfield  {title} {\bibinfo {title}
  {{Cosmology Intertwined II: The hubble constant tension}},\ }\href
  {https://doi.org/10.1016/j.astropartphys.2021.102605} {\bibfield  {journal}
  {\bibinfo  {journal} {Astroparticle Physics}\ }\textbf {\bibinfo {volume}
  {131}},\ \bibinfo {eid} {102605} (\bibinfo {year} {2021})},\ \Eprint
  {https://arxiv.org/abs/2008.11284} {arXiv:2008.11284 [astro-ph.CO]}
  \BibitemShut {NoStop}%
\bibitem [{\citenamefont {Sch\"oneberg}\ \emph {et~al.}(2022)\citenamefont
  {Sch\"oneberg}, \citenamefont {Franco~Abell\'an}, \citenamefont
  {P\'erez~S\'anchez}, \citenamefont {Witte}, \citenamefont {Poulin},\ and\
  \citenamefont {Lesgourgues}}]{Schoneberg:2021qvd}%
  \BibitemOpen
  \bibfield  {author} {\bibinfo {author} {\bibfnamefont {N.}~\bibnamefont
  {Sch\"oneberg}}, \bibinfo {author} {\bibfnamefont {G.}~\bibnamefont
  {Franco~Abell\'an}}, \bibinfo {author} {\bibfnamefont {A.}~\bibnamefont
  {P\'erez~S\'anchez}}, \bibinfo {author} {\bibfnamefont {S.~J.}\ \bibnamefont
  {Witte}}, \bibinfo {author} {\bibfnamefont {V.}~\bibnamefont {Poulin}},\ and\
  \bibinfo {author} {\bibfnamefont {J.}~\bibnamefont {Lesgourgues}},\
  }\bibfield  {title} {\bibinfo {title} {{The H0 Olympics: A fair ranking of
  proposed models}},\ }\href {https://doi.org/10.1016/j.physrep.2022.07.001}
  {\bibfield  {journal} {\bibinfo  {journal} {Phys. Rept.}\ }\textbf {\bibinfo
  {volume} {984}},\ \bibinfo {pages} {1} (\bibinfo {year} {2022})},\ \Eprint
  {https://arxiv.org/abs/2107.10291} {arXiv:2107.10291 [astro-ph.CO]}
  \BibitemShut {NoStop}%
\bibitem [{\citenamefont {Knox}\ and\ \citenamefont
  {Millea}(2020)}]{Knox:2019rjx}%
  \BibitemOpen
  \bibfield  {author} {\bibinfo {author} {\bibfnamefont {L.}~\bibnamefont
  {Knox}}\ and\ \bibinfo {author} {\bibfnamefont {M.}~\bibnamefont {Millea}},\
  }\bibfield  {title} {\bibinfo {title} {{Hubble constant
  hunter\textquoteright{}s guide}},\ }\href
  {https://doi.org/10.1103/PhysRevD.101.043533} {\bibfield  {journal} {\bibinfo
   {journal} {Phys. Rev. D}\ }\textbf {\bibinfo {volume} {101}},\ \bibinfo
  {pages} {043533} (\bibinfo {year} {2020})},\ \Eprint
  {https://arxiv.org/abs/1908.03663} {arXiv:1908.03663 [astro-ph.CO]}
  \BibitemShut {NoStop}%
\bibitem [{\citenamefont {Lynch}\ \emph
  {et~al.}(2024{\natexlab{a}})\citenamefont {Lynch}, \citenamefont {Knox},\
  and\ \citenamefont {Chluba}}]{DESIfollowup}%
  \BibitemOpen
  \bibfield  {author} {\bibinfo {author} {\bibfnamefont {G.~P.}\ \bibnamefont
  {Lynch}}, \bibinfo {author} {\bibfnamefont {L.}~\bibnamefont {Knox}},\ and\
  \bibinfo {author} {\bibfnamefont {J.}~\bibnamefont {Chluba}},\ }\href@noop {}
  {\bibfield  {journal} {\bibinfo  {journal} {unpublished}\ } (\bibinfo {year}
  {2024}{\natexlab{a}})}\BibitemShut {NoStop}%
\bibitem [{\citenamefont {Ge}\ \emph {et~al.}(2023)\citenamefont {Ge},
  \citenamefont {Cyr-Racine},\ and\ \citenamefont {Knox}}]{Ge:2022qws}%
  \BibitemOpen
  \bibfield  {author} {\bibinfo {author} {\bibfnamefont {F.}~\bibnamefont
  {Ge}}, \bibinfo {author} {\bibfnamefont {F.-Y.}\ \bibnamefont {Cyr-Racine}},\
  and\ \bibinfo {author} {\bibfnamefont {L.}~\bibnamefont {Knox}},\ }\bibfield
  {title} {\bibinfo {title} {{Scaling transformations and the origins of light
  relics constraints from cosmic microwave background observations}},\ }\href
  {https://doi.org/10.1103/PhysRevD.107.023517} {\bibfield  {journal} {\bibinfo
   {journal} {Phys. Rev. D}\ }\textbf {\bibinfo {volume} {107}},\ \bibinfo
  {pages} {023517} (\bibinfo {year} {2023})},\ \Eprint
  {https://arxiv.org/abs/2210.16335} {arXiv:2210.16335 [astro-ph.CO]}
  \BibitemShut {NoStop}%
\bibitem [{\citenamefont {Zhang}\ and\ \citenamefont
  {Frieman}(2023)}]{Zhang:2022ujw}%
  \BibitemOpen
  \bibfield  {author} {\bibinfo {author} {\bibfnamefont {J.}~\bibnamefont
  {Zhang}}\ and\ \bibinfo {author} {\bibfnamefont {J.~A.}\ \bibnamefont
  {Frieman}},\ }\bibfield  {title} {\bibinfo {title} {{Mirror dark sector
  solution of the Hubble tension with time-varying fine-structure constant}},\
  }\href {https://doi.org/10.1103/PhysRevD.107.043529} {\bibfield  {journal}
  {\bibinfo  {journal} {Phys. Rev. D}\ }\textbf {\bibinfo {volume} {107}},\
  \bibinfo {pages} {043529} (\bibinfo {year} {2023})},\ \Eprint
  {https://arxiv.org/abs/2211.03236} {arXiv:2211.03236 [astro-ph.CO]}
  \BibitemShut {NoStop}%
\bibitem [{\citenamefont {Greene}\ and\ \citenamefont
  {Cyr-Racine}(2023)}]{Greene:2023cro}%
  \BibitemOpen
  \bibfield  {author} {\bibinfo {author} {\bibfnamefont {K.~L.}\ \bibnamefont
  {Greene}}\ and\ \bibinfo {author} {\bibfnamefont {F.-Y.}\ \bibnamefont
  {Cyr-Racine}},\ }\bibfield  {title} {\bibinfo {title} {{Thomson scattering:
  one rate to rule them all}},\ }\href
  {https://doi.org/10.1088/1475-7516/2023/10/065} {\bibfield  {journal}
  {\bibinfo  {journal} {JCAP}\ }\textbf {\bibinfo {volume} {10}},\ \bibinfo
  {pages} {065}},\ \Eprint {https://arxiv.org/abs/2306.06165} {arXiv:2306.06165
  [astro-ph.CO]} \BibitemShut {NoStop}%
\bibitem [{\citenamefont {Greene}\ and\ \citenamefont
  {Cyr-Racine}(2024)}]{Greene:2024qis}%
  \BibitemOpen
  \bibfield  {author} {\bibinfo {author} {\bibfnamefont {K.}~\bibnamefont
  {Greene}}\ and\ \bibinfo {author} {\bibfnamefont {F.-Y.}\ \bibnamefont
  {Cyr-Racine}},\ }\bibfield  {title} {\bibinfo {title} {{A Ratio-Preserving
  Approach to Cosmological Concordance}},\ }\href@noop {} {\  (\bibinfo {year}
  {2024})},\ \Eprint {https://arxiv.org/abs/2403.05619} {arXiv:2403.05619
  [astro-ph.CO]} \BibitemShut {NoStop}%
\bibitem [{\citenamefont {Shimon}\ and\ \citenamefont
  {Rephaeli}(2020)}]{Shimon:2020dvb}%
  \BibitemOpen
  \bibfield  {author} {\bibinfo {author} {\bibfnamefont {M.}~\bibnamefont
  {Shimon}}\ and\ \bibinfo {author} {\bibfnamefont {Y.}~\bibnamefont
  {Rephaeli}},\ }\bibfield  {title} {\bibinfo {title} {{Parameter interplay of
  CMB temperature, space curvature, and expansion rate}},\ }\href
  {https://doi.org/10.1103/PhysRevD.102.083532} {\bibfield  {journal} {\bibinfo
   {journal} {Phys. Rev. D}\ }\textbf {\bibinfo {volume} {102}},\ \bibinfo
  {pages} {083532} (\bibinfo {year} {2020})},\ \Eprint
  {https://arxiv.org/abs/2009.14417} {arXiv:2009.14417 [astro-ph.CO]}
  \BibitemShut {NoStop}%
\bibitem [{\citenamefont {Prabhu}\ \emph {et~al.}(2024)\citenamefont {Prabhu}
  \emph {et~al.}}]{Prabhu:2024qix}%
  \BibitemOpen
  \bibfield  {author} {\bibinfo {author} {\bibfnamefont {K.}~\bibnamefont
  {Prabhu}} \emph {et~al.},\ }\bibfield  {title} {\bibinfo {title} {{Testing
  the $\mathbf{\Lambda}$CDM Cosmological Model with Forthcoming Measurements of
  the Cosmic Microwave Background with SPT-3G}},\ }\href@noop {} {\  (\bibinfo
  {year} {2024})},\ \Eprint {https://arxiv.org/abs/2403.17925}
  {arXiv:2403.17925 [astro-ph.CO]} \BibitemShut {NoStop}%
\bibitem [{\citenamefont {Sunyaev}\ and\ \citenamefont
  {Chluba}(2009)}]{Sunyaev2009}%
  \BibitemOpen
  \bibfield  {author} {\bibinfo {author} {\bibfnamefont {R.}~\bibnamefont
  {Sunyaev}}\ and\ \bibinfo {author} {\bibfnamefont {J.}~\bibnamefont
  {Chluba}},\ }\bibfield  {title} {\bibinfo {title} {Signals from the epoch of
  cosmological recombination - karl schwarzschild award lecture 2008},\ }\href
  {https://doi.org/10.1002/asna.200911237} {\bibfield  {journal} {\bibinfo
  {journal} {Astronomische Nachrichten}\ }\textbf {\bibinfo {volume} {330}},\
  \bibinfo {pages} {657 } (\bibinfo {year} {2009})}\BibitemShut {NoStop}%
\bibitem [{\citenamefont {Chluba}\ and\ \citenamefont
  {Ali-Haimoud}(2016)}]{Chluba2016Cosmospec}%
  \BibitemOpen
  \bibfield  {author} {\bibinfo {author} {\bibfnamefont {J.}~\bibnamefont
  {Chluba}}\ and\ \bibinfo {author} {\bibfnamefont {Y.}~\bibnamefont
  {Ali-Haimoud}},\ }\bibfield  {title} {\bibinfo {title} {{CosmoSpec: Fast and
  detailed computation of the cosmological recombination radiation from
  hydrogen and helium}},\ }\href {https://doi.org/10.1093/mnras/stv2691}
  {\bibfield  {journal} {\bibinfo  {journal} {Mon. Not. Roy. Astron. Soc.}\
  }\textbf {\bibinfo {volume} {456}},\ \bibinfo {pages} {3494} (\bibinfo {year}
  {2016})},\ \Eprint {https://arxiv.org/abs/1510.03877} {arXiv:1510.03877
  [astro-ph.CO]} \BibitemShut {NoStop}%
\bibitem [{\citenamefont {Chluba}\ and\ \citenamefont
  {Sunyaev}(2008)}]{Chluba:2007zz}%
  \BibitemOpen
  \bibfield  {author} {\bibinfo {author} {\bibfnamefont {J.}~\bibnamefont
  {Chluba}}\ and\ \bibinfo {author} {\bibfnamefont {R.~A.}\ \bibnamefont
  {Sunyaev}},\ }\bibfield  {title} {\bibinfo {title} {{Is there need and
  another way to measure the Cosmic Microwave Background temperature more
  accurately?}},\ }\href {https://doi.org/10.1051/0004-6361:20078200}
  {\bibfield  {journal} {\bibinfo  {journal} {Astron. Astrophys.}\ }\textbf
  {\bibinfo {volume} {478}},\ \bibinfo {pages} {L27} (\bibinfo {year}
  {2008})},\ \Eprint {https://arxiv.org/abs/0707.0188} {arXiv:0707.0188
  [astro-ph]} \BibitemShut {NoStop}%
\bibitem [{\citenamefont {Chluba}\ and\ \citenamefont
  {Sunyaev}(2009)}]{Chluba:2008aw}%
  \BibitemOpen
  \bibfield  {author} {\bibinfo {author} {\bibfnamefont {J.}~\bibnamefont
  {Chluba}}\ and\ \bibinfo {author} {\bibfnamefont {R.~A.}\ \bibnamefont
  {Sunyaev}},\ }\bibfield  {title} {\bibinfo {title} {{Pre-recombinational
  energy release and narrow features in the CMB spectrum}},\ }\href
  {https://doi.org/10.1051/0004-6361/200809840} {\bibfield  {journal} {\bibinfo
   {journal} {Astron. Astrophys.}\ }\textbf {\bibinfo {volume} {501}},\
  \bibinfo {pages} {29} (\bibinfo {year} {2009})},\ \Eprint
  {https://arxiv.org/abs/0803.3584} {arXiv:0803.3584 [astro-ph]} \BibitemShut
  {NoStop}%
\bibitem [{\citenamefont {{Chluba}}(2010)}]{Chluba2010dm}%
  \BibitemOpen
  \bibfield  {author} {\bibinfo {author} {\bibfnamefont {J.}~\bibnamefont
  {{Chluba}}},\ }\bibfield  {title} {\bibinfo {title} {{Could the cosmological
  recombination spectrum help us understand annihilating dark matter?}},\
  }\href {https://doi.org/10.1111/j.1365-2966.2009.15957.x} {\bibfield
  {journal} {\bibinfo  {journal} {\mnras}\ }\textbf {\bibinfo {volume} {402}},\
  \bibinfo {pages} {1195} (\bibinfo {year} {2010})},\ \Eprint
  {https://arxiv.org/abs/0910.3663} {arXiv:0910.3663 [astro-ph.CO]}
  \BibitemShut {NoStop}%
\bibitem [{\citenamefont {{Hart}}\ and\ \citenamefont
  {{Chluba}}(2023)}]{Hart2023CRR}%
  \BibitemOpen
  \bibfield  {author} {\bibinfo {author} {\bibfnamefont {L.}~\bibnamefont
  {{Hart}}}\ and\ \bibinfo {author} {\bibfnamefont {J.}~\bibnamefont
  {{Chluba}}},\ }\bibfield  {title} {\bibinfo {title} {{Using the cosmological
  recombination radiation to probe early dark energy and fundamental constant
  variations}},\ }\href {https://doi.org/10.1093/mnras/stac3697} {\bibfield
  {journal} {\bibinfo  {journal} {\mnras}\ }\textbf {\bibinfo {volume} {519}},\
  \bibinfo {pages} {3664} (\bibinfo {year} {2023})},\ \Eprint
  {https://arxiv.org/abs/2209.12290} {arXiv:2209.12290 [astro-ph.CO]}
  \BibitemShut {NoStop}%
\bibitem [{\citenamefont {{Sathyanarayana Rao}}\ \emph
  {et~al.}(2015)\citenamefont {{Sathyanarayana Rao}}, \citenamefont
  {{Subrahmanyan}}, \citenamefont {{Udaya Shankar}},\ and\ \citenamefont
  {{Chluba}}}]{Mayuri2015Detection}%
  \BibitemOpen
  \bibfield  {author} {\bibinfo {author} {\bibfnamefont {M.}~\bibnamefont
  {{Sathyanarayana Rao}}}, \bibinfo {author} {\bibfnamefont {R.}~\bibnamefont
  {{Subrahmanyan}}}, \bibinfo {author} {\bibfnamefont {N.}~\bibnamefont {{Udaya
  Shankar}}},\ and\ \bibinfo {author} {\bibfnamefont {J.}~\bibnamefont
  {{Chluba}}},\ }\bibfield  {title} {\bibinfo {title} {{On the Detection of
  Spectral Ripples from the Recombination Epoch}},\ }\href
  {https://doi.org/10.1088/0004-637X/810/1/3} {\bibfield  {journal} {\bibinfo
  {journal} {\apj}\ }\textbf {\bibinfo {volume} {810}},\ \bibinfo {eid} {3}
  (\bibinfo {year} {2015})}\BibitemShut {NoStop}%
\bibitem [{\citenamefont {{Desjacques}}\ \emph {et~al.}(2015)\citenamefont
  {{Desjacques}}, \citenamefont {{Chluba}}, \citenamefont {{Silk}},
  \citenamefont {{de Bernardis}},\ and\ \citenamefont
  {{Dor{\'e}}}}]{Vincent2015}%
  \BibitemOpen
  \bibfield  {author} {\bibinfo {author} {\bibfnamefont {V.}~\bibnamefont
  {{Desjacques}}}, \bibinfo {author} {\bibfnamefont {J.}~\bibnamefont
  {{Chluba}}}, \bibinfo {author} {\bibfnamefont {J.}~\bibnamefont {{Silk}}},
  \bibinfo {author} {\bibfnamefont {F.}~\bibnamefont {{de Bernardis}}},\ and\
  \bibinfo {author} {\bibfnamefont {O.}~\bibnamefont {{Dor{\'e}}}},\ }\bibfield
   {title} {\bibinfo {title} {{Detecting the cosmological recombination signal
  from space}},\ }\href {https://doi.org/10.1093/mnras/stv1291} {\bibfield
  {journal} {\bibinfo  {journal} {\mnras}\ }\textbf {\bibinfo {volume} {451}},\
  \bibinfo {pages} {4460} (\bibinfo {year} {2015})},\ \Eprint
  {https://arxiv.org/abs/1503.05589} {arXiv:1503.05589 [astro-ph.CO]}
  \BibitemShut {NoStop}%
\bibitem [{\citenamefont {Chluba}\ \emph {et~al.}(2019)\citenamefont {Chluba}
  \emph {et~al.}}]{Chluba2019Spectral}%
  \BibitemOpen
  \bibfield  {author} {\bibinfo {author} {\bibfnamefont {J.}~\bibnamefont
  {Chluba}} \emph {et~al.},\ }\bibfield  {title} {\bibinfo {title} {{Spectral
  Distortions of the CMB as a Probe of Inflation, Recombination, Structure
  Formation and Particle Physics}},\ }\href@noop {} {\bibfield  {journal}
  {\bibinfo  {journal} {\baas}\ }\textbf {\bibinfo {volume} {51}},\ \bibinfo
  {eid} {184} (\bibinfo {year} {2019})},\ \Eprint
  {https://arxiv.org/abs/1903.04218} {arXiv:1903.04218 [astro-ph.CO]}
  \BibitemShut {NoStop}%
\bibitem [{\citenamefont {Hart}\ \emph {et~al.}(2020)\citenamefont {Hart},
  \citenamefont {Rotti},\ and\ \citenamefont {Chluba}}]{Hart2020Sensitivity}%
  \BibitemOpen
  \bibfield  {author} {\bibinfo {author} {\bibfnamefont {L.}~\bibnamefont
  {Hart}}, \bibinfo {author} {\bibfnamefont {A.}~\bibnamefont {Rotti}},\ and\
  \bibinfo {author} {\bibfnamefont {J.}~\bibnamefont {Chluba}},\ }\bibfield
  {title} {\bibinfo {title} {{Sensitivity forecasts for the cosmological
  recombination radiation in the presence of foregrounds}},\ }\href
  {https://doi.org/10.1093/mnras/staa2255} {\bibfield  {journal} {\bibinfo
  {journal} {Mon. Not. Roy. Astron. Soc.}\ }\textbf {\bibinfo {volume} {497}},\
  \bibinfo {pages} {4535} (\bibinfo {year} {2020})},\ \Eprint
  {https://arxiv.org/abs/2006.04826} {arXiv:2006.04826 [astro-ph.CO]}
  \BibitemShut {NoStop}%
\bibitem [{\citenamefont {Chluba}\ \emph {et~al.}(2021)\citenamefont {Chluba}
  \emph {et~al.}}]{Chluba2019Voyage}%
  \BibitemOpen
  \bibfield  {author} {\bibinfo {author} {\bibfnamefont {J.}~\bibnamefont
  {Chluba}} \emph {et~al.},\ }\bibfield  {title} {\bibinfo {title} {{New
  horizons in cosmology with spectral distortions of the cosmic microwave
  background}},\ }\href {https://doi.org/10.1007/s10686-021-09729-5} {\bibfield
   {journal} {\bibinfo  {journal} {Exper. Astron.}\ }\textbf {\bibinfo {volume}
  {51}},\ \bibinfo {pages} {1515} (\bibinfo {year} {2021})},\ \Eprint
  {https://arxiv.org/abs/1909.01593} {arXiv:1909.01593 [astro-ph.CO]}
  \BibitemShut {NoStop}%
\bibitem [{\citenamefont {Adame}\ \emph
  {et~al.}(2024{\natexlab{a}})\citenamefont {Adame} \emph
  {et~al.}}]{DESI:2024uvr}%
  \BibitemOpen
  \bibfield  {author} {\bibinfo {author} {\bibfnamefont {A.~G.}\ \bibnamefont
  {Adame}} \emph {et~al.} (\bibinfo {collaboration} {DESI}),\ }\bibfield
  {title} {\bibinfo {title} {{DESI 2024 III: Baryon Acoustic Oscillations from
  Galaxies and Quasars}},\ }\href@noop {} {\  (\bibinfo {year}
  {2024}{\natexlab{a}})},\ \Eprint {https://arxiv.org/abs/2404.03000}
  {arXiv:2404.03000 [astro-ph.CO]} \BibitemShut {NoStop}%
\bibitem [{\citenamefont {Adame}\ \emph
  {et~al.}(2024{\natexlab{b}})\citenamefont {Adame} \emph
  {et~al.}}]{DESI:2024lzq}%
  \BibitemOpen
  \bibfield  {author} {\bibinfo {author} {\bibfnamefont {A.~G.}\ \bibnamefont
  {Adame}} \emph {et~al.} (\bibinfo {collaboration} {DESI}),\ }\bibfield
  {title} {\bibinfo {title} {{DESI 2024 IV: Baryon Acoustic Oscillations from
  the Lyman Alpha Forest}},\ }\href@noop {} {\  (\bibinfo {year}
  {2024}{\natexlab{b}})},\ \Eprint {https://arxiv.org/abs/2404.03001}
  {arXiv:2404.03001 [astro-ph.CO]} \BibitemShut {NoStop}%
\bibitem [{\citenamefont {Adame}\ \emph
  {et~al.}(2024{\natexlab{c}})\citenamefont {Adame} \emph
  {et~al.}}]{DESI:2024mwx}%
  \BibitemOpen
  \bibfield  {author} {\bibinfo {author} {\bibfnamefont {A.~G.}\ \bibnamefont
  {Adame}} \emph {et~al.} (\bibinfo {collaboration} {DESI}),\ }\bibfield
  {title} {\bibinfo {title} {{DESI 2024 VI: Cosmological Constraints from the
  Measurements of Baryon Acoustic Oscillations}},\ }\href@noop {} {\  (\bibinfo
  {year} {2024}{\natexlab{c}})},\ \Eprint {https://arxiv.org/abs/2404.03002}
  {arXiv:2404.03002 [astro-ph.CO]} \BibitemShut {NoStop}%
\bibitem [{\citenamefont {Lynch}\ \emph
  {et~al.}(2024{\natexlab{b}})\citenamefont {Lynch}, \citenamefont {Knox},\
  and\ \citenamefont {Chluba}}]{Lynch:2024hzh}%
  \BibitemOpen
  \bibfield  {author} {\bibinfo {author} {\bibfnamefont {G.~P.}\ \bibnamefont
  {Lynch}}, \bibinfo {author} {\bibfnamefont {L.}~\bibnamefont {Knox}},\ and\
  \bibinfo {author} {\bibfnamefont {J.}~\bibnamefont {Chluba}},\ }\bibfield
  {title} {\bibinfo {title} {{DESI and the Hubble tension in light of modified
  recombination}},\ }\href@noop {} {\  (\bibinfo {year}
  {2024}{\natexlab{b}})},\ \Eprint {https://arxiv.org/abs/2406.10202}
  {arXiv:2406.10202 [astro-ph.CO]} \BibitemShut {NoStop}%
\bibitem [{\citenamefont {Lewis}\ \emph {et~al.}(2000)\citenamefont {Lewis},
  \citenamefont {Challinor},\ and\ \citenamefont {Lasenby}}]{Lewis:1999bs}%
  \BibitemOpen
  \bibfield  {author} {\bibinfo {author} {\bibfnamefont {A.}~\bibnamefont
  {Lewis}}, \bibinfo {author} {\bibfnamefont {A.}~\bibnamefont {Challinor}},\
  and\ \bibinfo {author} {\bibfnamefont {A.}~\bibnamefont {Lasenby}},\
  }\bibfield  {title} {\bibinfo {title} {{Efficient computation of CMB
  anisotropies in closed FRW models}},\ }\href {https://doi.org/10.1086/309179}
  {\bibfield  {journal} {\bibinfo  {journal} {Astrophys. J.}\ }\textbf
  {\bibinfo {volume} {538}},\ \bibinfo {pages} {473} (\bibinfo {year}
  {2000})},\ \Eprint {https://arxiv.org/abs/astro-ph/9911177}
  {arXiv:astro-ph/9911177} \BibitemShut {NoStop}%
\bibitem [{\citenamefont {McCarthy}\ \emph {et~al.}(2022)\citenamefont
  {McCarthy}, \citenamefont {Hill},\ and\ \citenamefont
  {Madhavacheril}}]{McCarthy:2021lfp}%
  \BibitemOpen
  \bibfield  {author} {\bibinfo {author} {\bibfnamefont {F.}~\bibnamefont
  {McCarthy}}, \bibinfo {author} {\bibfnamefont {J.~C.}\ \bibnamefont {Hill}},\
  and\ \bibinfo {author} {\bibfnamefont {M.~S.}\ \bibnamefont
  {Madhavacheril}},\ }\bibfield  {title} {\bibinfo {title} {{Baryonic feedback
  biases on fundamental physics from lensed CMB power spectra}},\ }\href
  {https://doi.org/10.1103/PhysRevD.105.023517} {\bibfield  {journal} {\bibinfo
   {journal} {Phys. Rev. D}\ }\textbf {\bibinfo {volume} {105}},\ \bibinfo
  {pages} {023517} (\bibinfo {year} {2022})},\ \Eprint
  {https://arxiv.org/abs/2103.05582} {arXiv:2103.05582 [astro-ph.CO]}
  \BibitemShut {NoStop}%
\bibitem [{\citenamefont {Hill}\ \emph {et~al.}(2022)\citenamefont {Hill},
  \citenamefont {Calabrese}, \citenamefont {Aiola}, \citenamefont {Battaglia},
  \citenamefont {Bolliet}, \citenamefont {Choi}, \citenamefont {Devlin},
  \citenamefont {Duivenvoorden}, \citenamefont {Dunkley}, \citenamefont
  {Ferraro}, \citenamefont {Gallardo}, \citenamefont {Gluscevic}, \citenamefont
  {Hasselfield}, \citenamefont {Hilton}, \citenamefont {Hincks}, \citenamefont
  {Hlo\ifmmode~\check{z}\else \v{z}\fi{}ek}, \citenamefont {Koopman},
  \citenamefont {Kosowsky}, \citenamefont {La~Posta}, \citenamefont {Louis},
  \citenamefont {Madhavacheril}, \citenamefont {McMahon}, \citenamefont
  {Moodley}, \citenamefont {Naess}, \citenamefont {Natale}, \citenamefont
  {Nati}, \citenamefont {Newburgh}, \citenamefont {Niemack}, \citenamefont
  {Page}, \citenamefont {Partridge}, \citenamefont {Qu}, \citenamefont
  {Salatino}, \citenamefont {Schillaci}, \citenamefont {Sehgal}, \citenamefont
  {Sherwin}, \citenamefont {Sif\'on}, \citenamefont {Spergel}, \citenamefont
  {Staggs}, \citenamefont {Storer}, \citenamefont {van Engelen}, \citenamefont
  {Vavagiakis}, \citenamefont {Wollack},\ and\ \citenamefont
  {Xu}}]{Hill:2021yec}%
  \BibitemOpen
  \bibfield  {author} {\bibinfo {author} {\bibfnamefont {J.~C.}\ \bibnamefont
  {Hill}}, \bibinfo {author} {\bibfnamefont {E.}~\bibnamefont {Calabrese}},
  \bibinfo {author} {\bibfnamefont {S.}~\bibnamefont {Aiola}}, \bibinfo
  {author} {\bibfnamefont {N.}~\bibnamefont {Battaglia}}, \bibinfo {author}
  {\bibfnamefont {B.}~\bibnamefont {Bolliet}}, \bibinfo {author} {\bibfnamefont
  {S.~K.}\ \bibnamefont {Choi}}, \bibinfo {author} {\bibfnamefont {M.~J.}\
  \bibnamefont {Devlin}}, \bibinfo {author} {\bibfnamefont {A.~J.}\
  \bibnamefont {Duivenvoorden}}, \bibinfo {author} {\bibfnamefont
  {J.}~\bibnamefont {Dunkley}}, \bibinfo {author} {\bibfnamefont
  {S.}~\bibnamefont {Ferraro}}, \bibinfo {author} {\bibfnamefont {P.~A.}\
  \bibnamefont {Gallardo}}, \bibinfo {author} {\bibfnamefont {V.}~\bibnamefont
  {Gluscevic}}, \bibinfo {author} {\bibfnamefont {M.}~\bibnamefont
  {Hasselfield}}, \bibinfo {author} {\bibfnamefont {M.}~\bibnamefont {Hilton}},
  \bibinfo {author} {\bibfnamefont {A.~D.}\ \bibnamefont {Hincks}}, \bibinfo
  {author} {\bibfnamefont {R.}~\bibnamefont {Hlo\ifmmode~\check{z}\else
  \v{z}\fi{}ek}}, \bibinfo {author} {\bibfnamefont {B.~J.}\ \bibnamefont
  {Koopman}}, \bibinfo {author} {\bibfnamefont {A.}~\bibnamefont {Kosowsky}},
  \bibinfo {author} {\bibfnamefont {A.}~\bibnamefont {La~Posta}}, \bibinfo
  {author} {\bibfnamefont {T.}~\bibnamefont {Louis}}, \bibinfo {author}
  {\bibfnamefont {M.~S.}\ \bibnamefont {Madhavacheril}}, \bibinfo {author}
  {\bibfnamefont {J.}~\bibnamefont {McMahon}}, \bibinfo {author} {\bibfnamefont
  {K.}~\bibnamefont {Moodley}}, \bibinfo {author} {\bibfnamefont
  {S.}~\bibnamefont {Naess}}, \bibinfo {author} {\bibfnamefont
  {U.}~\bibnamefont {Natale}}, \bibinfo {author} {\bibfnamefont
  {F.}~\bibnamefont {Nati}}, \bibinfo {author} {\bibfnamefont {L.}~\bibnamefont
  {Newburgh}}, \bibinfo {author} {\bibfnamefont {M.~D.}\ \bibnamefont
  {Niemack}}, \bibinfo {author} {\bibfnamefont {L.~A.}\ \bibnamefont {Page}},
  \bibinfo {author} {\bibfnamefont {B.}~\bibnamefont {Partridge}}, \bibinfo
  {author} {\bibfnamefont {F.~J.}\ \bibnamefont {Qu}}, \bibinfo {author}
  {\bibfnamefont {M.}~\bibnamefont {Salatino}}, \bibinfo {author}
  {\bibfnamefont {A.}~\bibnamefont {Schillaci}}, \bibinfo {author}
  {\bibfnamefont {N.}~\bibnamefont {Sehgal}}, \bibinfo {author} {\bibfnamefont
  {B.~D.}\ \bibnamefont {Sherwin}}, \bibinfo {author} {\bibfnamefont
  {C.}~\bibnamefont {Sif\'on}}, \bibinfo {author} {\bibfnamefont {D.~N.}\
  \bibnamefont {Spergel}}, \bibinfo {author} {\bibfnamefont {S.~T.}\
  \bibnamefont {Staggs}}, \bibinfo {author} {\bibfnamefont {E.~R.}\
  \bibnamefont {Storer}}, \bibinfo {author} {\bibfnamefont {A.}~\bibnamefont
  {van Engelen}}, \bibinfo {author} {\bibfnamefont {E.~M.}\ \bibnamefont
  {Vavagiakis}}, \bibinfo {author} {\bibfnamefont {E.~J.}\ \bibnamefont
  {Wollack}},\ and\ \bibinfo {author} {\bibfnamefont {Z.}~\bibnamefont {Xu}},\
  }\bibfield  {title} {\bibinfo {title} {Atacama cosmology telescope:
  Constraints on prerecombination early dark energy},\ }\href
  {https://doi.org/10.1103/PhysRevD.105.123536} {\bibfield  {journal} {\bibinfo
   {journal} {Phys. Rev. D}\ }\textbf {\bibinfo {volume} {105}},\ \bibinfo
  {pages} {123536} (\bibinfo {year} {2022})}\BibitemShut {NoStop}%
\bibitem [{\citenamefont {Hinton}\ \emph {et~al.}(2020)\citenamefont {Hinton},
  \citenamefont {Howlett},\ and\ \citenamefont {Davis}}]{Hinton:2019nky}%
  \BibitemOpen
  \bibfield  {author} {\bibinfo {author} {\bibfnamefont {S.~R.}\ \bibnamefont
  {Hinton}}, \bibinfo {author} {\bibfnamefont {C.}~\bibnamefont {Howlett}},\
  and\ \bibinfo {author} {\bibfnamefont {T.~M.}\ \bibnamefont {Davis}},\
  }\bibfield  {title} {\bibinfo {title} {{Barry and the BAO Model
  Comparison}},\ }\href {https://doi.org/10.1093/mnras/staa361} {\bibfield
  {journal} {\bibinfo  {journal} {Mon. Not. Roy. Astron. Soc.}\ }\textbf
  {\bibinfo {volume} {493}},\ \bibinfo {pages} {4078} (\bibinfo {year}
  {2020})},\ \Eprint {https://arxiv.org/abs/1912.01175} {arXiv:1912.01175
  [astro-ph.CO]} \BibitemShut {NoStop}%
\bibitem [{\citenamefont {Carter}\ \emph {et~al.}(2020)\citenamefont {Carter},
  \citenamefont {Beutler}, \citenamefont {Percival}, \citenamefont {DeRose},
  \citenamefont {Wechsler},\ and\ \citenamefont {Zhao}}]{Carter:2019ulk}%
  \BibitemOpen
  \bibfield  {author} {\bibinfo {author} {\bibfnamefont {P.}~\bibnamefont
  {Carter}}, \bibinfo {author} {\bibfnamefont {F.}~\bibnamefont {Beutler}},
  \bibinfo {author} {\bibfnamefont {W.~J.}\ \bibnamefont {Percival}}, \bibinfo
  {author} {\bibfnamefont {J.}~\bibnamefont {DeRose}}, \bibinfo {author}
  {\bibfnamefont {R.~H.}\ \bibnamefont {Wechsler}},\ and\ \bibinfo {author}
  {\bibfnamefont {C.}~\bibnamefont {Zhao}},\ }\bibfield  {title} {\bibinfo
  {title} {{The impact of the fiducial cosmology assumption on BAO distance
  scale measurements}},\ }\href {https://doi.org/10.1093/mnras/staa761}
  {\bibfield  {journal} {\bibinfo  {journal} {Mon. Not. Roy. Astron. Soc.}\
  }\textbf {\bibinfo {volume} {494}},\ \bibinfo {pages} {2076} (\bibinfo {year}
  {2020})},\ \Eprint {https://arxiv.org/abs/1906.03035} {arXiv:1906.03035
  [astro-ph.CO]} \BibitemShut {NoStop}%
\bibitem [{\citenamefont {Sherwin}\ and\ \citenamefont
  {White}(2019)}]{Sherwin:2018wbu}%
  \BibitemOpen
  \bibfield  {author} {\bibinfo {author} {\bibfnamefont {B.~D.}\ \bibnamefont
  {Sherwin}}\ and\ \bibinfo {author} {\bibfnamefont {M.}~\bibnamefont
  {White}},\ }\bibfield  {title} {\bibinfo {title} {{The Impact of Wrong
  Assumptions in BAO Reconstruction}},\ }\href
  {https://doi.org/10.1088/1475-7516/2019/02/027} {\bibfield  {journal}
  {\bibinfo  {journal} {JCAP}\ }\textbf {\bibinfo {volume} {02}},\ \bibinfo
  {pages} {027}},\ \Eprint {https://arxiv.org/abs/1808.04384} {arXiv:1808.04384
  [astro-ph.CO]} \BibitemShut {NoStop}%
\bibitem [{\citenamefont {Bernal}\ \emph {et~al.}(2020)\citenamefont {Bernal},
  \citenamefont {Smith}, \citenamefont {Boddy},\ and\ \citenamefont
  {Kamionkowski}}]{Bernal:2020vbb}%
  \BibitemOpen
  \bibfield  {author} {\bibinfo {author} {\bibfnamefont {J.~L.}\ \bibnamefont
  {Bernal}}, \bibinfo {author} {\bibfnamefont {T.~L.}\ \bibnamefont {Smith}},
  \bibinfo {author} {\bibfnamefont {K.~K.}\ \bibnamefont {Boddy}},\ and\
  \bibinfo {author} {\bibfnamefont {M.}~\bibnamefont {Kamionkowski}},\
  }\bibfield  {title} {\bibinfo {title} {{Robustness of baryon acoustic
  oscillation constraints for early-Universe modifications of $\Lambda$CDM
  cosmology}},\ }\href {https://doi.org/10.1103/PhysRevD.102.123515} {\bibfield
   {journal} {\bibinfo  {journal} {Phys. Rev. D}\ }\textbf {\bibinfo {volume}
  {102}},\ \bibinfo {pages} {123515} (\bibinfo {year} {2020})},\ \Eprint
  {https://arxiv.org/abs/2004.07263} {arXiv:2004.07263 [astro-ph.CO]}
  \BibitemShut {NoStop}%
\bibitem [{\citenamefont {Anselmi}\ \emph {et~al.}(2023)\citenamefont
  {Anselmi}, \citenamefont {Starkman},\ and\ \citenamefont
  {Renzi}}]{PhysRevD.107.123506}%
  \BibitemOpen
  \bibfield  {author} {\bibinfo {author} {\bibfnamefont {S.}~\bibnamefont
  {Anselmi}}, \bibinfo {author} {\bibfnamefont {G.~D.}\ \bibnamefont
  {Starkman}},\ and\ \bibinfo {author} {\bibfnamefont {A.}~\bibnamefont
  {Renzi}},\ }\bibfield  {title} {\bibinfo {title} {Cosmological forecasts for
  future galaxy surveys with the linear point standard ruler: Toward consistent
  bao analyses far from a fiducial cosmology},\ }\href
  {https://doi.org/10.1103/PhysRevD.107.123506} {\bibfield  {journal} {\bibinfo
   {journal} {Phys. Rev. D}\ }\textbf {\bibinfo {volume} {107}},\ \bibinfo
  {pages} {123506} (\bibinfo {year} {2023})}\BibitemShut {NoStop}%
\bibitem [{\citenamefont {O'Dwyer}\ \emph {et~al.}(2020)\citenamefont
  {O'Dwyer}, \citenamefont {Anselmi}, \citenamefont {Starkman}, \citenamefont
  {Corasaniti}, \citenamefont {Sheth},\ and\ \citenamefont
  {Zehavi}}]{PhysRevD.101.083517}%
  \BibitemOpen
  \bibfield  {author} {\bibinfo {author} {\bibfnamefont {M.}~\bibnamefont
  {O'Dwyer}}, \bibinfo {author} {\bibfnamefont {S.}~\bibnamefont {Anselmi}},
  \bibinfo {author} {\bibfnamefont {G.~D.}\ \bibnamefont {Starkman}}, \bibinfo
  {author} {\bibfnamefont {P.-S.}\ \bibnamefont {Corasaniti}}, \bibinfo
  {author} {\bibfnamefont {R.~K.}\ \bibnamefont {Sheth}},\ and\ \bibinfo
  {author} {\bibfnamefont {I.}~\bibnamefont {Zehavi}},\ }\bibfield  {title}
  {\bibinfo {title} {Linear point and sound horizon as purely geometric
  standard rulers},\ }\href {https://doi.org/10.1103/PhysRevD.101.083517}
  {\bibfield  {journal} {\bibinfo  {journal} {Phys. Rev. D}\ }\textbf {\bibinfo
  {volume} {101}},\ \bibinfo {pages} {083517} (\bibinfo {year}
  {2020})}\BibitemShut {NoStop}%
\bibitem [{\citenamefont {Anselmi}\ \emph {et~al.}(2019)\citenamefont
  {Anselmi}, \citenamefont {Corasaniti}, \citenamefont {Sanchez}, \citenamefont
  {Starkman}, \citenamefont {Sheth},\ and\ \citenamefont
  {Zehavi}}]{PhysRevD.99.123515}%
  \BibitemOpen
  \bibfield  {author} {\bibinfo {author} {\bibfnamefont {S.}~\bibnamefont
  {Anselmi}}, \bibinfo {author} {\bibfnamefont {P.-S.}\ \bibnamefont
  {Corasaniti}}, \bibinfo {author} {\bibfnamefont {A.~G.}\ \bibnamefont
  {Sanchez}}, \bibinfo {author} {\bibfnamefont {G.~D.}\ \bibnamefont
  {Starkman}}, \bibinfo {author} {\bibfnamefont {R.~K.}\ \bibnamefont
  {Sheth}},\ and\ \bibinfo {author} {\bibfnamefont {I.}~\bibnamefont
  {Zehavi}},\ }\bibfield  {title} {\bibinfo {title} {Cosmic distance inference
  from purely geometric bao methods: Linear point standard ruler and
  correlation function model fitting},\ }\href
  {https://doi.org/10.1103/PhysRevD.99.123515} {\bibfield  {journal} {\bibinfo
  {journal} {Phys. Rev. D}\ }\textbf {\bibinfo {volume} {99}},\ \bibinfo
  {pages} {123515} (\bibinfo {year} {2019})}\BibitemShut {NoStop}%
\end{thebibliography}%

\clearpage
\appendix

\section{\label{app:emulator_appendix}Emulator details}

Here we provide additional details regarding the \texttt{CONNECT} settings used when creating our final \modrec\ emulator, as well as additional error metrics used to assess its accuracy.

\subsection{\texttt{CONNECT} settings}

In order to achieve the precision required for the \sptthreeg\ forecasting in Sec.~\ref{sec:Forecasts}, we had to adjust the default \texttt{CLASS} settings used by \texttt{CONNECT} when generating the training data. \citet{Bolliet:2023sst} provide settings for \texttt{CLASS} which match ultra-precise  \texttt{CAMB} \citep{Lewis:1999bs} calculations for the multipole range considered in that paper, i.e. with $\ell_{\rm max} = 11000$ \citep{McCarthy:2021lfp, Hill:2021yec}. We found that for our purposes, we could relax some of theses settings to further optimize the computational cost per Einstein-Boltzmann evaluation. The final precision settings used were:

\begin{itemize}
    \item \texttt{accurate\_lensing=1.0} 
    \item  \texttt{k\_max\_tau0\_over\_l\_max=15.00}
    \item  \texttt{P\_k\_max\_1/Mpc=500.00}
    \item  \texttt{perturbations\_sampling\_stepsize=0.05}
    \item  \texttt{start\_sources\_at\_tau\_c\_over\_tau\_h=0.004}
    \item  \texttt{non\_linear=hmcode}
    \item  \texttt{eta\_0=0.603}
    \item  \texttt{c\_min=3.130}
\end{itemize}
These settings were found to match the ultra-precise \texttt{CAMB} settings to within $0.25\%$ for the TT, TE, EE and $\phi \phi$ power spectra with the fiducial cosmological parameters and standard recombination process. Due to the prohibitive cost of generating ultra-precise power spectra for a range of recombination histories, we did not assess the impact of modified recombination on this agreement. Using these settings, a \texttt{CLASS} computation takes $\mathcal{O}(10^2{\rm s})$ on a single laptop using 8 CPUs. 

Our choices for the boundaries of the initial Latin hypercube are shown in Table~\ref{tab:cube_boundaries}. For the standard cosmological parameters, the boundaries are chosen to cover the $\pm 10\sigma$ ranges around the \planck\ mean value, but are in most cases much greater than this. For the control point parameters, bounds are chosen to be larger than prior bounds described in Sec.~\ref{sec:Methodology}.

\begin{table}[t]
\caption{\label{tab:cube_boundaries}
Boundaries of the initial cube used to generate training data for the emulator. }
\begin{ruledtabular}
\begin{tabular}{lcc}
\textrm{Parameter} & \textrm{Lower bound} & \textrm{Upper bound} \\
\colrule
$\omega_b$ & 0.01768 & 0.02707 \\
$\omega_{cdm}$ & 0.1028 & 0.1374 \\
$n_s$ & 0.8784 & 1.053\\
$\tau_{reio}$ & 0.02760 & 0.1000 \\
$\ln(10^{10}A_s)$ & 2.837 & 3.257 \\
$H_0$ & 50.00 & 90.00 \\
\colrule
$\tilde{q}_1$ & -0.001305 & 0.01305 \\
$\tilde{q}_2$ & -0.003562 & 0.03562 \\
$\tilde{q}_3$ & -0.02026 & 0.2026 \\
$\tilde{q}_4$ & -0.1043 & 0.9883 \\
$\tilde{q}_5$ & -0.3225 & 0.7700 \\
$\tilde{q}_6$ & -0.6459 & 0.4467 \\
$\tilde{q}_7$ & -0.9191 & 0.1735 \\
\end{tabular}
\end{ruledtabular}
\end{table}

\subsection{Additional considerations}

\begin{figure*}[t]
    \centering
    \includegraphics{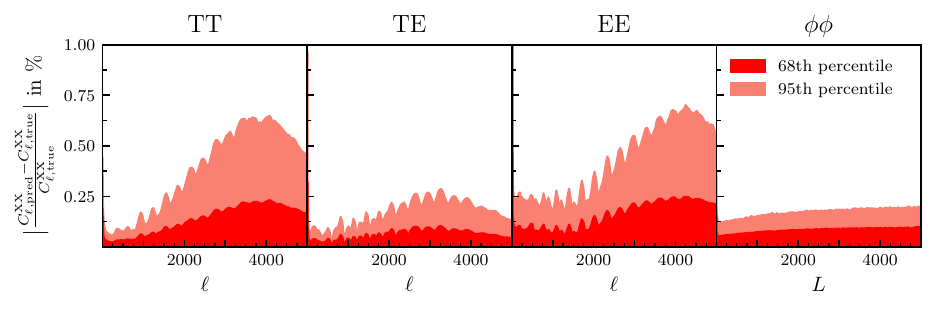}
    \caption{\label{fig:cl_percentile_errors} Error percentiles for the TT, TE, and EE angular power spectra for the \modrec\ model on a test set comprising of 5\% of the models randomly selected after the iterative process described in the text. The emulator does not use the test data during training. The error is defined relative to the true value as computed with \texttt{CLASS}, except for the TE spectrum, where the denominator is $\sqrt{C_{\ell, {\rm true}}^{TT} C_{\ell, {\rm true}}^{EE}}$.}
\end{figure*}

\begin{figure*}[t]
    \centering
    \includegraphics{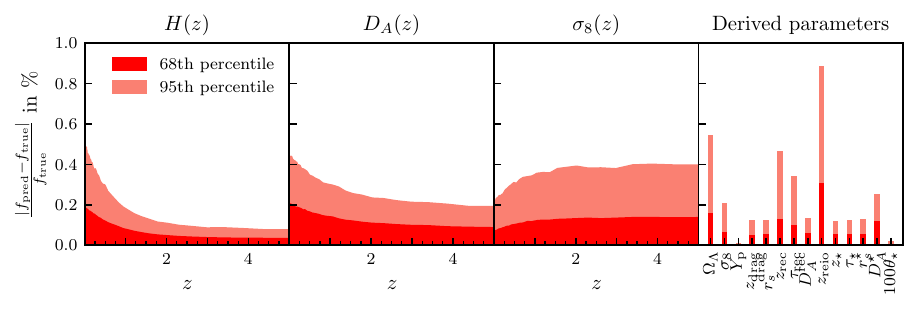}
    \caption{\label{fig:z_percentile_errors} Error percentiles for the redshift dependent and derived quantities from our final emulator.}
\end{figure*}

\begin{figure*}[t]
    \centering
    \includegraphics{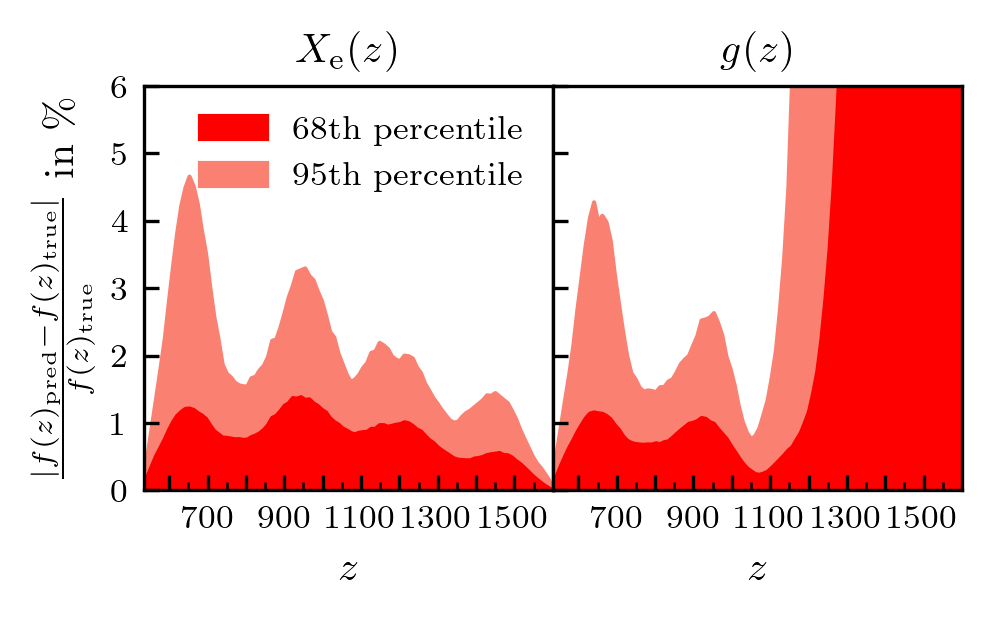}
    \caption{\label{fig:rec_errors} Error percentiles for the recombination quantities. For $\Xe(z)$, the 68th percentile errors are mostly below 1\% over the redshift range in question. For $g(z)$, the large errors at $z\gtrsim 1050$ are because the visibility function is very close to zero. }
\end{figure*}

The agreement between constraints obtained using the emulator and the reweighted constraints using \texttt{CLASS} are shown in Fig.~\ref{fig:reweighted}. Here we discuss additional considerations that we found to be important when assessing emulator accuracy. The high temperature MCMC chain generated during a single iterative phase did not fully explore the region of parameter space containing the $T=1$ posterior. To increase the amount of training data in these undersampled regions, we introduced priors in subsequent iterations that forced training data to be generated from these areas. Because of these priors, successive iterations would often explore different parts of the parameter space, preventing termination of the iterative process based on \texttt{CONNECT}'s default internal convergence criteria. 

As such, we instituted two checks to decide when to terminate the data generation process, and we subsequently validated an emulator trained on the compelete training set using the test in Fig.~\ref{fig:reweighted}. First, we required that the 95th percentile errors in the CMB power spectra were below 1\% up to $\ell = 4000$. These errors, as well as errors for the other quantities reproduced by our emulator, are presented in Figs.~\ref{fig:cl_percentile_errors}, ~\ref{fig:z_percentile_errors}, and ~\ref{fig:rec_errors}.

The percentile errors do not contain information about where in parameter space errors might be concentrated. To identify such regions, our second check was to visually assess heatmap plots similar to Fig.~\ref{fig:reweighted} and ensure there were no concentrated regions where $\Delta \chi^2$ for the test set was large. If there were such regions, we introduced priors that would limit the next iteration to the area of poor accuracy, increasing the amount of training data covering that region. As a result of these interventions, our test set was not optimally distributed over the posterior probability. However, we found that this method was generally successful in eliminating regions of high error. \clearpage

\changed{\section{\label{app:BAO_appendix}Robustness of BAO measurements with modified recombination}}

\changed{In this appendix, we assess the robustness of BAO measurements in the case that a modified recombination era is actually present, but not accounted for during the BAO analyses.}

\changed{\subsection{Impact of modified recombination on the BAO feature}}

\changed{Modern BAO measurements are typically obtained from galaxy surveys by fitting a pre-computed template of the BAO feature, either in configuration space or Fourier space, to observations \citep[see Appendix A of][]{eBOSS:2020yzd}. In such template-based methods, generally the observed linear matter power spectrum is modeled as a smooth component and an oscillatory component containing the BAO feature, such that $P(k) = P_{\rm sm}(k) O_{\rm lin}(k)$, where both components contain various nuisance parameters which capture effects beyond changes in the BAO scale. This template is computed using an assumed cosmological model and fiducial parameters. The exact analytical form of the the template used can vary between analyses as can the decomposition method  \citep[for a comparison, see][]{Hinton:2019nky}, but in general template-based approaches are robust to wrong assumptions regarding the fiducial cosmology \citep{Carter:2019ulk} and density field reconstruction \citep{Sherwin:2018wbu}.}

\changed{\citet{Bernal:2020vbb} demonstrated that the template-based method described above is also robust to a variety of early-universe modifications of \lcdm\ such as early dark energy and additional light relics. They noted that, if the true value of $r_s^{\rm drag}$ differs from the fiducial value used to compute the template, then the true and fiducial $O_{\rm lin}(k)$ will be identical upon rescaling $k \to k/(r_s^{\rm drag}/r_s^{\rm drag, fid})$. This is easily accommodated in template-based approaches, so if a model only changes the value of $r_s^{\rm drag}$, the template-based BAO measurements are expected to be robust. However, if a modified recombination era is actually present, and it affects the BAO feature in $O_{\rm lin}(k)$ beyond a change in $r_s^{\rm drag}$, the resulting BAO angle determinations may be biased.}

\changed{To assess whether this, we partially extend the analysis of \citep{Bernal:2020vbb} to the \modrec\ model. We first isolate the BAO feature by decomposing $P(k) = P_{\rm sm}(k) O_{\rm lin}(k)$ and then vary the recombination history to examine the impact of these changes on $O_{\rm lin}$, after rescaling $k \to k/(r_s^{\rm drag}/r_s^{\rm drag, fid})$. See Section IV of \citep{Bernal:2020vbb} for a full description of this method.}

\changed{As we only consider BAO data in combination with CMB data, we limit the variations of the recombination history to those allowed by \planck. Additionally, instead of considering variations in individual control points, we instead consider the correlated variations found in the \planck-only chain. We bin all points in the \planck-only chain according to their value of $r_s^{\rm drag}$, between $137.5 \text{ Mpc} \le r_s^{\rm drag} \le 154.5 \text{ Mpc}$, with unit bin width. These limits represent the 95\% confidence interval in the \planck-only chain. Within each bin, we compute the mean $\tilde{q}_i$, and adopt those values as the representative ionization history for that bin. Keeping the cosmological parameters fixed to their fiducial values, we compute the effect on $O_{\rm lin}(k)$ of each of these alternative recombination histories.}

\begin{figure}[t!]
    \centering
    \includegraphics{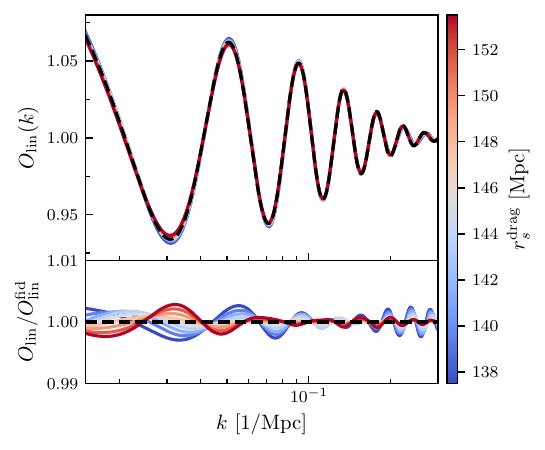}
    \caption{\label{fig:bao_robustness} \changed{\textit{Top panel:} Changes to the BAO feature in the linear matter power spectrum, $O_{\rm lin}(k)$, induced by a modified recombination epoch. Models from the \planck-only chain are binned according to their value of $r_s^{\rm drag}$, and the matter power spectrum is computed using the fiducial cosmological parameters with the mean recombination history in each bin. We have rescaled $k \to k/(r_s^{\rm drag}/r_s^{\rm drag, fid})$ for each model to isolate changes beyond a change in the sound horizon. The selected models represent the $95\%$ confidence region in the \planck-only chain. \textit{Bottom panel:} The ratio of the modified $O_{\rm lin}(k)$ to the fiducial for each of the models.} }
\end{figure}

\changed{The result of this is shown in Fig.~\ref{fig:bao_robustness}, after we have rescaled $k \to k/(r_s^{\rm drag}/r_s^{\rm drag, fid})$. Adjusting $r_s^{\rm drag}$ accounts for the majority of the variation in the BAO feature induced by varying the recombination history, with residual differences in the peak amplitudes. As noted in \citep{Bernal:2020vbb}, the peak amplitudes are affected by the nonlinear evolution of the density field, but these nonlinearities are often marginalized over in analyses. The amplitude of these variations are less than 1\%, which is smaller than the analogous variations found by \citep{Bernal:2020vbb} when varying $\omega_{\rm cdm }$ and $\omega_b$, for which they found no appreciable bias in the subsequent BAO determinations. Therefore we do not carry out the full analysis presented in \citet{Bernal:2020vbb} to estimate the resulting biases, and conclude that template-based BAO determinations are robust to the changes in the matter power spectrum induced by the \modrec\ model.}

\changed{Additional complications may arise if the cosmological parameters are allowed to simultaneously vary with the ionization history, due in part to the increased distance in parameter space from the fiducial model \citep[e.g.][]{PhysRevD.107.123506}. However, we note that \citet{Bernal:2020vbb} likewise found no bias if the true $\omega_{\rm cdm}$ and $\omega_{\rm b}$ are significantly different from the fiducial value used in the template in the range $.09 \leq \omega_{\rm cdm} \leq .16$ and $.015 \leq \omega_{\rm b} \leq .03$, assuming \lcdm. This range is broader than what is compatible with CMB data within the \modrec\ model. We have left the full analysis, including estimating biases when simultaneously varying cosmological parameters, outside the scope of this work. If significant biases are found, an analysis using BAO determinations which mitigate these potential biases (for example, using geometric methods \citep{PhysRevD.101.083517, PhysRevD.99.123515}) might enable more robust conclusions.}

\changed{\subsection{Considerations for RSD measurements}}

\changed{Similar considerations arise for redshift-space distortion measurements, which are also obtained by fitting precomputed templates to observations. In particular, the overall normalization of the template is fixed by the parameters $b \sigma_8$ and $f \sigma_8$, where $b$ is the galaxy bias and $f$ is the linear growth rate. This is particularly relevant as the \modrec\ model, constrained by \planck\ data, allows a wider range of values for $\sigma_8$, with $\sigma_8 = 0.818^{+0.037}_{-0.033}$, compared to $\sigma_8 = 0.8111\pm 0.0060$ for \lcdm. This possibly leads to systematic error in the RSD measurements due to the dependence of these determinations on the fiducial cosmology used to compute the template, which can be very different than true cosmology if a modified recombination era is actually present.}

\changed{Despite this, we find that the $f \sigma_8$ measurements have extremely little constraining power, so we do not carry out a full analysis of their robustness. If they are removed, we find negligible shifts in the cosmological parameters, and no difference in their covariance. The largest shift is for $H_0$, which changes by $-1.7\%$, and all other shift by less than $1\%$. If future RSD measurements become more constraining, a full analysis of their robustness in a modified recombination scenario may be necessary.}

\end{document}